\documentclass[12pt,preprint]{aastex}
\def\gs{\mathrel{\raise0.35ex\hbox{$\scriptstyle >$}\kern-0.6em
\lower0.40ex\hbox{{$\scriptstyle \sim$}}}}
\def\ls{\mathrel{\raise0.35ex\hbox{$\scriptstyle <$}\kern-0.6em
\lower0.40ex\hbox{{$\scriptstyle \sim$}}}}

\shorttitle{CO lines of IR-luminous galaxies and AGN}
\shortauthors{Papadopoulos,  van der Werf, Isaak, \& Xilouris}

\begin{document}

\title{CO Spectral Line Energy Distributions of IR-luminous galaxies and AGN}

\author{Padeli \ P.\ Papadopoulos}
\affil{Argelander-Institut f\"ur Astronomie,  Auf dem H\"ugel 71,  D-53121 Bonn,  Germany}
\email{padeli@astro.uni-bonn.de}

\author{Paul van der Werf}
\affil{Leiden Observatory, Leiden University, P.O. Box 9513, 2300 RA Leiden, The Netherlands}
\email{pvdwerf@strw.leidenuniv.nl}

\author{Kate Isaak}
\affil{School of Physics and Astronomy, University of Wales, Cardiff, CF24 3YB, UK}
\email{Kate.Isaak@astro.cf.ac.uk}

\and

\author{Emmanuel M. Xilouris}
\affil{Institute of Astronomy and Astrophysics, National Observatory of Athens, P. Penteli, 15236 Athens, Greece}
\email{xilouris@astro.noa.gr}

\begin{abstract}

We  report on  new sensitive  CO J=6--5  line observations  of several
luminous  infrared Galaxies  (LIRGs: L$_{\rm  IR}$(8--1000$\mu $m$)\ga
$10$^{11}$\,L$_{\odot}$),   36\%  (8/22)   of  them   ULIRGs  (L$_{\rm
IR}$$>$10$^{12}$\,L$_{\odot}$),  and  two   powerful  local  AGN:  the
optically  luminous QSO  PG\,1119+120, and  the powerful  radio galaxy
3C\,293 using the James Clerk Maxwell Telescope (JCMT) on Mauna Kea in
Hawaii.  We combine these observations with existing low-J CO data and
dust emission Spectral Energy Distributions (SEDs) in the far-infrared
- submillimetre from the literature to constrain the properties of the
star-forming  ISM in  these systems.   We  then build  the first  {\it
local}  CO Spectral  Line Energy  Distributions (SLEDs)  for  the {\it
global}  molecular gas  reservoirs  that reach  up  to high  J-levels.
These CO  SLEDs are  neither biased by  strong lensing  (which affects
many of those constructed for high-redshift galaxies), nor suffer from
undersampling  of  CO-bright  regions   (as  most  current  high-J  CO
observations  of   nearby  extended  systems  do).   We   find:  1)  a
significant influence of  dust optical depths on the  high-J CO lines,
suppressing the J=6--5  line emission in some of  the most IR-luminous
LIRGs, 2)  low global CO  line excitation possible even  in vigorously
star-forming  systems,   3)  the  first  case   of  a  shocked-powered
high-excitation CO SLED  in the radio galaxy 3C\,293  where a powerful
jet-ISM interaction  occurs, and 4) unusually highly  excitated gas in
the  optically powerful  QSO PG\,1119+120.   In Arp\,220  and possibly
other (U)LIRGs  very faint CO J=6--5  lines can be  attributed to {\it
significant dust optical depths  at short submm wavelengths} immersing
those lines  in a  strong dust continuum,  and also causing  the C$^+$
line  luminosity deficit  often observed  in such  extreme starbursts.
Re-analysis of the CO line ratios available for submillimeter galaxies
(SMGs) suggests  that similar  dust opacities may  be also  present in
these high-redshift starbursts, with genuinely low-excitation of large
amounts of {\it SF-quiescent} gas the only other possibility for their
often  low  CO  (high-J)/(low-J)  line  ratios.   We  then  present  a
statistical   method  of  separating   these  two   almost  degenerate
possibilities,  and  show  that  high  dust optical  depths  at  submm
wavelengths  can impede  the  diagnostic potential  of submm/IR  lines
(e.g.  starburst  versus AGN  as gas excitation  agents), which  is of
particular importance  for the  upcoming observations of  the Herschel
Space Observatory and the era of~ALMA.

\end{abstract}


\section{Introduction}

Obtaining unbiased Spectral Energy  Distributions (SEDs) for local and
distant galaxy  populations provides a  crucial yardstick by  which to
compare their properties and  eventually relate such populations along
evolutionary    paths    within    any    given    galaxy    evolution
framework.  Unbiased  Spectral  Line Energy  Distributions  (hereafter
SLEDs) of  the rotational  transitions of molecules  such as  CO, HCN,
HCO$^+$ are of particular importance since:

\begin{itemize}

\item they can trace the mass distribution of the star formation fuel,
      the molecular  gas, across its considerable  range of properties
      ($\rm  n$$\sim   $($10^2$--$10^7$)\,cm$^{-3}$,  $\rm  T_k$$\sim  $
      (10--200)\,K)

\item   relative   molecular   line   strengths   are   in   principle
      extinction-free   probes  of   molecular   gas  properties   and
      AGN-versus-starburst  as  power sources  of  IR luminosities  of
      galaxies (Meijerink \& Spaans 2005; Meijerink, Spaans, \& Israel
      2006)

\item  imaging  their emission  distribution  and  velocity fields  at
      mm/submm  wavelengths yields  unique dynamical  mass  probes of
      deeply  dust-enshrouded  star-forming   galaxies  and  QSO  host
      galaxies across the Universe (e.g.  Walter et al. 2004; Greve et
      al. 2005; Tacconi et al. 2006)

\item cooling  and thus the  thermodynamic state of the  molecular gas
      (heated  by stellar  far-UV  light, cosmic  rays, and  turbulent
      motions)  is regulated by  [C\,II], [O\,II]  and high-J  CO line
      emission, which in turn may efficiently regulate the local Jeans
      mass and the stellar IMF (Elmegreen, Klessen, \& Wilson 2008).

\end{itemize}

\noindent
Much of the  star formation in the distant  Universe occurs in heavily
dust-enshrouded  IR-luminous systems  (e.g.  Smail,  Ivison,  \& Blain
1997),  challenging to  image  at optical  wavelengths  even with  the
current 8-10 meter-class  telescopes. Thus it may well  be that in the
upcoming era of ALMA molecular  lines will replace optical and even IR
lines as the  most potent probes of galaxy structure  and of the power
sources that  drive galaxy evolution across cosmic  epoch.  The recent
discovery  of  significant dust  optical  depths  {\it  even at  submm
wavelengths} in Arp\,220 (a local prototype of dust-enshrouded extreme
starbursts)  by Sakamoto et  al.  2008  further highlights  how deeply
obscured such extreme star-forming systems can~be.

Local templates  are key in  understanding not only the  properties of
local galaxy populations,  but also of those in  the distant Universe.
To this end we have undertaken  a large multi-J CO and HCN line survey
of  local   Luminous  Infrared  Galaxies   (LIRGs)  --  dust-obscured,
star-forming   systems  with  SFR$\sim   $(10--100)\,$\rm  M_{\odot}\,
yr^{-1}$  (eg.   Sanders  \&  Ishida  2004), for  which  the  dominant
fraction of  the bolometric  luminosity is in  the rest-frame  IR with
$\rm  L_{IR}$(8-1000)\,$\mu  $m)$\ga  $$\rm 10^{11}\,L_{\odot}$.   The
prodigious star  formation events in  LIRGs/ULIRGs, most often  due to
dissipative galaxy interactions/mergers  of gas-rich progenitors, make
them  the best local  analogues of  the submillimeter  galaxies (SMGs)
(Tacconi et al.  2006;  Iono et al.  2009), dust-enshrouded starbursts
at  high redshifts  with even  higher star  formation rates  ($\rm SFR
\sim$$ \rm  10^3\,M_{\odot}\,yr^{-1}$) and the sites  of a significant
part  of cosmic  star formation  history (e.g.   Hughes et  al.  1998;
LeFloch et  al.  2009).   Finally we observed  the hosts of  two local
AGN: the powerful radio galaxy 3C\,293, and the optically luminous but
radio  quiet  QSO  PG\,1119+120, as  part  of  a  pilot study  of  the
molecular  gas excitation in  the presence  of a  bona fide  AGN.  The
latter  remain as  the  most  effective beacons  of  the most  distant
galaxies  where molecular  lines  have been  detected  (Walter et  al.
2004),  and can  be the  cause  of distinct  molecular gas  excitation
conditions   in  high   redshift  quasars   (Schleicher,   Spaans,  \&
Klessen~2010).

In this paper  we report on sensitive CO  J=6--5 line measurements for
 galaxies in our  survey.  This allows: a) a  first systematic glimpse
 of their high-excitation molecular gas phase, b) constraints on their
 CO SLEDs, and c)  direct comparisons with starburst and AGN-dominated
 systems at  high redshifts where  predominantly only high-J  CO lines
 are  currently available (e.g.   Solomon \&  Vanden Bout  2005; Omont
 2007).   Throughout this work  we adopt  a flat  $\Lambda $-dominated
 cosmology    with    $\rm    H_0=71\,$km\,s$^{-1}$\,Mpc$^{-1}$    and
 $\Omega_{\rm m}=0.27$, and  calculate luminosity distances using the
 NED calculator developed by Wright (2006).

\section{Sample  selection, observations and data reduction}

Our  sample  of   LIRGs  where  $\rm  L_{IR}$=L(8--1000\,$\mu  $m)$\ga
$10$^{11}$\,L$_{\odot}$  as estimated  from  all 4  {\it IRAS}  bands,
(e.g.  Sanders \& Mirabel 1996), of which 36\% (8/22) are ULIRGs ($\rm
L_{IR}$$>$10$^{12}$\,L$_{\odot}$), is drawn from two CO J=1--0 surveys
of LIRGs by Sanders et al.  1991, and Solomon et al.  1997 (themselves
drawn from the {\it IRAS} BGS flux-limited sample with $\rm f_{60\,\mu
m}$$>$5.24\,Jy; Soifer et al.  1987,  1989; Sanders et al.  2003).  We
then  imposed two  additional  criteria, namely:  a)  z$\la $0.1  (the
maximum redshift  for which the JCMT  B-band receivers can  tune to CO
J=3--2),  and  b) compact  CO-emitting  regions  (sizes obtained  from
interferometric images in Sanders et al.  1988; Planesas et al.  1991;
Wang et  al.  1991; Downes \&  Solomon 1998; Bryant  \& Scoville 1999;
Evans et al.  2000, 2001, 2002) so that one or two telescope pointings
can record the total line flux up to the highest line frequency of the
survey (i.e.  690 GHz of the CO J=6--5 line).  The availability of the
CO  J=1--0   line  flux   ($\rm  E_{10}/k_B$$\sim  $5.5\,K   and  $\rm
n_{crit}$$\sim  $410\,cm$^{-3}$)  for  all  selected LIRGs  allows  an
inventory of  {\it all} the CO-rich molecular  gas phases irrespective
of their average  excitation state, and is thus  a prerequisite if the
fraction of  the star-forming  dense and warm  molecular gas is  to be
determined. The latter is constrained from high-J CO line luminosities
and  reliable global  CO (high-J)/(1--0)  line ratios  (and  hence our
second criterion for source compactness).

Finally, along with the aforementioned LIRG sample, two local powerful
AGN were also included for multi-J CO observations as a pilot study of
the possible effects of the  AGN on the global molecular gas reservoir
of their hosts, and to  guide future efforts towards a large molecular
line  survey of AGN-selected  rather than  starburst-selected systems.
The two AGN chosen are the powerful F-R\,II radio galaxy 3C\,293 whose
powerful jets strongly interact with the ambient neutral gas (Morganti
et  al.   2003),  and  the  optically luminous  but  radio  quiet  QSO
PG\,1119+120.   Both have  single dish  and interferometric  CO J=1--0
line measurements that find  its emission concentrated in regions with
$\rm  \theta  _{CO}$$\la  $5$''$--7$''$  around their  active  nucleus
(Evans et  al.  1999,  2001, 2005), and  thus satisfy  our compactness
criterion.

\subsection{Observations and data reduction}

We conducted the CO J=6--5 measurements  as part of our multi-J CO and
HCN line survey  of LIRGs in the local Universe.   The first CO J=6--5
(691.473\,GHz)  line   measurements  with  the   James  Clerk  Maxwell
Telescope  (JCMT)\footnote{The   James  Clerk  Maxwell   Telescope  is
operated by  The Joint Astronomy Centre  on behalf of  the Science and
Technology Facilities  Council of the United  Kingdom, the Netherlands
Organisation  for  Scientific  Research,  and  the  National  Research
Council of  Canada.}  atop Mauna  Kea (Hawaii) were conducted  for the
luminous ULIRG/QSO Mrk\,231  and the LIRG Arp\,193, using  the old W/D
band  (620--710\,GHz)  receiver  (operating  in  SSB mode)  on  20  of
February and 22 of April  2005 respectively, under very dry conditions
($\rm \tau _{220\,GHz}$$\la  $0.035).  The typical system temperatures
were   $\rm  T_{sys}$$\sim  $(3700--5500)\,K   (including  atmospheric
absorption).   The DAS  spectrometer was  used at  its widest  mode of
1.8\,GHz ($\rm  \sim $780\,km\,s$^{-1}$  at 690\,GHz), and  rapid beam
switching at $\rm \nu _{chop}$=2\,Hz  with azimuthal throws of $ 60''$
resulted in remarkably  flat baselines.  The beam size  at 691\,GHz is
$\rm \Theta  _{HPBW}$=8$''$. Good pointing  with such narrow  beams is
crucial and was checked every 45--60\,mins using differential pointing
with receiver B3(350\,GHz).  The latter technique can access many more
compact sources  in the sky in  order to conduct  pointing checks than
direct  pointing with  the W/D  receiver,  and was  found accurate  to
within $\rm  \sigma _r$$\sim $2.6$''$ (rms).   The aperture efficiency
for these periods was  found to be $\rm \eta^* _a$=0.25.\footnote{This
is higher  than the value adopted  by Papadopoulos et al  2007, and we
deduced it  from a  larger number of  planetary measurements  for that
period.}

The main set  of CO J=6--5 observations were  conducted during several
periods in January, February, March and May 2009 with the upgraded W/D
receiver and its  new SIS mixers (effectively the  same type that will
equip the ALMA telescopes  at this waveband) that dramatically enhance
its performance.  The resulting low receiver temperatures allowed very
sensitive observations with  typical $\rm T_{sys}\sim $(1500--3000)\,K
(including  atmospheric transmission)  for  $\rm \tau_{220\,GHz}$$\sim
$0.035--0.06.   Dual  channel  operation  (two  polarization  channels
aligned within  $\la $1$''$) further  enhanced the W/D  band observing
capability of  the JCMT.  The new  ACSIS spectrometer was  used at its
widest mode of 1.8\,GHz ($\sim $780\,km\,s$^{-1}$ at 690\,GHz), and in
a few cases  (e.g. Arp\,220) two separate tunnings  per object with an
effective  bandwidth of  $\sim  $3.2\,GHz ($\sim  $1390\,km\,s$^{-1}$)
ensured   adequate   coverage   of   CO   lines   with   FWZI$\sim   $
800--1000\,km\,s$^{-1}$.     Rapid     beam    switching    at    $\rm
\nu_{chop}$=4\,Hz (continuum mode) with a throw of $30''$ (in azimuth)
yielded flat baselines under most circumstances.  Finally the pointing
was checked  every 45--60\,mins by observing compact  sources with W/D
receiver  itself, and  differential  pointing with  the A\,3  receiver
(230\,GHz),  yielding rms residual  error radius  of $\rm  \sigma _r$$
\sim  $$2.2''$  (see Figure~1).   The  aperture  efficiency for  these
periods is $\rm \eta^* _a$=0.32, and a flux calibration uncertainty of
$\sim  $25\%  (encompassing  $\rm  \eta^* _a$  variations  and  3-load
calibration uncertainties) was  estimated from observations of several
strong spectral line standards at $\sim $690\,GHz.

A final  set of CO  J=6--5 observations were obtained  recently during
several days  between 12 and 23  of January, and 5-6  of February 2010
under  dry weather  conditions ($\rm  \tau _{220GHz}$$\la  $0.05) that
yielded   typical   system    temperatures   of   $\rm   T_{sys}$$\sim
$(1400--2800)\,K. The spectrometer setup was identical to that used in
our 2009 observations with two  separate tunnings used to cover a wide
line in  a few objects  (e.g.  Mrk\,273, NGC\,6240).   Observations of
Mars yielded an aperture efficiency of $\rm \eta^* _a$=0.35, while the
overal calibration uncertainty remained  within $\sim $25\%.  The same
beam switching  scheme was employed  while pointing was  checked every
hour and its accuracy remained within the range shown in Figure~1.

\subsection{Line intensity estimates}

 We inspected each 10-min  spectrum for baseline ripples and intensity
``spikes'' in individual channels.   The edited CO J=6--5 spectra were
then co-added  and are  shown in Figure~2,  overlaid (mostly)  with CO
J=3--2 lines.   The velocity-integrated line flux  densities were then
estimated from  those spectra, after subtraction  of linear baselines,
using

\begin{equation}
\rm \rm S_{line}=\int _{\Delta V} S_{\nu } dV = \frac{8 k_B}{\eta ^{*}
_a  \pi  D^2}  K_c  (x)\int   _{\Delta  V}  T^*  _A  dV=  \frac{\Gamma
(Jy/K)}{\eta ^{*} _a} K_c(x) \int _{\Delta V} T^* _A dV,
\end{equation}

\noindent
where $\rm  \Gamma _{JCMT}$=15.62 and  $\eta ^{*} _a$ is  the aperture
efficiency.   The factor  $\rm K_c  (x)$=$\rm  x^2/(1-e^{-x^2})$, with
$\rm  x=\theta _{co}/(1.2\theta  _{HPBW})$ and  $\theta _{co}$=(source
diameter) accounts for the geometric  coupling of a gaussian main beam
to  a  finite-sized  disk-like  source  when  $\rm  \theta  _{co}$  is
available.

\subsubsection{Observing compact sources with narrow beams: a bias}

 For point-like sources, even  with the accurate tracking and pointing
achievable by enclosed  telescopes such as the JCMT,  the residual rms
pointing errors and  the narrow beams of large  mm/submm telescopes at
high frequencies can lead to a substantial systematic reduction of the
measured fluxes  of compact sources.   In Papadopoulos et  al. (2009)
this is shown to be

\begin{equation}
\rm G(\sigma_r)=1+8\,ln2\left(\frac{\sigma_r}{\sqrt{2}\,
\Theta _{HPBW}}\right)^2,
\end{equation}

\noindent
(see also Condon 2001) where  $\sigma _r/\sqrt{2}$ is the rms pointing
error   per   direction  (for   identical   error  distributions   per
coordinate).  Thus in Equation 1,  when $\rm \theta _{co} \leq 2\sigma
_r$,  $\rm  K_c(x)\rightarrow   G(\sigma  _r)$,  since  the  geometric
coupling  correction   is  then   overtaken  by  the   pointing  error
correction.

\subsection{New observations of Arp\,220: a revised CO J=6--5 line flux}

The archetypal starburst Arp\,220  has been reobserved during our 2010
observing  campaign, and  a  significantly larger  ($\sim $2.6  times)
integrated line  flux has been  measured, along with a  different line
profile.  The latter now agrees  well with those from high-density gas
tracers such  as HCN  J=3--2 (see Figure  3) which places  the highest
density gas  at the high velocity component  corresponding to emission
from  the  eastern nucleus  (see  Greve  et  al. 2009  and  references
therein).  We  thus conclude that  the first CO J=6--5  measurement of
Arp\,220  reported  by  Papadopoulos  et  al.  2010  has  likely  been
affected by  a pointing  offset of the  order of $\rm  \sigma _r$$\sim
$4.7$''$ ($\sim $3.3$''$ per direction). This is clearly an outlier of
the  distributions shown  in Figure  1 but  is certainly  possible and
accounts  also for  the  different  CO J=6--5  line  profile shown  in
Papadopoulos  et al.  2010  ($\rm \sigma  _{r}$$\sim $4.7$''$  in this
system of two  CO-bright nuclei $\sim$1$''$ apart can  yield a maximum
apparent  emission asymmetry  of S(V$_1$)/S(V$_2$)$\sim  $0.54 between
two intrinsically equal emission contributions at the velocity centers
V$_1$ and V$_2$ of the  two nuclei). Finally our fast-switching scheme
allows an independent estimate of the adjacent dust continuum from the
line-free part  of the  spectrum, which we  find to be  $\rm S_{434\mu
m}$=$(6.0\pm  1.5)$\,Jy, in  good agreement  with the  value  given by
Dunne \& Eales 2001 rather  than that in the JCMT calibrator database.
The ramifications from this revision are discussed in section 3.2.

\subsection{CO J=6--5 line and IR continuum luminosities}

 The  sources, their  coordinates and  CO J=6--5  line  flux densities
(using Equation 1)  are tabulated in Table 1.   The corresponding line
luminosities are  estimated~using

\begin{equation}
\rm  L^{'} _{co}  =  \int  _{\Delta  V} \int_{A_s}  \Delta T_{b}\,da\,dV  =
\frac{c^2}{2 k_B \nu  ^2 _{co,rest}} \left(\frac{D^2 _L}{1+z}\right)
\int _{\Delta V} S_{\nu}\,dV,
\end{equation}

\noindent
where $\rm  \Delta T_{b}$ is the  continuum-subtracted line brightness
 temperature emerging  at the  source frame, $\rm  \Delta V$  and $\rm
 A_s$ are the total linewidth and line-emitting source area, $\rm D_L$
 is  the  luminosity  distance,  and  $\rm \nu  _{co,rest}$  the  line
 rest-frame frequency.   Substituting and converting  to astrophysical
 units yields

\begin{equation}
\rm L^{'} _{co} = 3.25\times 10^7\,(1+z)^{-1} 
\left(\frac{\nu _{co,rest}}{GHz}\right)^{-2} \left(\frac{D_L}{Mpc}\right)^2
 \left(\frac{\int _{\Delta V} S_{\nu}\,dV}{Jy\,km\,s^{-1}}\right)\, K\, km\,s^{-1}\,pc^2,
\end{equation}

\noindent
while the  conversion to ordinary luminosity  units ($\rm L_{\odot}$),
used for the CO SLEDs, is

\begin{equation}
\rm L_{co}=\frac{8\pi k_B\nu ^3 _{co,rest}}{c^3}\,L^{'} _{co}
=3.18\times 10^4\left(\frac{\nu _{co,rest}}{100\,GHz}\right)^3
\left[\frac{L^{'} _{co}}{10^{9}\,K\,km\,s^{-1}\,pc^2}\right]L_{\odot}.
\end{equation}

\noindent
In Table~2  we tabulate the total  IR luminosity L$_{\rm  IR}$ of each
system, with  any contribution from a warm  AGN-heated dust subtracted
when possible.  In that table  we also list  the IR luminosity  due to
star formation  L$^{\rm (SF)} _{\rm  IR}$ (see section~4), as  well as
the $\rm R_{HCN/CO}$=$\rm L^{'} _{HCN(1-0)}/L^{'} _{CO(1-0)}$ and $\rm
R_{65/32}$=$\rm L^{'}  _{CO(6-5)}/L^{'} _{CO(3-2)}$ line  ratios.  The
latter is  currently  the most  widely  available one for  conducting
comparisons between the excitation conditions in our local LIRG sample
and   high-z  starbursts,   though   for  some   high-z  systems   the
``neighboring''  J=7--6  transition  rather  than J=6--5  is  the  one
available (Tacconi et al.  2006).

\section{Molecular gas  in LIRGs: the star forming phase}

The star formation  rates of LIRGs are fueled  by molecular gas masses
of  $\sim  $  (10$^{9}$--10$^{10}$)\,$\rm  M_{\odot}$ (Tinney  et  al.
1990;  Sanders, Scoville  \& Soifer  1991; Solomon  et al.   1997), at
efficiencies reaching up to what  could be a maximum possible value of
SFE=$\rm  L_{IR}/M(H_2)$$\sim $500\,$\rm L_{\odot}/M_{\odot}$,  set by
the radiative  feedback of  massive stars on  the dust mixed  with the
molecular gas accreted by the star forming sites (Scoville 2004). Such
high SFEs  are one the many  indicators of the  extreme ISM conditions
found  in such  systems, with  several studies  (e.g.  Solomon  et al.
1992;  Aalto et  al.  1995;  Yao  et al.   2003; Greve  et al.   2009)
revealing also  a qualitatively different  molecular ISM than  that of
quiescent  gas-rich  systems.   In  LIRGs a  diffuse  ($\rm  n(H_2)\la
10^3\,cm^{-3}$) non  self-gravitating phase (a likely  product of high
gas  pressures  inducing   an  intercloud  HI$\rightarrow$H$_2$  phase
transition,  and/or tidal  disruption of  molecular clouds  in merging
systems) dominates the CO  J=1--0, 2--1 line emission.  Nevertheless a
dense phase  with $\rm n(H_2)$$\ga$10$^{4}$\,cm$^{-3}$  often contains
most of  the molecular  gas mass present,  especially in  ULIRGs (e.g.
Solomon  et al.   1992;  Gao  \& Solomon  2004),  dominating their  CO
J+1$\rightarrow  $J, J+1$\geq $3  and HCN  line emission.   The latter
phase is  warm, $\rm T_{kin}$$\sim  $(40--100)\,K, a result  of strong
far-UV  photoelectric  and/or turbulent  gas  heating,  its high-J  CO
transitions will  thus be very  luminous and expected to  be excellent
unobscured  ``markers''   of  dust-enshrouded  starburst   regions  of
galaxies over most cosmic epochs (e.g.  Walter \& Carilli~2008).

\subsection{CO J=6--5 in LIRGs: evidence for dust-affected CO lines}

In Figure 4  the $\rm R_{65/32}$ ratio is  plotted versus $\rm L_{IR}$
 for all the  LIRGs in Table~1, and any local  or distant galaxies for
 which this or  a similar (high-J)/(low-J) {\it global}  CO line ratio
 is  available.   The lightly-shaded  area  marks  the  range of  $\rm
 R_{65/32}$  values expected  for the  dense and  warm gas  phase that
 fuels  star  formation  in  starbursts, determined  using  our  Large
 Velocity Gradient  (LVG) molecular line radiative  transfer code for:
 $\rm  \langle  n(H_2)\rangle  $=(10$^4$--10$^6$)\,cm$^{-3}$ and  $\rm
 T_{kin}$=(30--100)\,K,   which    encompasses   the   conditions   of
 star-forming  molecular  gas.   In  this determination  of  the  $\rm
 R_{65/32}$   range  we   consider  only   self-gravitating   gas,  as
 appropriate for any star-forming gas phase, where

\begin{equation}
\rm K_{vir}=\frac{\left(dV/dr\right)_{obs}}{\left(dV/dr\right)_{virial}}\sim 
1.54\frac{[CO/H_2]}{\sqrt{\alpha}\Lambda _{co}}\left[\frac{\langle n(H_2)\rangle}{10^3\,
 cm^{-3}}\right]^{-1/2}\sim 1,
\end{equation}

\noindent
 (e.g.  Goldsmith  2001;  Greve  et  al.    2009).   Values  of  $\rm
 K_{vir}$$\gg$1  correspond  to unbound  gas  motions ($\alpha  $$\sim
 $1-2.5  depending  on  the   assumed  cloud  density  profile).   The
 parameter $\rm  \Lambda _{co}$=$\rm[CO/H_2]/(dV/dr)$ along  with $\rm
 T_k$ and $\rm \langle n(H_2)\rangle  $ classify the LVG solutions (we
 assume a Galactic abundance $\rm [CO/H_2]$=10$^{-4}$).

% For n=10^4 cm^{-3}:
% R=0.43: Tk=30K, Lco=0.10000E-04(km/s/pc) **Kvir=4.87/a^0.5**
% R=0.61: Tk=30K, Lco=0.31623E-04(km/s/pc) **Kvir=1.54/a^0.5**
% R=0.69: Tk=30K, Lco=0.10000E-03(km/s/pc) **Kvir=0.48/a^0.5**
%--------
% R=0.55: Tk=40K, Lco=0.10000E-04(km/s/pc) **Kvir=4.87/a^0.5**
% R=0.71: Tk=40K, Lco=0.31623E-04(km/s/pc) **Kvir=1.54/a^0.5**
% R=0.77: Tk=40K, Lco=0.10000E-03(km/s/pc) **Kvir=0.48/a^0.5**
%--------
% R=0.62: Tk=50K, Lco=0.10000E-04(km/s/pc) **Kvir=4.87/a^0.5**
% R=0.76: Tk=50K, Lco=0.31623E-04(km/s/pc) **Kvir=1.54/a^0.5**
% R=0.81: Tk=50K, Lco=0.10000E-03(km/s/pc) **Kvir=0.48/a^0.5**
%--------
% R=0.72: Tk=100K,Lco=0.10000E-04(km/s/pc) **Kvir=4.87/a^0.5**
% R=0.84: Tk=100K,Lco=0.31623E-04(km/s/pc) **Kvir=1.54/a^0.5**
% R=0.89: Tk=100K,Lco=0.10000E-03(km/s/pc) **Kvir=0.48/a^0.5**
%------------------------------------------------------------
% For n=10^6 cm^{-3}: 
% R=0.73: Tk=30K, Lco=0.10000E-05(km/s/pc) **Kvir=4.87/a^0.5**
% R=0.73: Tk=30K, Lco=0.31623E-05(km/s/pc) **Kvir=1.54/a^0.5**
% R=0.73: Tk=30K, Lco=0.10000E-04(km/s/pc) **Kvir=0.48/a^0.5**
%--------
% R=0.92: Tk=100K,Lco=0.10000E-05(km/s/pc) **Kvir=4.87/a^0.5**
% R=0.92: Tk=100K,Lco=0.31623E-05(km/s/pc) **Kvir=1.54/a^0.5**
% R=0.92: Tk=100K,Lco=0.10000E-04(km/s/pc) **Kvir=0.48/a^0.5**

From Figure 4  it is clear that for several  LIRGs the $\rm R_{65/32}$
  ratio  {\it falls  well below  the range  expected  for star-forming
  gas,} and this happens in some of the most extreme starbursts in the
  local (e.g.   Arp\,220: $\rm  R_{65/32}$$\sim $0.2) or  the distant
  (e.g.  HR\,10:  $\rm R_{65/32}\sim$0.053--0.08) Universe\footnote{If
  a  ``neighboring'' CO  line  ratio rather  than  $\rm R_{65/32}$  is
  avalaible for a high-z object, we obtain a range for $\rm R_{65/32}$
  by  using  the available  one  to  obtain  LVG solutions  over  $\rm
  T_k$=30--110\,K,                     $\rm                    \langle
  n(H_2)\rangle$=$10^2$--$10^7$\,cm$^{-3}$,      and     with     $\rm
  K_{vir}$$\sim $1.  For HR\,10: $\rm R_{54/21}$=0.16$\rightarrow
  $$\rm  R_{65/32}\sim$0.053--0.08)}.  On  the other  hand  all nearby
   galaxies clearly  seggregate into two groups  with the star-forming
   ones  exhibiting consistently  high $\rm  R_{65/32}$  ratios ($\sim
   $0.8--0.9), whilst the SF-quiescent  ones have much lower CO ratios
   ($\sim $0.12--0.24)  reflecting a much less excited  state of their
   molecular gas.  Thus  in some local LIRGs as well  as some of their
   high-z counterparts, $ R_{65/32}$ {\it  can be as low or even lower
   than  in  SF-quiescent systems.}   This  contradiction between  the
   excitation  conditions of  the star-forming  molecular gas  and the
   emergent  faint  CO  J=6--5   line  emission  (and  thus  low  $\rm
   R_{65/32}$)  found  in some  LIRGs  becomes  even  more acute  when
   additional  information  from other  transitions  tracing the  same
   phase as CO  J=6--5 (e.g.  those of HCN) is  used to further narrow
   the expected  range of $\rm R_{65/32}$.  The  (U)LIRGs Arp\,220 and
   NGC\,6240 are  excellent such examples, with a  multi-J line survey
   of several  heavy rotor molecules available to  constraint the $\rm
   R_{65/32}$ ratio (see section 3.2).  For all other local LIRGs much
   less such  information is available (and  is virtually non-existent
   for their high redshift  counterparts), with only two small surveys
   of HCN  J=1--0, 3--2  (Graci\'a-Carpio et al.   2008; Krips  et al.
   2008) currently providing the most uniform line datasets probing the
   star-forming   phase   locally.   These   yield   an  average   HCN
   (3--2)/(1--0)   brightness  temperature   ratio  of   $\rm  \langle
   R_{32/10}(HCN)\rangle$$ \sim $0.55,  reaching up to $\sim $0.8--0.9
   for the dense  gas found in starbursts such as  M\,82 (Krips et al.
   2008) and Arp\,220, NGC\,6240 (Greve  et al.  2009).  Adopting $\rm
   R_{32/10}(HCN)$=0.55  and $\rm  K_{vir}(HCN)$$\sim  $1 (implemented
   for an assumed  $\rm [HCN/H_2]$=2$\times $10$^{-8}$) as constraints
   on our  LVG code yields $\rm T_{kin}$=(35--65)\,K  and $\rm \langle
   n(H_2)\rangle $  =(3$\times $10$^4$--10$^5$)\,cm$^{-3}$, conditions
   that are  indeed typical for  the dense star-forming  molecular gas
   (e.g.  G\"uesten et al.  1993;  Mao et al.  2000) and corresponding
   to  an  even narrower  range  of  $\rm R_{65/32}$$\sim  $0.70--0.85
   (cross-hatched area in Figure~4), excluding several more~LIRGs.

In the absence of dominant  low-excitation molecular gas mass in LIRGs
(a real possibility which is discussed in section 5.1), non-negligible
dust optical depths  at short submm wavelengths provide  {\it the only
other  mechanism}   that  can  suppress  the  large   CO  J=6--5  line
luminosities expected from their star-forming molecular gas.  This has
been recently demonstrated in  the archetypal ULIRG\,Arp220, where its
average dust optical depth of $\rm \tau _{860\mu m}$$\sim $1 (Sakamoto
et al.  2008)  has been shown to be responsible for  its very faint CO
J=6--5 line (Papadopoulos et al.  2010).  Our current results indicate
that such ISM conditions  may occur often in extreme starburst~systems
throughout the Universe.

\subsection{The special cases of Arp\,220 and NGC\,6240}

A  recent  multi-J  HCN, HCO$^+$  and  CS  line  survey of  these  two
archetypal ULIRGs by Greve et  al. 2009 allows placing their CO J=6--5
line  luminosities in  the best  context currently  possible  for such
systems.  Moreover  the significantly  higher CO J=6--5  luminosity we
measured for Arp\,220 than that  reported by Papadopoulos et al.  2010
necessitates  a revisit  of the  analysis  for this  ULIRG.  From  the
findings of  the aforementioned line survey it  is immediately obvious
that for  the dense and warm  gas phase of  $\rm \langle n(H_2)\rangle
$$\sim $(10$^5$--10$^6$)\,cm$^{-3}$ and $\rm T_{k}$$\sim $(50--120)\,K
dominating   the   gaseous  ISM   in   these   two  starbursts:   $\rm
R_{65/32}$$\sim  $0.85--0.93.  This  confines it  well into  the upper
(cross-hatched) range of $\rm R_{65/32}$  values in Figure 4, and well
above  the observed  values for  both systems.   For Arp\,220  the new
higher  value of  $\rm  R_{65/32}$=$0.21\pm 0.06$  remains $\sim  $3-6
smaller  than that  expected for  its dense  gas phase,  and  thus the
evidence  for   significant  dust   optical  depths  at   short  submm
wavelengths quenching its CO J=6--5 line emission (Papadopoulos et al.
2010) remains strong.   Regarding the  dust emission itself,  even for
the larger value of $\rm  S_{434\mu m}$$\sim $6\,Jy and $\rm S_{860\mu
m}$=$(0.55\pm 0.082)$\,Jy  (Sakamoto et al.  2008),  the same analysis
as in  Papadopoulos et  al.  2010 yields  $\rm \tau  _{860\mu m}$$\sim
$0.27.  The  latter is  smaller than that  deduced by Sakamoto  et al.
2008,  but  remains  more  than  enough  (with  a  corresponding  $\rm
\tau_{434\mu m}$$\sim  $1) to  quench the CO  J=6--5 line  emission in
Arp\,220 by  immersing it into  a nearly blackbody continuum  at short
submm wavelengths.   For NGC\,6240  a similar discordance  between the
$\rm R_{65/32}$  expected for its  HCN/CS-bright gas and  its measured
$\rm R_{65/32}$=0.26 (Table 2) exists.

Given the  large impact that high  dust optical depths  at short submm
wavelengths will have  to a large number of  issues (see discussion in
4.3 and  5), it is  worth asking whether  a large ``boost'' of  the CO
J=3--2 rather  than a suppresion of  the CO J=6--5  line luminosity by
dust can  be responsible  for the low  $\rm R_{65/32}$ ratios  in some
ULIRGs.  This can happen only  if the diffuse non self-gravitating gas
phase  found in  these  systems contributes  substantially  to the  CO
J=3--2 emission  (but not  to CO J=6--5).   For Arp\,220  the detailed
study by Greve et.  al.  yields a diffuse phase with $\rm n(H_2)$$\sim
$(1--3)$\times  $10$^2$\,cm$^{-3}$, $\rm  T_{k}$$\ga  $40\,K and  $\rm
\Lambda          _{CO}$$\sim         $(1--3)$\times         $10$^{-5}$
(km\,s$^{-1}$\,pc$^{-1}$)$^{-1}$,  with   negligible  CO  J=3--2  line
emission. This is  also obvious from the fact that  even the CO J=2--1
transition appears globally subthermally excited in this system.  Thus
{\it a diffuse  gas phase cannot be responsible for  the very low $\rm
R_{65/32}$ value in Arp\,220.}  For NGC\,6240 this is less clear since
its     diffuse     phase:     $\rm    n(H_2)$$\sim     $(1--3)$\times
$10$^3$\,cm$^{-3}$,   $\rm   T_{k}$$\ga   $40\,K  and   $\rm   \Lambda
_{CO}$$\sim $(1--3)$\times $10$^{-6}$ (km\,s$^{-1}$\,pc$^{-1}$)$^{-1}$
can contribute  to CO  J=3--2 while having  negligible CO  J=6--5 line
luminosity. This demonstrates the ``degeneracy'' between dust-affected
and genuinely  low-excitation CO SLEDs,  and the need for  multi-J and
multi-species    molecular   line    observations    with   excitation
characteristics spanning  the entire  range of molecular  gas physical
conditions in  LIRGs.  For  an earlier such  study, conducted  for the
archetypal  ULIRG/QSO Mrk\,231,  that finds  luminous CO  J=6--5, 4--3
line emission from  the same gas phase as that  emitting HCN lines see
Papadopoulos, Isaak, \& van der Werf 2007.

\section{Global dust emission SEDs in LIRGs: fundamental limitations }

Dust emission SEDs could  in principle provide independent constraints
on  the average dust  optical depths  and thus  help settle  the issue
whether these  can affect molecular  line emission at short  submm and
far-IR  wavelengths.   Moreover,  particularly well-sampled  SEDs  are
available  for ULIRGs because  of their  early importance  in deciding
their dominant  power source, i.e.  AGN or starburst (e.g.   Haas 2001
and references  therein).  We start  by first assuming  optically thin
dust emission  SEDs with $\rm \tau  _{\lambda} \ll 1$  over the entire
$\rm \lambda $$ \sim$(10--1200)\,$\mu $m range, and a superposition of
three such SEDs

\begin{equation}
\rm S_{\nu } = \frac{1+z}{D^2 _L}\times \left[\kappa _d(\nu) M_{d,c}
\left(B_{\nu }(T_c)+ \frac{M_{d,SF}}{M_{d,c}}B_{\nu}(T_{SF})+
 \frac{M_{d,AGN}}{M_{d,c}}B_{\nu}(T_{AGN})\right)\right].
\end{equation}

\noindent
This characterizes:  a) an AGN-heated dust phase  ($\rm T_{AGN}$, $\rm
M_{d,AGN}$)  (estimated  so  that  its contribution  can  be  reliably
subtracted from the global SED),  b) a star formation (SF)-heated dust
($\rm T_{SF}$, $\rm  M_{d,SF}$), and c) a cold  cirrus-type dust ($\rm
T_c$,  $\rm M_{d,c}$).   In all  cases  the important  850$\mu $m  and
1.2\,mm fluxes  have also been  included (when available),  after been
corrected for CO J=3--2, 2--1 line contamination.  We also correct for
any non-thermal  radio continuum contributions, and  a dust emissivity
law  $\rm   \kappa  _d(\nu  )  =   (\nu/\nu  _{\circ})^{\beta}  \kappa
_d(\nu_{\circ})$ with $\beta  $=2 is adopted as the  most suitable for
{\it global} dust  emission SEDs of LIRGs (Dunne  \& Eales 2001).  The
SED fits  are then used to  estimate the total-AGN  ($\rm L_{IR}$) and
SF-related  ($\rm L^{(SF)}  _{IR}$)  dust IR  luminosities (Table  2).
When a lack  of long-wavelength data prevents constraints  on the cold
dust component,  the fits are  performed by setting  $\rm T_{c}$=22\,K
(the average value obtained from  all the other SEDs where enough data
allowed its estimate).   Dust masss and temperatures as  well as H$_2$
gas mass  (obtained from CO  J=1--0 data reported in  the literature),
and the corresponding gas/dust ratios  are listed in Table 3.  For all
$\rm M(H_2)$  estimates we used  a CO-H$_2$ conversion factor  of $\rm
X_{CO}$=$\rm       M(H_2)/L^{'}      _{CO(1-0)}$=1\,$\rm      M_{\odot
}\,(K\,km\,s^{-1}\,pc^2)^{-1}$ (Downes \& Solomon~1998).

\subsection{Dust in ULIRGs: cold or optically thick?}

From Table  3 it  becomes apparent that  cold dust mass  obtained from
typical multi-component SED  fits always contains most of  the mass in
LIRGs  with  $\rm   M_{d,c}/M_{d,SF}$$\sim  $5-100,  even  in  extreme
starbursts such as Arp\,220 and  Mrk\,231.  This is a direct result of
the far-IR/submm ``excess'' found in their SEDs once submm data became
available  (Lisenfeld et  al.  2000;  Dunne  \& Eales  2001), and  was
suspected even  when only IR data  were available because  of the high
gas/dust ratios  deduced with  respect to the  Milky Way  (Devereux \&
Young  1991).  Sensitive submm  imaging has  conclusively demonstrated
the presence of  massive and extended cold dust mass  in some LIRGs by
{\it spatially} separating  its emission from that of  a more luminous
SF-heated  dust in  nuclear  regions (e.g.   Papadopoulos \&  Seaquist
1999, Thomas et al.~2001).  From Figure  5 it would then seem that the
cold dust mass  dominates across the entire L$_{\rm  IR}$ range of our
sample and up to ULIRGs.  Nevertheless there are some serious problems
with this scenario, at least in extreme starburst systems, namely

\begin{itemize}

\item The bulk of the molecular gas mass in most ULIRGs is in a dense,
       warm, and presumably star-forming phase (i.e.  $\rm n(H_2)$$\ga
       $10$^5$\,cm$^{-3}$, $\rm  T_{dust}$$\ga $40\,K), and  thus most
       of  the concomitant  dust should  also  be warm  (at such  high
       densities $\rm T_k$$\sim $$\rm T_{dust}$).

\item The $\rm M(H_2)/M_{dust}$ ratio obtained from classical dust SED
       fits of such  objects can be very low:  $\sim $4--20 (Table 3),
       thus requiring most  of the gas in ULIRGs to  be atomic (with a
       presumably Galactic gas/dust ratio) so that the global gas/dust
       ratio can attain Galactic  values ($\sim $140--190, Sodroski et
       al.   1994).   However ULIRGs  are  found  to be  exceptionally
       H$_2$-rich with $\la  30\%$ of the gas in  atomic form (Mirabel
       \& Sanders 1989, for the $\rm X_{CO}$ factor used in our work).

\item Submm  imaging of  ULIRGs with the  SMA (Sakamoto et  al.  2008;
   Wilson et al.  2008) found most of the dust emission emanating from
   very compact  regions ($\la $100-150\,pc),  of hot dust  ($\rm \sim
   $100--140\,K), and with $\rm M(H_2)/M_{dust}$ close to the Galactic
   value.

\end{itemize}

These issues could be resolved if the far-IR/submm ``excess'' found in
 the dust SEDs of most LIRGs  is, in some of them, of different origin
 rather than cold  dust emission.  Indeed such ``excess''  can be also
 generated  by emission  that is  optically thick  at far-IR  and even
 short submm wavelengths. Recently this has been shown for the compact
 molecular  gas  disks  in  Arp\,220  (e.g.  Sakamoto  et  al.   2008,
 Matsushita  et al.   2009), and  is indeed  expected if  most  of the
 molecular gas in  ULIRGs is in a very dense  state.  For $\rm \langle
 n(H_2)\rangle  $$\sim  $(10$^5$--10$^6$)\,cm$^{-3}$  (e.g.   Solomon,
 Radford, \& Downes 1990; Greve et al.  2009), solar metalicities, and
 thickness  of $\rm  h$$\sim  $(40--60)\,pc (e.g.   Downes \&  Solomon
 1998) such molecular  gas disks, if  not very clumpy, will  have dust
 optical depths of

\begin{equation}
\rm \tau _{d}(\lambda) \sim 1.66\left(\frac{\lambda}{400\,\mu m}\right)^{-2}
\left[\frac{h \langle n(H_2)\rangle (cos\theta)^{-1}}{10^{25}\,cm^{-2}}\right]
\sim (2.05-30.7)\times \left(\frac{\lambda}{400\,\mu m}\right)^{-2} (cos\theta)^{-1},
\end{equation}

\noindent
($\theta  $ is  the inclination  angle, $\theta  $=0:  face-on).  Such
extreme gas  configurations are the  results of mergers  and/or strong
dynamical interactions  found in all ULIRGs (Sanders  \& Ishida 2004),
essentially packing  the entire molecular  gas supply of  two gas-rich
spirals within a few  hundred parsecs.  Their emergent IR luminosities
will then be less than proportional to their dust mass reservoirs, and
can be further absorbed by outer dust distributions that are optically
thick at  IR/far-IR wavelengths (Condon  et al.  1991; Solomon  et al.
1997).  For  such very dense gas  disks the dust  mass estimates using
submm  fluxes   and  temperatures   obtained  from  global   SED  fits
(e.g. Dunne at al.  2000) can be significant {\it overestimates} since
an outer  cooler dust distribution,  while not containing the  bulk of
the mass, dominates most of the observed SED (see also 4.2).  This may
explain why submm interferometric imaging can recover a Galactic value
for M(H$_2$)/M$_{\rm dust}$ in ULIRGs  (e.g. Wilson et al. 2008) while
global SED fits can yield  significantly lower ones (Table 3; Dunne \&
Eales  2001)\footnote{The M(H$_2$)/M$_{\rm  dust}$ ratios  deduced for
ULIRGs in Dunne  \& Eales 2001 must be divided by  5 if an appropriate
$\rm X_{CO}$ conversions  factor is used for such  systems.  Then from
their Table 10: M(H$_2$)/M$\rm _{dust}$(Arp\,220)$\sim 8$}.

% Only  high  resolution   interferometric  mm/submm  imaging  of  dust
%emission can see ``past''  such outer dust distributions, identify the
%hot ($\sim  $100--140\,K) compact dusty  staburst regions as  the true
%main contributors to  the dust submm fluxes of  ULIRGs (e.g.  Sakamoto
%et al.   2008), yielding the  correct, lower dust masses  and gas/dust
%ratios than those obtained using global SEDs.

\subsection{Optically thick dust emission: a new degeneracy}

Optically thick dust emission in ULIRGs at far-IR wavelengths has been
first  noticed by  Condon  et al.   1991  using interferometric  radio
continuum  imaging  to identify  the  true  sizes  of their  starburst
regions.  To  explore its effect  on emergent dust emission  we follow
Lisenfeld  et  al.  2000  and  replace  the  SF-heated and  cold  dust
components inside the brackets of Equation 7~with

\begin{equation}
\rm L_{\nu }=\kappa _d(\nu) B_{\nu }(T_{d,\tau})\times 
\left(\frac{1-e^{-\tau (\nu)}}{\tau(\nu )}\right) M_{d,\tau}
\end{equation}

\noindent
where $\rm \tau (\nu ) = \tau _{\circ} (\nu/\nu_{\circ})^2$.  In Table
3 we list the results of these dust SED fits, and in Figure 6 show the
distribution  of the  deduced optical  depths for  all the  LIRGs with
available submm  data from  the literature.  Substantial  dust optical
depths  are found  at 100$\mu  $m ($\sim  $4--21) for  most  LIRGs, in
accord with earlier results (e.g.   Solomon et al.  1997; Lisenfeld et
al.  2000).   Even NGC\,1068, a  vigorously star-forming LIRG  where a
massive     and    extended     cold    dust     reservoir    with$\rm
T_{d}$$\sim$(10--15)\,K is revealed via submm imaging (Papadopoulos \&
Seaquist  1999), can  have  its non-AGN  global  dust emission  fitted
equally well  with a single-temperature but optically  thick SED.  The
latter  would then erroneously  interpret its  cold dust  as optically
thick emission  with $\rm \tau  _{100\mu m}$$\sim $4.6.  Thus  the SED
fits  using Equations  7  and 9  are  degenerate, with  both cold  and
optically  thick dust  emission  capable of  producing a  far-IR/submm
emission ``excess''.  This  $\rm \tau_{IR}$-$\rm T_{d}$ degeneracy can
lead to  a misinterpretation of the  true state of the  dust in LIRGs,
and  cannot be  broken even  for well-sampled  SEDs such  as  those of
Mrk\,231 and Arp\,220 (Figure 7).  Only sensitive mm/submm imaging can
overcome it  by: a)  spatially separating the  cold from the  warm and
much more luminous SF-heated dust, and b) identifying any warm compact
dust/gas regions  (with potentially significant  submm optical depths)
``nested''  inside more  extended cooler  dust distributions  that are
optically thick at far-IR wavelengths.  In Arp\,220 such submm imaging
has  identified hot  and spatially  extended  (i.e.  starburst-heated)
dust   reservoirs,    with   2r$\sim   $100\,pc,    $\rm   T_{h}$$\sim
$(100--180)\,K and $\rm \tau _{h}(850\mu m)$$\sim $1 (Eckart \& Downes
2007;  Sakamoto  et  al.   2008),  within  its  extended  cooler  dust
distribution  with  $\rm   T_{w}$$\sim  $(45--60)\,K  and  significant
optical  depths  at the  far-IR  ($\rm  \tau  _w(100\mu m)$$\sim  $10;
Lisenfeld et al.  2000).  A  simple expression for the emergent SED of
such a dust configuration

\begin{equation}
\rm S_{\nu } = \frac{1+z}{4\pi D^2 _L}\left[1+\frac{B_{\nu}(T_h)}{B_{\nu}(T_w)}
\frac{M_{d,h}}{M_{d,w}}\frac{F[\tau_h(\nu)]}{F[\tau_w(\nu)]}e^{-\tau_w(\nu)/2}\right]
k_{d}(\nu)M_{d,w} B_{\nu}(T_{w}) F[\tau _w(\nu )],
\end{equation}

\noindent
where   $\rm  F(\tau)=(1-e^{-\tau})/\tau$,   $\rm   \tau  (\nu)$=$\tau
_{\circ}(\nu/\nu_{\circ})^2$, can easily show the inadequacy of global
dust  emission SEDs  in disentangling  the true  dust  components.  In
Figure  8, after  setting $\rm  T_{w}$=40\,K, $\rm  T_{h}$=120\,K $\rm
\tau_{\circ,  h}(850\mu m)$$\sim $1,  and $\rm  \tau_{\circ, w}(100\mu
m)$$\sim   $10,   $\rm   M_{d,w}$=10$^8$\,M$_{\odot}$  we   plot   the
aforementioned SED which makes apparent that hot and compact starburst
regions  can remain  inconspicuous in  the global  dust  emission SED,
apart from  an emission contribution  in the mm/sub-mm domain  where a
cold  dust  reservoir  would   also  contribute  (and  thus  the  $\rm
\tau_{IR}$-$\rm T_{d}$ degeneracy).  This is a result of the high dust
temperatures in  the compact regions (which makes  their emission peak
at  wavelengths  where  the   outer  dust  distribution  remains  very
absorptive  i.e.   at  IR/far-IR  wavelengths), and  their  high  dust
optical depths over almost the entire observed range of a typical SED.
The latter  diminishes the emission contribution of  those compact and
warm  starburst regions well  below from  being proportional  to their
dust mass content.

\subsection{The [C\,II] line luminosity deficit in ULIRGs: a dust optical depth effect?}

One  of the  most unexpected  discoveries made  by {\it  ISO}  was the
so-called [CII] line  luminosity deficit for some of  the most extreme
starbursts  in the  local Universe  (Malhotra et  al. 1997;  Luhman et
al. 1998,  2003). This was deemed  all the more  perplexing for ULIRGs
since the  [CII] fine structure line  at 157.74\,$\mu $m  is the major
ISM cooling line and would  thus be expected to be particularly bright
in such extreme star-forming  systems.  Various explanations have been
offered  for this  such  as :  a)  absorption by  a  cooler dust  {\it
foreground}  ``screen'', b)  saturation of  the [CII]  line luminosity
(while  the IR continuum  rises unabated)  predicted for  high density
PDRs or PDRs in which the local UV radiation field is high and the gas
density  is low,  c) soft  far-UV  radiation fields  (see Malhotra  et
al. 1997;  Luhman et al. 1998  for a review).   Significant submm dust
optical depths  for the bulk  of the dust  mass suggests a  simple and
common  cause  for  the  [CII]  and the  high-J  CO  lines  luminosity
deficiency in  some ULIRGs since  in an almost featureless  black body
dust continuum,  no lines are  expected from a  concomitant isothermal
gas  reservoir.  In  particular the  weakness  of the  CO J=6--5  line
argues against  the most  prominent alternative explanation  given for
the [CII]  line luminosity deficit:  the dense and/or  strongly far-UV
illuminated PDRs where  in fact such high-J CO lines  ought to be very
luminous.  Finally even modest submm  dust optical depths such as $\rm
\tau  _{400\mu   m}$$\sim$0.2  (from  Equation  8   and  $\rm  \langle
n(H_2)\rangle$=$10^4$\,$\rm  cm^{-3}$, $\rm  h$=40\,pc)  correspond to
$\rm \tau^{(c)} _{158\mu  m}$$\sim $1.3 (for $\beta $=2)  for the dust
continuum  at  the  rest  frequency  of the  [CII]  line,  yielding  a
F($\tau$)=$(1-e^{-\tau})/\tau$=0.56  reduction in  line  strength with
respect to the optically thin case.

\section{CO SLEDs in LIRGs: effects of dust and AGN}

%We have now  shown that high dust optical  depths at submm wavelengths
% can  occur for  some LIRGs  (frequently  used as  local analogues  of
% high-z starbursts) severely surpressing  their high-J CO or any other
% high frequency line  emission (e.g.  C$^+$ at 158$\mu  $m) from their
% starburst-fueling dense  molecular gas.   In Figure 8  we demonstrate
% such effects on  the CO SLED of Arp\,220, the  first case where these
% have been shown  to be significant (Papadopoulos et  al.  2009).  The
% ``intrinsic'' highly-excited  CO SLED expected for  its massive dense
% and warm  gas phase (Greve et  al.  2009) is shown  overlaid with two
% dust-affected   ones,   estimated  from   the   unaffected  CO   line
% luminosities  $\rm L_{J+1,J}$  and $\rm  L^{(d)} _{J+1,J}  = e^{-\tau
% _d(\nu_{J+1,J})}\times L  _{J+1,J}$ (valid  for an isothermal  mix of
% gas and dust).   We adopted an emissivity law of  $\beta =2$ and $\rm
% \tau _{850\mu m}$=0.2,  1 for the rest frequency  of CO J=3--2 ($\sim
% $350\,GHz),  values   that  bracket  the  range   revealed  by  submm
% interferometry  ($\rm  \tau _{850\mu  m}$$\sim  $1;  Sakamoto et  al.
% 2008),  and  global  dust  SED  fits  ($\rm  \tau  _{100\mu  m}$$\sim
% $11.6$\Rightarrow  \tau  _{850\mu m}$$\sim  $0.16;  Lisenfeld et  al.
% 2000 for $\beta=2$).

High  dust optical  depths at  short submm  wavelengths in  some LIRGs
 (worryingly  amongst  those frequently  used  as  local analogues  of
 high-z starbursts) can significantly suppress high-J CO or indeed any
 other high-frequency line emission from their starburst-fueling dense
 molecular gas  distributions. In Figure 9 we  demonstrate the effects
 of dust  optical depth  on the  CO SLED of  Arp\,220, the  first case
 where  these can  be significant  (Papadopoulos et  al.   2010).  The
 highly-excited ``intrinsic''  CO SLED  derived for its  massive dense
 and warm  gas phase  (Greve et  al.  2009) is  shown overlaid  by two
 SLEDS in  which the  effects of different  dust optical depth  on the
 intrinsic  CO line  luminosities $\rm  L_{J+1,J}$ are  inserted using
 $\rm   L^{(d)}  _{J+1,J}$=$\rm   e^{-\tau   _d(\nu_{J+1,J})}\times  L
 _{J+1,J}$ (valid for an isothermal  mix of gas and dust).  We adopted
 an emissivity law  of $\beta $=2 and $\rm  \tau _{850\mu m}$=0.2, 1.0
 at the rest  frequency of CO J=3--2 ($\sim  $350\,GHz).  These values
 bracket  the  range  of  submm  optical  depths  deduced  from  submm
 interferometry  ($\rm  \tau _{850\mu  m}$$\sim  $1;  Sakamoto et  al.
 2008),  and global  IR/submm/mm  dust SED  fits  ($\rm \tau  _{100\mu
 m}$$\sim 11.6$$\Rightarrow \tau _{850\mu m}$$\sim $0.16; Lisenfeld et
 al.  2000 for $\beta$=2).

% Arp220 SLEDs
% L(Lsol)=3.18x10^4 (f_rest/100GHz)^3 [L'/(10^9 L_l)], L_l=K km/s pc^2
% Sco(1-0)dV = (419 +/- 36) Jy km/s
% DL=78 Mpc, z=0.0182, Lco(1-0)=(6.12+/-0.52)x10^9  K km/s pc^2
% ****Lco(1-0)(sol) = 0.30x10^6 Lsol***** OBSERVED
% ****Lco(tau=0.2)  ~ 0.30x10^6 Lsol*****
% ****Lco(tau=1.0)  = 0.33x10^6 Lsol*****

% ``Intrinsic'' SLED
% Conditions adopted:  Tk=50 K, n=10^5 cm^{-3}, <Kvir>=5
%                      Lco=3x10^{-6}
%
%                      L(sol)        L(tau_350=0.2)  L(tau_350=1)
%                  L_10 = 0.32x10^6    0.32x10^6     0.32x10^6  
% r_21 = 0.96      L_21 = 2.46x10^6    2.25x10^6     1.58x10^6
% r_32 = 0.91      L_32 = 7.86x10^6    6.44x10^6     2.89x10^6
% r_43 = 0.86      L_43 = 1.76x10^7    1.23x10^7     2.97x10^6
% r_54 = 0.80      L_54 = 3.20x10^7    1.84x10^7     1.98x10^6
% r_65 = 0.75      L_65 = 5.18x10^7    2.33x10^7     9.48x10^5
% r_76 = 0.70      L_76 = 7.68x10^7    2.58x10^7     3.32x10^5
% r_87 = 0.63      L_87 = 1.03x10^8    2.48x10^7     8.40x10^4
% r_98 = 0.51      L_98 = 1.19x10^8    1.97x10^7     1.47x10^4
% r_109= 0.30      L_109= 9.6 x10^7    1.04x10^7     1.43x10^3
% r_1110=0.09      L1110= 3.8 x10^7    2.58x10^6      55.07

From  Figure 9  is apparent  that even  the small  submm  dust optical
depths found  from global  dust SED of  ULIRGs can have  a substantial
impact  on   their  CO  SLEDs   from  J=5--4  and  beyond.    In  such
dust-affected CO  SLEDs the J-level  of the line luminosity  peak will
not reflect a true global gas excitation turnover and thus {\it deeply
dust-enshrouded compact  starbursts can have  misleadingly ``cool'' CO
SLEDs.}  This  can particularly affect molecular  line observations of
high-z galaxies where mostly high-J CO lines are currently accessible.
In such systems, unlike local LIRGs whose molecular gas properties can
be  constrained by  many  low frequency  molecular  lines (where  dust
optical  depths  remain  negligible),  separating dust  optical  depth
effects from genuinely low-excitation gas can be~difficult.

\subsection{Low CO line excitation: the case of Arp\,193}

The physical  conditions of star-forming molecular gas  used to deduce
 its ``intrinsic'' CO SLED in LIRGs (section 3.1) are typical, however
 significant   deviations   do    exist.    Arp\,193   with   a   star
 formation-related  IR luminosity  of $\rm  L^{(SF)} _{IR}$=2.2$\times
 $10$^{11}$\,L$_{\odot}$   is    notable   in   having    the   lowest
 (4--3)/(1--0), (3--2)/(1--0)  HCN line ratios:  $\rm r_{43}(HCN)$$\la
 $0.08 (2$\sigma $),  $\rm r_{32}(HCN)$=0.22 (Papadopoulos 2007; Krips
 et  al.   2008),  as  well  as  a faint  CO  J=6--5  line  with  $\rm
 R_{65/32}$$\la  $0.08. These  are indicative  of very  low excitation
 conditions  for the  bulk of  its  molecular gas  reservoir while  by
 comparison  NGC\,6240,   a  galaxy  with  a   similar  $\rm  L^{(SF)}
 _{IR}$$\sim    $3.8$\times    $10$^{11}$\,L$_{\odot}$,    has    $\rm
 r_{43}(HCN)$=0.6  and  $\rm r_{32}(HCN)$=0.8  (Greve  et al.   2009),
 indicative of a much denser and warmer gas phase.  Radiative transfer
 LVG  modeling of  HCN and  CO (6--5)/(3--2)  line ratios  in Arp\,193
 (constrained also by $\rm  K_{vir}$$\sim $1) are consistent only with
 low   density  $\rm  n(H_2)$$\sim   $100\,cm$^{-3}$  and   warm  $\rm
 T_{k}$=(85--110)\,K  gas.  These are  highly atypical  conditions for
 either  vigorously  star-forming  LIRGs  (where the  ISM  is  usually
 dominated  by  dense {\it  and}  warm  molecular  gas), or  quiescent
 galaxies (where molecular gas has similarly low densities but is much
 colder, $\rm  T_{kin}$$\sim $15\,K),  and demonstrate that  {\it even
 vigorously  star  forming  systems   can  have  low  high-J  CO  line
 excitation for the bulk of their gas.}  The corresponding CO SLED for
 the  dense  gas in  Arp\,193  already turns  over  at  CO J=3--2,  in
 contrast to  the HCN-bright star-forming phase of  NGC\,6240 which is
 expected to emit  copiously up to CO J=10--9  (Figure~10).  It should
 be noted that besides similar  $\rm L^{(SF)} _{IR}$, these LIRGs both
 have  $\rm M(H_2)$$\sim$(0.5--1)$\times 10^{10}$$\rm  M_{\odot}$ (for
 the CO-H$_2$ conversion factor of  Downes \& Solomon 1998), global SF
 efficiencies   $\rm    L^{(SF)}   _{IR}/M(H_2)$$\sim   $50\,L$_{\odot
 }$/M$_{\odot }$, and similar dense gas mass fractions as indicated by
 $\rm  R_{HCN/CO}$=$\rm  L^{'}  _{HCN(1-0)}/L^{'}  _{CO(1-0)}$  ($\sim
 $0.05 for Arp\,193, and $\sim  $0.08 for NGC\,6240).  This shows that
 observables commonly  available for LIRGs  such as IR  luminosity, CO
 J=1--0,  and HCN  J=1--0 lines,  may  provide no  indication of  very
 different molecular gas excitation  conditions and CO SLEDs, at least
 on an individual object~basis.

The rapid  evolution of starburst  events can in principle  have short
periods during which  a (U)LIRG stays IR-luminous while  its dense gas
phase is strongly dispersed and/or depleted, by the formation of stars
and their concentrated and coherent mechanical feedback (Loenen 2009).
This could place LIRGs such as  NGC\,6240 and Arp\,193 at the two ends
of  such  a short  evolutionary  track,  but  observations of  similar
objects  are needed to  decide such  issues.  Finally  we note  that a
population of near-IR selected  gas-rich galaxies with ULIRG-levels of
star  formation ($\sim  $100--150\,$\rm M_{\odot}\,yr^{-1}$)  but very
low  (comparable to  the  Milky-Way) levels  of  global molecular  gas
excitation,  and star-formation  efficiency similar  to  Arp\,193, has
been  recently  revealed  at  high  redshifts  (Daddi  et  al.   2008;
Dannerbauer et al.~2009).

\subsection{Suppressed high-J CO lines: dust absorption versus low gas excitation}

%The two possibilities of significant  dust submm optical depths or low
% molecular gas excitation as  responsible for quenching global CO line
% ratios  in LIRGs can  be hard  to tell  apart without  the additional
% information from other molecular lines, typically much weaker than CO
% (e.g.   HCN).   The  insidious  aspect  of this  degeneracy  is  that
% dust-affected  CO SLEDs  of  extreme starbursts  can appear  ``cold''
% despite  intense star forming  activity simply  because their  low CO
% (high-J)/(low-J)  line ratios would  be attributed  to low-excitation
% (i.e.    non   star-forming)   molecular   gas   rather   than   dust
% absorption. Only  line observations of heavy rotor  molecules such as
% HCN,  HCO$^+$,  and  CS,  at frequencies  $<$350\,GHz  (i.e.   mostly
% unaffected  by  potentially  large   dust  optical  depths  at  submm
% wavelengths) can ``break'' this degeneracy.  Unfortunately such lines
% can be  difficult to detect  routinely, even in bright  nearby LIRGs,
% since even  the brightest such  line: HCN J=1--0, typically  has $\rm
% L^{'}  _{HCN(1-0)}$$\sim  $(1/5--1/30)$\rm  \times L^{'}  _{CO(1-0)}$
% (e.g.   Gao  \&  Solomon  2004).  It is  clear  that  obtaining  such
% additional diagnostic information  for distant galaxies will required
% the sensitivity of ALMA.

Once both  possibilities of large  dust optical depths at  short submm
 wavelengths and low gas  excitation have been recognized as affecting
 emergent CO SLEDs, they can  be hard to tell apart without additional
 information from  other lines  tracing the dense  star-forming phase.
 These are low-J transitions (thus not affected by dust) of HCN or CS,
 but can  be difficult to  detect on a  routine basis, even  in bright
 nearby LIRGs,  since the brightest  such line: HCN  J=1--0, typically
 has  $\rm  L^{'}  _{HCN(1-0)}$$\sim $(1/5--1/30)$\times  $$\rm  L^{'}
 _{CO(1-0)}$ (e.g.  Gao \& Solomon  2004).  An incidious aspect of the
 aforementioned degeneracy is that some of the most extreme starbursts
 in the  Universe can show  ``cold'' CO SLEDs despite  hosting intense
 star forming  activity simply  because their low  CO (high-J)/(low-J)
 line ratios are attributed to low-excitation (i.e.  non star-forming)
 molecular gas  rather than dust absorption.  As  already discussed in
 sections  4.1  and  4.2,   global  dust  emission  cannot  provide  a
 definitive distinction  either, even  with well-sampled dust  SEDs in
 the crucial IR/submm range.

In Figure 11 we show  a possible statistical distinction between these
 two possibilities using  CO J=1--0 (the main bulk  molecular gas mass
 tracer), HCN  J=1--0 (the most  widely used dense gas  tracer), along
 with   CO   J=6--5,    3--2.    The   $\rm   R_{HCN/CO}$=$\rm   L^{'}
 _{HCN(1-0)}/L^{'}_{CO(1-0)}$ ratio (Table 2) is considered here as an
 approximate measure  of the dense molecular gas  mass fraction (since
 $\rm n_{crit}(CO)$$\sim  $410\,cm$^{-3}$ and $\rm n_{crit}(HCN)$$\sim
 $2$\times $10$^5$\,cm$^{-3}$), with little sensitivity to temperature
 since for both  CO and HCN J=1--0: $\rm  E_{10}/k_B\sim $4--5\,K. The
 low  frequencies of $\sim  $115\,GHz (CO  J=1--0) and  $\sim $87\,GHz
 (HCN J=1--0), ensure that $\rm R_{HCN/CO}$ remains unaffected by dust
 extinction.  From Figure 11  becomes apparent that several LIRGs have
 very  high $\rm  R_{HCN/CO}$ ($\ga  0.15$)  {\it and}  very low  $\rm
 R_{65/32}$$\la  $0.2   ratios,  quite  unlike   typical  star-forming
 galaxies  with much  lower IR  luminosities (and  thus  SFRs).  Thus,
 unless one is willing to entertain the possibility of massive amounts
 of dense but otherwise SF-idle molecular gas in these (mostly) merger
 systems,  {\it   such  LIRGs  are   likely  to  be   starbursts  with
 dust-suppressed high-J CO lines.}

Considering   star  formation   as  a   process  directly   fueled  by
self-gravitating  dense  gas means  that  star  formation switches  on
rapidly  in  the dense  HCN-bright  phase,  with  its typically  short
dynamical  times $\rm  t_{dyn}$$\sim $10$^5$\,yrs  (much  shorter than
Giant    Molecular   Cloud    evolutionary    timescales   of    $\sim
$10$^6$--10$^7$\,yrs).    This   likely    yields   the   tight   $\rm
L_{IR}$--$\rm  L_{HCN(1-0)}$ correlation in  galaxies (Gao  \& Solomon
2004), extending  ``down'' to individual GMCs and  spanning 7-8 orders
of magnitude in $\rm L_{IR}$ (Wu  et al.  2005).  It also provides the
basis  for our  aforementioned CO-HCN  line ratio  diagnostic  and its
ability  to   ``weed-out''  dust-affected  CO   SLEDs  from  genuinely
low-excitation ones.  In the few cases where multi-J dense gas tracers
(e.g.  HCN, CS) and dust SEDs  are available to constrain the state of
the  dense  gas  in  LIRGs,  it  is  always  found  in  the  warm  and
self-gravitating ``corner'' of  the available parameter space, typical
of  star-forming gas  (Mao et  al.  2000;  Papadopoulos et  al.  2007;
Greve et al.  2009), and where $\rm R_{65/32}$$\ga $0.45. Moreover for
the lowest  possible temperature of  $\rm T_{k}$$\sim $10\,K  in dense
starless  cores deep  in GMCs  (regulated by  cosmic rays  rather than
photons), a minimum

\begin{equation}
\rm R_{65/32}=\frac{E_{65}}{E_{32}} \left(\frac{e^{E_{32}/k_BT_k}-1}{e^{E_{65}/k_BT_k}-1}\right)=
2\left(\frac{e^{16.6/T_k}-1}{e^{33.2/T_k}-1}\right) \sim 0.32, 
\end{equation}

\noindent
is expected for the optically  thick CO emission (shown also in Figure
4 as a dotted  line).  Thus even for the unlikely case  of a high $\rm
R_{HCN/CO}$ ratio  due to  a massive and  dense but  otherwise SF-idle
cold gas  phase, this  is the lowest  $\rm R_{65/32}$  value possible.
Values  of $\rm  R_{65/32}$$\la  $0.3  would then  imply  either a
dominant gas phase that necessarily has $\rm \langle n(H_2)\rangle
$$\la  $10$^4$\,cm$^{-3}$ (and  thus  also low  $\rm R_{HCN/CO}$),  or
dust-suppression of the CO J=6--5  line luminosity. In the second case
such dust-affected LIRGs  would then populate the area  marked by $\rm
R_{65/32}$$\la$0.3  {\it  and} $\rm  R_{HCN/CO}$$\ga  $0.15, shown  in
Figure 11 to be containing several  objects.

% IRAS 00057+4021: L_HCN(1-0)/L_co(1-0)= unavailable
% IRAS 02483+4302: L_HCN(1-0)/L_co(1-0)= unavailable
% VIIZw31        : L_HCN(1-0)/L_co(1-0)= 0.057+/-0.012
% Arp55          : L_HCN(1-0)/L_co(1-0)= 0.048+/-0.010
% Mrk 231        : L_HCN(1-0)/L_co(1-0)= 0.290+/-0.07
% IRAS 09320+6134: L_HCN(1-0)/L_co(1-0)= 0.21+/-0.05
% IRAS 10173+0828: L_HCN(1-0)/L_co(1-0)= unavailable
% IRAS 10565+2448: L_HCN(1-0)/L_co(1-0)= 0.19+/-0.035
% PG 1119+120    : L_HCN(1-0)/L_co(1-0)= unavailable
% IRAS 12112+0305: L_HCN(1-0)/L_co(1-0)= 0.15+/-0.035
% Arp 193        : L_HCN(1-0)/L_co(1-0)= 0.054+/-0.010
% 3C 293         : L_HCN(1-0)/L_co(1-0)= unavailable
% Zw49           : L_HCN(1-0)/L_co(1-0)= unavailable
% Arp220         : L_HCN(1-0)/L_co(1-0)= 0.18+/-0.031
% IRAS 17208-0014: L_HCN(1-0)/L_co(1-0)= 0.14+/-0.025
% NGC 7469       : L_HCN(1-0)/L_co(1-0)= 0.066+/-0.011

%Literature of L_HCN/CO ratios of systems with CO 6-5/3-2

% NGC 253:   L_HCN(1-0)/L_co(1-0)=0.059, R_65/32=0.88+/-0.32
% NGC6946:   L_HCN(1-0)/L_co(1-0)=0.053, R_65/32=0.22+/-0.07 (Bayet et al. 2006)
% M82    :   L_HCN(1-0)/L_co(1-0)=0.053, R_65/32=0.89+/-0.30
% IC342  :   L_HCN(1-0)/L_co(1-0)=0.050, R_65/32=0.24+/-0.07 (Bayet et al. 2006)
% M83    :   L_HCN(1-0)/L_co(1-0)=0.043, R_65/32=0.83+/-0.20 (Bayet et al. 2006)

\subsection{AGN effects on the high-J CO lines}

The brightest CO J=6--5 lines in  our sample are measured in the hosts
of two powerful AGNs, the  optically luminous QSO PG\,1119+120 and the
FR\,II  radio galaxy  3C\,293 known  for a  particularly  powerful jet
(Floyd et al.  2006).  These are  also the only two objects where $\rm
R_{65/32}$$>$1, possible  only for CO line emission  that is optically
thin {\it and} remains well-excited  up to J=6--5, which requires very
warm and dense molecular gas.   This becomes apparent from its maximum
value, achieved in the LTE optically thin limit,

\begin{equation}
\rm R^{(thin)} _{65/32}(LTE)=\frac{J(\nu_{65}, T_k)}{J(\nu_{32}, T_k)}
\left(\frac{\tau _{65}}{\tau _{32}}\right)=
\left(\frac{\nu_{65}}{\nu _{32}}\right)^2\,e^{-E_{52}/k_BT_k}=4\,e^{-66.35/T_k}, 
\end{equation}

\noindent
where                            $\rm                           J(\nu,
T_k)=(h\nu/k_B)\left[exp\left(h\nu/k_BT_k\right)-1\right]^{-1}$    (the
CMB  has been  omitted for  simplicity).  For  PG\,1119+120  even $\rm
R_{65/32}$=2 (i.e.  $\sim $(the measured value)--$\sigma$) yields $\rm
T_k=96\,K$, which  is rather high for  the bulk of  its molecular gas.
Submm  interferometric  imaging   has  recently  uncovered  such  high
brightness  temperatures for the  dust (and  thus for  the concomitant
molecular gas  since $\rm  T_{kin}$$\ga $$\rm T_{dust}$)  in Arp\,220,
but its  molecular gas  reservoir consists of  very dense gas  with CO
line optical  depths that  are nowhere near  the optically  thin limit
(Greve et  al.  2009).  Comprehensive LVG  radiative transfer modeling
of  such  a  high CO  (6--5)/(3--2)  ratio,  made  over a  larger  gas
temperature   range  of   $\rm  T_{k}$=(15--310)\,K   (to   allow  the
possibility  of very  warm gas)  yields  its best  solutions for  $\rm
n(H_2)$$\sim  $10$^5$\,cm$^{-3}$  and   $\rm  T_k$$\ga  $140\,K  (with
solutions improving  right up to 310\,K).  Such  high gas temperatures
could in principle be the  result of bulk irradiation of the molecular
gas  by X-rays  emanating  from  a central  AGN  creating giant  X-ray
Dominated  Regions   (XDRs),  rather  than   starburst-induced  far-UV
irradiated Photon  dominated Regions  (PDRs) (see Meijerink  \& Spaans
2005).   Very high  CO  line excitation  in  AGN hosts  that could  be
attributed  to   X-ray  luminous   AGN  has  been   recently  observed
(Papadopoulos  et al.   2008),  but unfortunately  no  X-ray data  are
available for PG\,1119+120 in particular, and the cause of its large CO
J=6--5 luminosity remains unknown.

\subsection{CO lines in 3C\,293:  a shock-powered CO SLED?}

Large amounts of  molecular gas ($\sim $$10^9$--$10^{10}$$\rm M_{\odot
}$)  have  been  discovered  in IR-luminous  powerful  radio  galaxies
locally as well as at high redshifts (e.g. Evans et al. 1999, 2005; De
Breuck et al.  2005).  In 3C\,293, shocks emanating from a very strong
jet-ISM interaction  (Emonts et  al.  2005) may  be responsible  for a
galaxy-wide  shock-induced  CO line  excitation  (Papadopoulos et  al.
2008).   A ``trademark''  of this  excitation mechanism  could  be its
capability of producing highly excited mid-J and high-J CO lines, {\it
even  in  the  absence  of  high  star  formation  rates,}  since  the
shock-induced turbulent  heating would deposit  most of its  energy on
the  gas  leaving  the  concomitant  dust  mass  reservoir  much  less
affected.  The high luminosity CO  J=4--3 and J=6--5 lines in 3C\,293,
with  (4--3)/(3--2), (6--5)/(3--2)  ratios of  $\rm R_{43/32}$=2.3$\pm
$0.91  and $\rm  R_{65/32}$=1.3$\pm $0.54,  seem to  bear this  out by
representing  some  of the  highest  levels  of  global molecular  gas
excitation discovered  in our sample  but in a system  whose low-level
star   formation   ($\rm   SFR$$\la   $4\,M$_{\odot}$\,yr$^{-1}$)   is
insufficient to ``power'' them.

We  extended  our  LVG  modeling  of  the  extraordinarily  high  $\rm
R_{43/32}$ and  $\rm R_{65/32}$  ratios in 3C\,293  to cover  a larger
range  of $\rm  T_{k}$=(15--310)\,K and  allow also  $\rm K_{vir}$$>$1
rather than $\rm K_{vir}$$\sim $1.  We do so because turbulent heating
can drive molecular gas temperatures  well above those in star forming
regions  (where   typically  $\rm  T_{kin}\la   100\,K$),  while  $\rm
K_{vir}\ga  1$   widens  the  LVG  solution  search   to  include  non
self-gravitating gas  phases.  These are  now more probable  given the
large  kinetic energy injection  from the  powerful jets  driving very
impressive gas outflows  in 3C\,293 (Morganti et el.   2003; Emonts et
al.   2005), and  the  fact  that no  massive  star-forming (and  thus
self-gravitating) gas phase ``powering''  the luminous high-J CO lines
is  present  in  3C\,293.   The  characteristics of  the  various  LVG
solution groups are summarized in Table~4, where two facts immediately
stand out:  a) the  best solutions  are provided by  a very  warm $\rm
T_k$$\sim     $(130--310)\,K    and     dense     $\rm    n(H_2)$$\sim
$($10^4$--$10^5$)\,$\rm cm^{-3}$ gas phase, and b) most solutions with
$\rm T_k$$\la $80\,K  are ruled out since they  have $\rm K_{vir}$$\ll
$1  (marked with  asterisk  in Table~4).   Such  kinematic states  are
deemed  unphysical  since gas  motions  cannot  be  slower than  those
induced by gas self-gravity (only raising the $\rm [CO/H_2]$ abundance
by  factors of $\ga  $10 could  make $\rm  K_{vir}$$\sim $1).   On the
other  hand,  almost  all  of  the  good  solutions  found  have  $\rm
K_{vir}$$\sim $15--49, indicating highly  unbound gas states.  We must
also note  that in all solutions  listed in Table~4  the fit continues
improving well  past the $\rm  T_k$=310\,K limit of the  LVG parameter
grid,  while gas  densities remain  high  ($\sim $($10^4$--$10^5$)$\rm
cm^{-3}$).  On the other  hand the  non-AGN cool  dust SED  in 3C\,293
carries     little      luminosity     ($\rm     L_{IR}$$\sim$$3\times
10^{10}$\,L$_{\odot }$) and is typical for quiescent~ISM.

Thus in the  powerful radio galaxy 3C\,293 the  molecular gas seems to
be in  a surprisingly hot,  dense, and gravitationally  unbound state,
whilst  the bulk  of its  concomitant dust  reservoir  maintains lower
temperatures and has only  a modest IR luminosity.  Tellingly, similar
ISM conditions  have been reported for the  turbulent molecular clouds
in the Galactic Center (Rodriguez-Fern\'andez et al.  2001) where very
warm and  dense H$_2$ is  concomitant with cold dust  (Pierce-Price et
al.  2000), as well as in 3C\,326, another another FR\,II radio galaxy
(Ogle et  a.  2007).  In the  latter case jet-induced  shocks are also
suggested   as   the   culprit    for   a   LIRG-sized   mass   ($\sim
$10$^9$\,M$_{\odot}$)  of  very  warm  molecular gas  ($\rm  T_k$$\sim
$125--1000\,K),  despite  star  formation  levels  even  lower  ($\sim
$0.1$\rm M_{\odot} yr^{-1}$) than those found in 3C\,293.

%Indicatively, the  molecular gas  in LIRGs  (the closest
%other  case of  high excitation  when considering  bulk  molecular gas
%masses) typically  has $\rm T_k$$\sim $40--60\,K,  with $\rm T_{k}\sim
%T_{dust}$  (i.e.    CO  SLEDs  and  dust  emission   SEDs  are  almost
%isothermal),   while  the  dense   molecular  gas   is  found   to  be
%self-gravitating   ($\rm  K_{vir}$$\sim  $1)   as  expected   for  the
%star-forming phase of these  extreme star-forming systems (e.g.  Greve
%et al.   2009).

In  Figure 12  we show  the  range of  CO SLEDs  corresponding to  the
shocked-excited ISM environment of 3C\,293.  Clearly more observations
of high-J  CO lines, now  possible with the spaceborne  Herschel Space
Observatory,  are   needed  to   better  constrain  them   and  verify
shock-excited molecular  gas over large scales in  this powerful radio
galaxy  and  other similar  systems.   The  very  large molecular  gas
reservoirs  found around  radio galaxies  at high  redshifts,  and the
potential feedback role of their energetic jets in inducing (Klamer et
al.    2004)  or  hindering   (Ogle  et   al.   2007)   starbursts  in
galaxy-formation events at early cosmic epochs, makes a detailed study
of shock-energized  molecular gas  even more important.   Distinct and
accessible ``signatures'' of shocked  molecular gas states on SLEDs of
various molecules  would be particularly valuable in  the upcoming era
of~ALMA.

% 3C293 CO SLEDs: D_L=192 Mpc, z=0.044
% A: <n>=3x10^4 cm^{-3}, Tk=30K,  Lco=3x10^{-5}  km/s pc^{-1} (special min fit)
% B: <n>=3x10^4 cm^{-3}, Tk=95K,  Lco=10^{-6} km/s pc^{-1} (min fit)
% C: <n>=10^5 cm^{-3},   Tk=310K, Lco=10^{-6} km/s pc^{-1} (max fit)
% D: <n>=10^5 cm^{-3},   Tk=130K, Lco=10^{-6} km/s pc^{-1} (ordinary fit also max fit)
% E: <n>=3x10^3 cm^{-3}  Tk=130K, Lco=10^{-4} km/s pc^{-1} (Kvir=0.866/a^0.5, one
% of the few isolated, virial solution available from the fit of the lowest ratios
% possible)

% Sco(3-2)dV = 208 Jy km/s,---> Lco(3-2)'=1.99x10^9  L_l
% L(Lsol)=3.18x10^4 (f_rest/100GHz)^3 [L'/(10^9 L_l)], L_l=K km/s pc^2
% ***L_10(sol) = 2.62x10^6 Lsol***
         
%                                         A            E          D
%          A      E      D       L_10 = 1.14x10^5 - 9.15x10^4 - 4.13x10^4
% r_21  =0.93 -  1.14 - 2.12     L_21 = 8.49x10^5 - 8.35x10^5 - 7.00x10^5    
% r_32  =0.85 -  1.06 - 2.35     L_32 = 2.62x10^6 - 2.62x10^6 - 2.62x10^6
% r_43  =0.76 -  0.98 - 2.34     L_43 = 5.55x10^6 - 5.74x10^6 - 6.18x10^6
% r_54  =0.69 -  0.89 - 2.26     L_54 = 9.85x10^6 - 1.02x10^7 - 1.16x10^7
% r_65  =0.61 -  0.77 - 2.18     L_65 = 1.50x10^7 - 1.52x10^7 - 1.94x10^7
% r_76  =0.53 -  0.63 - 2.09     L_76 = 2.08x10^7 - 1.98x10^7 - 2.96x10^7
% r_87  =0.36 -  0.45 - 1.93     L_87 = 2.10x10^7 - 2.11x10^7 - 4.08x10^7
% r_98  =0.12 -  0.27 - 1.72     L_98 = 1.00x10^7 - 1.80x10^7 - 5.18x10^7
% r_109 =0.014-  0.13 - 1.42     L_109= 1.60x10^6 - 1.19x10^7 - 5.86x10^7
% r_1110=0.00096-0.04 - 0.82     L_1110=1.46x10^5 - 4.87x10^6 - 7.80x10^7

\subsection{The diagnostic power of IR/submm ISM lines revisited}

Extremely dense  and compact gaseous disks able  to attain significant
dust optical depths at short submm wavelengths in LIRG/AGN systems can
have   dramatic  effects  on   their  emergent   ISM  lines   used  as
AGN-versus-starburst diagnostics.   In the presence of  such gas disks
around AGN, ISM  line diagnostic using IR and  even molecular lines at
short submm  wavelengths (e.g.  very high-J CO  lines) to discriminate
between dust-enshrouded AGN-excited XDRs versus starburst-excited PDRs
can be become difficult to employ. Relative line strengths may then be
more indicative of relative  dust optical depths rather than intrinsic
gas excitation  properties and  their causes.  Interestingly  all four
LIRGs   with   $\rm   R_{65/32}\ga   0.5$   (i.e.    IRAS\,02483+4302,
VII\,Zw\,31,  PG\,1119+120,  and  Mrk\,231)  have  face-on  or  nearly
face-on molecular  gas disks,  indicated either by  high-resolution CO
interferometry  (Downes and Solomon  1998) and/or  by their  narrow CO
line widths (3C\,293  is the only exception).  This  would be expected
for gas disks where the smaller dust optical depths in face-on than in
highly inclined gas disks along  the line of sight would enhance their
(line)-(continuum) contrast.  Nevertheless nearly face-on systems with
low  $\rm R_{65/32}$  ratios  do exist  (e.g.  IRAS\,00057+4021),  and
systematic  CO J=6--5 observations  of optically-selected  QSO samples
(which will contain many more  face-on gas disks) would be valuable in
revealing such geometric~effects.

 \section{Conclusions} 

We report on our new sensitive CO J=6--5 observations of LIRGs and two
powerful AGN with the JCMT, part of a now completed multi-J CO and HCN
molecular line survey. Our findings are as follows:

\noindent
1. Large dust  optical depths at short submm  wavelengths can suppress
   the CO  J=6--5 line in some  extreme starbursts by  immersing it in
   strong  continuum   dust  emission.    This  can  yield   faint  CO
   J+1$\rightarrow  $J lines  for J+1$\geq  $6, even  in  ULIRGs whose
   large  amounts of  dense star-forming  molecular  are intrinsically
   luminous  in  such lines.   Such  high  optical  depths can  easily
   account for  the so-called [CII] line luminosity  deficity known to
   exists in such systems.

\noindent
2. Similar  conditions  may be  present  in  high redshift  galaxies,
   yielding deceptively ``cool'' CO  SLEDs at high frequencies even in
   extreme starbursts such as the submillimeter~galaxies.

\noindent
3. Global dust emission  SEDs cannot unambigiously distinguish between
   high  dust optical  depths  at far-IR/submm  wavelengths or  large
   amounts  of  cold  dust,  as   they  can  both  contribute  to  the
   far-IR/submm part of the dust emission SED in a similar fashion.

\noindent
4.  The  very  low   CO  {\it and}  HCN  line excitation in  Arp\,193
    demonstrates that low global  gas excitation remains possible even
    in vigorously  star-forming LIRGs.

\noindent
5. The suppressing  effects of  high dust optical  depths on  the high
    frequency  part of  CO  SLEDs are  difficult  to distinguish  from
    genuine  low  gas  excitation.  Low-frequency  ($<$350\,GHz)  line
    observations of  highly dipolar heavy rotor  molecules (e.g.  HCN,
    CS),  with  their  high   critical  densities  but  unhindered  by
    potentially  large  dust  extinctions  at submm  wavelengths,  can
    ``break'' this  degeneracy.  In  the simplest such  application we
    propose  that LIRGs  with  large  HCN/CO J=1--0  but  very low  CO
    (6--5)/(3--2) line ratios are  likely to have dust-affected rather
    than low-excitation CO SLEDs.

\noindent
6. Remarkably  high  CO  line   excitation,  above  that  typical  for
   star-forming gas, is  found for the hosts of  two prominent AGN, an
   optically bright  QSO (PG\,1119+120) and a  radio galaxy (3C\,293).
   The  latter could be  the first  known case  of a  shock-excited CO
   SLED, likely powered by a  strong jet-ISM interaction.  As a result
   much   of  its  large   molecular  gas   reservoir  is   hot  ($\rm
   T_{kin}$$>$100\,K) and dense ($\rm n(H_2)$$\geq $10$^4$\,cm$^{-3}$)
   while most  of its dust mass remains  ``cool'' ($\rm T_{dust}$$\sim
   $15\,K) typical of quiescent ISM with low star formation rates.

In summary,  the emerging picture of  CO SLEDs in LIRGs  and AGN seems
 diverse,  with high dust  optical depths  at short  submm wavelengths
 capable  of   ``quenching''  high-J  CO  line   emission  in  extreme
 starbursts while low global gas excitation remaining a possibility in
 such systems.   Finally, shocks and  possibly X-rays seem  capable of
 surpassing far-UV photons as  the main excitation contributors in AGN
 hosts.   The now  spaceborne  Herschel Space  Observatory is  ideally
 suited  for  fully   characterizing  these  environments  via  high-J
 molecular  line observations  of  local IR-bright  galaxies and  AGN,
 paving the way for ALMA and the study of such phenomena across cosmic
 epoch.

\acknowledgments We would  like to take this opportunity  to thank the
 entire superb  crew of people that  make JCMT a success  over so many
 years.  Special thanks to  the Telescope System Specialists: Benjamin
 Warrington, Jim Hoge, Jonathan  Kemp, and Jan Wouterloot for expertly
 assisting this project. Jessica  Dempsey and Per Friberg deserve much
 credit  for helping  commission the  W/D  receiver, as  well as  Iain
 Coulson  and   Antonio  Chrysostomou  for   supporting  and  flexibly
 allocating the  generous amounts of  time necessary to  conclude this
 project. We  would like to also  thank Holly Thomas  for assisting us
 with crucial aspects  of telescope pointing. Like all  wines from the
 island of Santorini, you are  all good.  Finally we thank the referee
 for useful suggestions  that much enhanced this work,  and would like
 to thank again the referee of  the paper by Papadopoulos et al.  2010
 whose  suggestions and  criticism lead  to the  re-measurement  of CO
 J=6--5 in Arp\,220 and the new value of its flux presented here.

{}

\clearpage

\begin{figure}
\epsscale{1.0}
\plotone{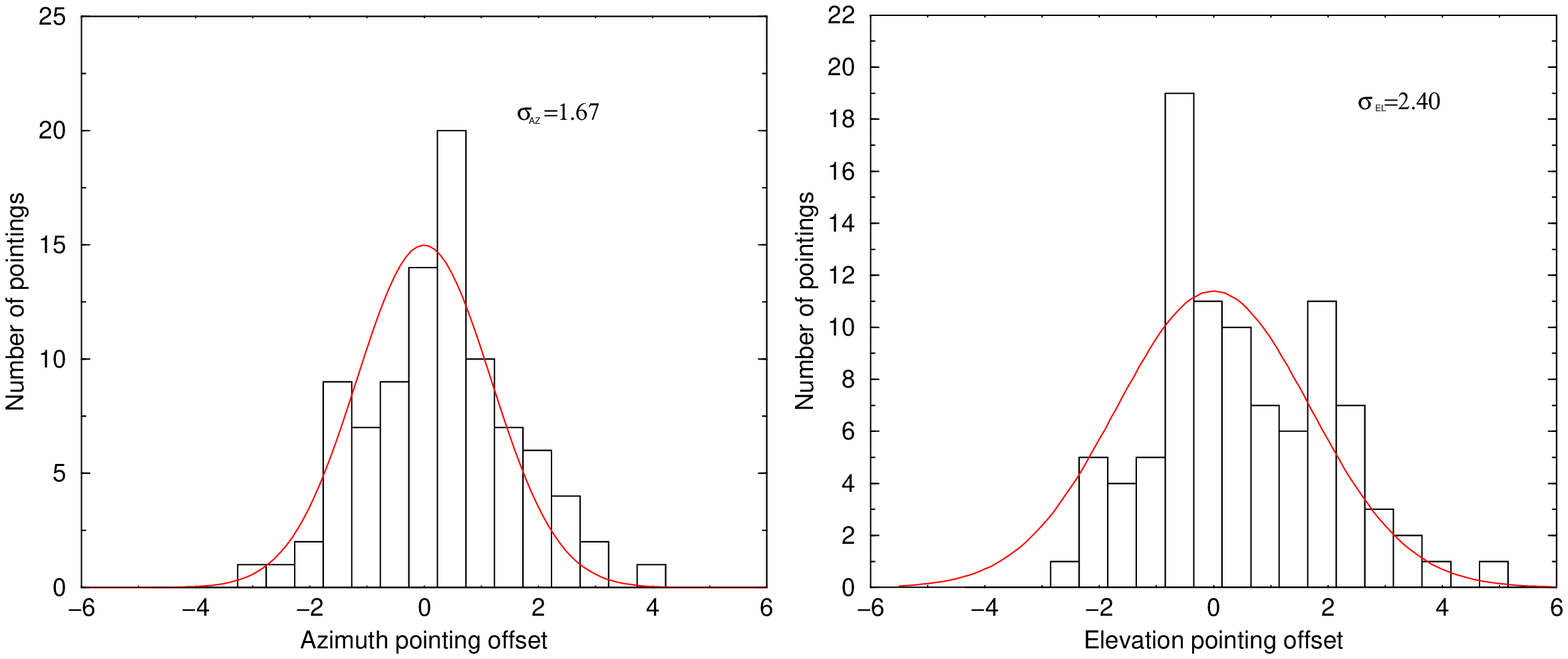}
%Figure 1
\caption{The  distributions of  pointing offsets  obtained  during all
 observing  periods   and  over   many  sky  sectors.    Offsets  with
 $\sigma$$\la $2$''$  are those  typical for real  source observations
 (larger values are typically  obtained after large changes in azimuth
 and elevation when the  telescope changed sky sectors between regular
 observing  sessions).  During  actual  source observations  typically
 $\rm  \sigma  _{az}$$\sim  $$\sigma  _{el}$$\sim $1.6$''$,  and  $\rm
 \sigma _r=(\sigma ^2 _{az}+\sigma ^2 _{el})^{1/2}$$\sim $2.2$''$.}
\end{figure}

\clearpage

\newpage

\begin{figure}
%Figure 2
\centering
\epsscale{0.28}
\plotone{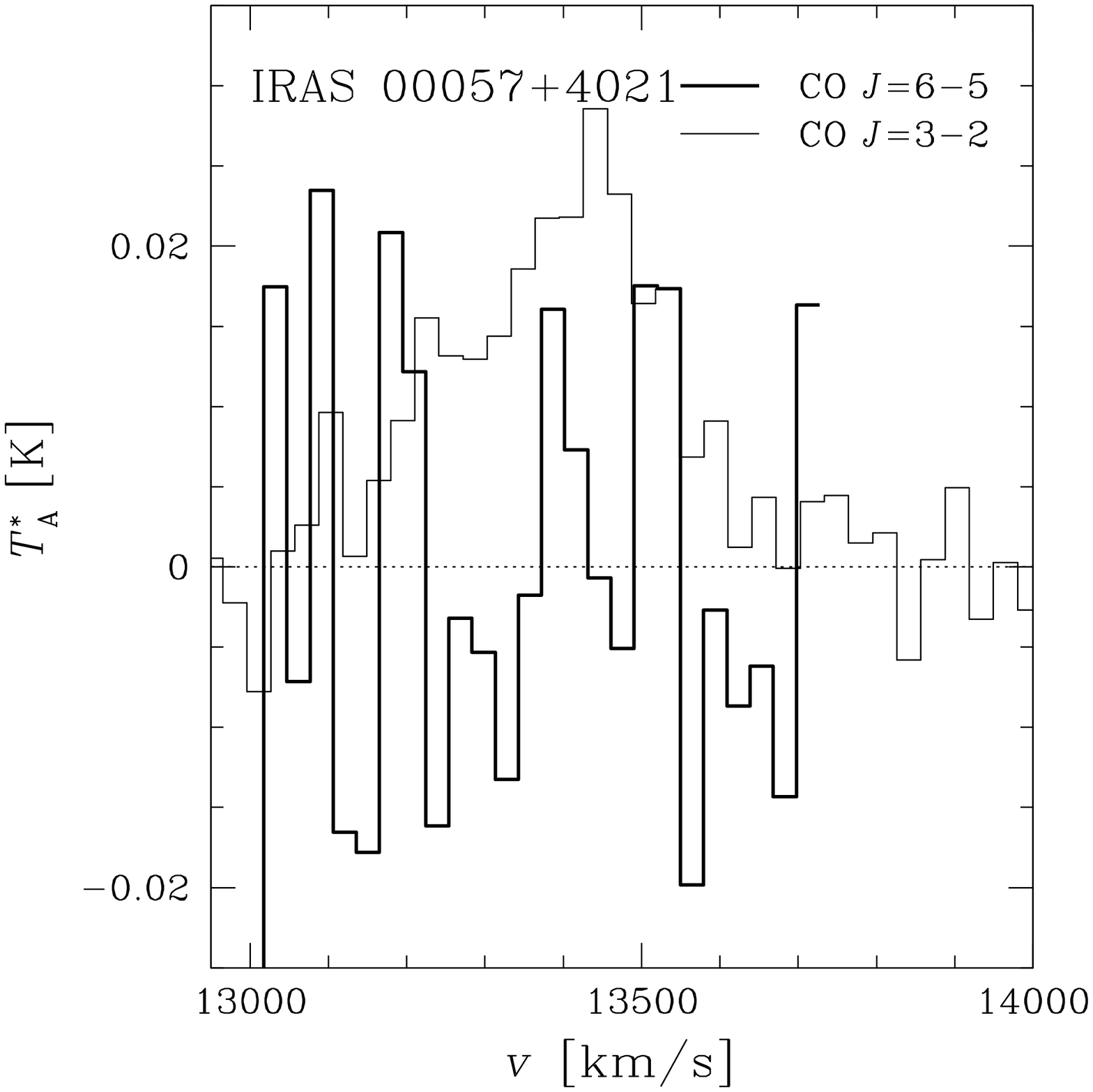} % CO32=CO65=30km/s
\plotone{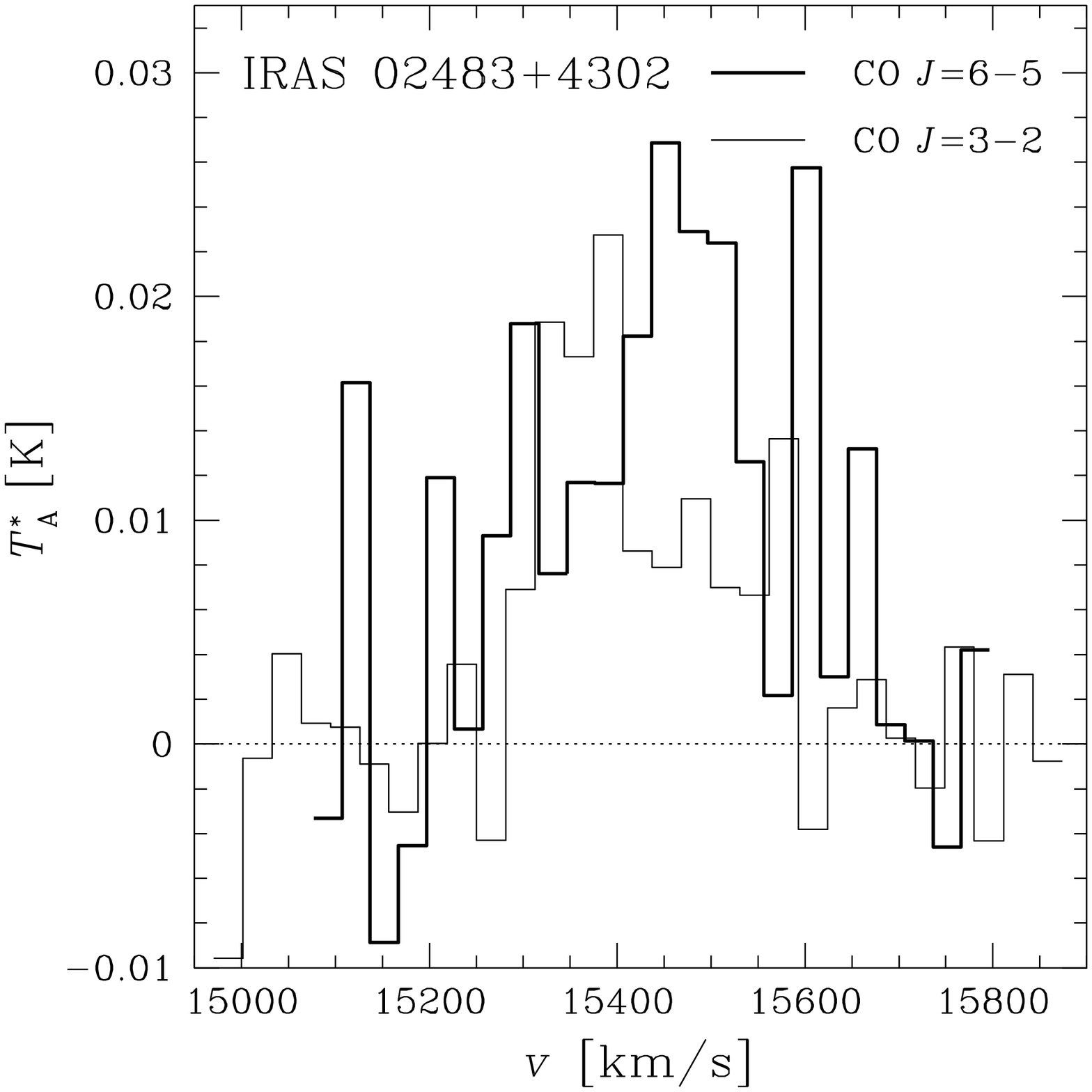} % CO32=CO65=30km/s
\plotone{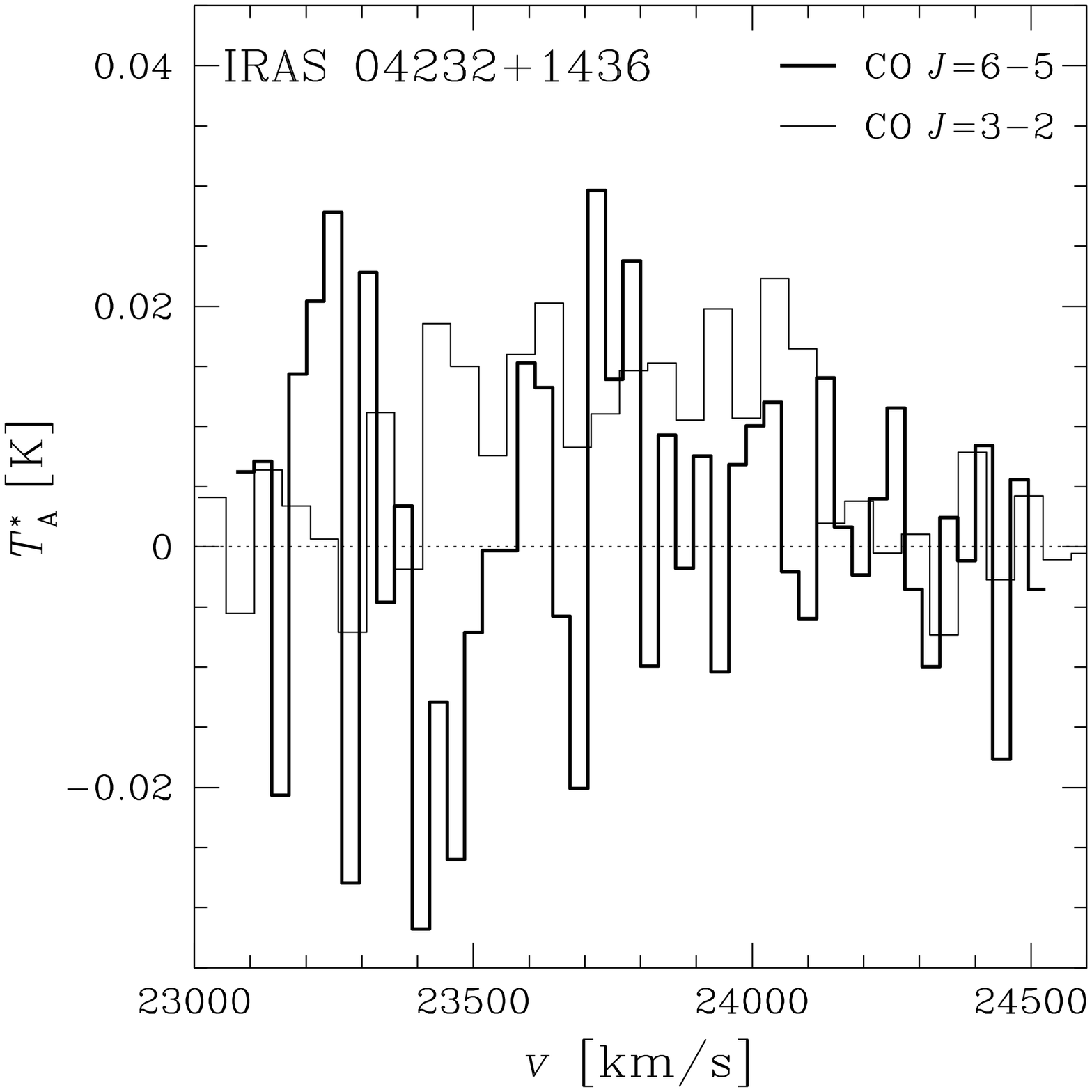} % CO32=50km/s, CO65=31km/s

\plotone{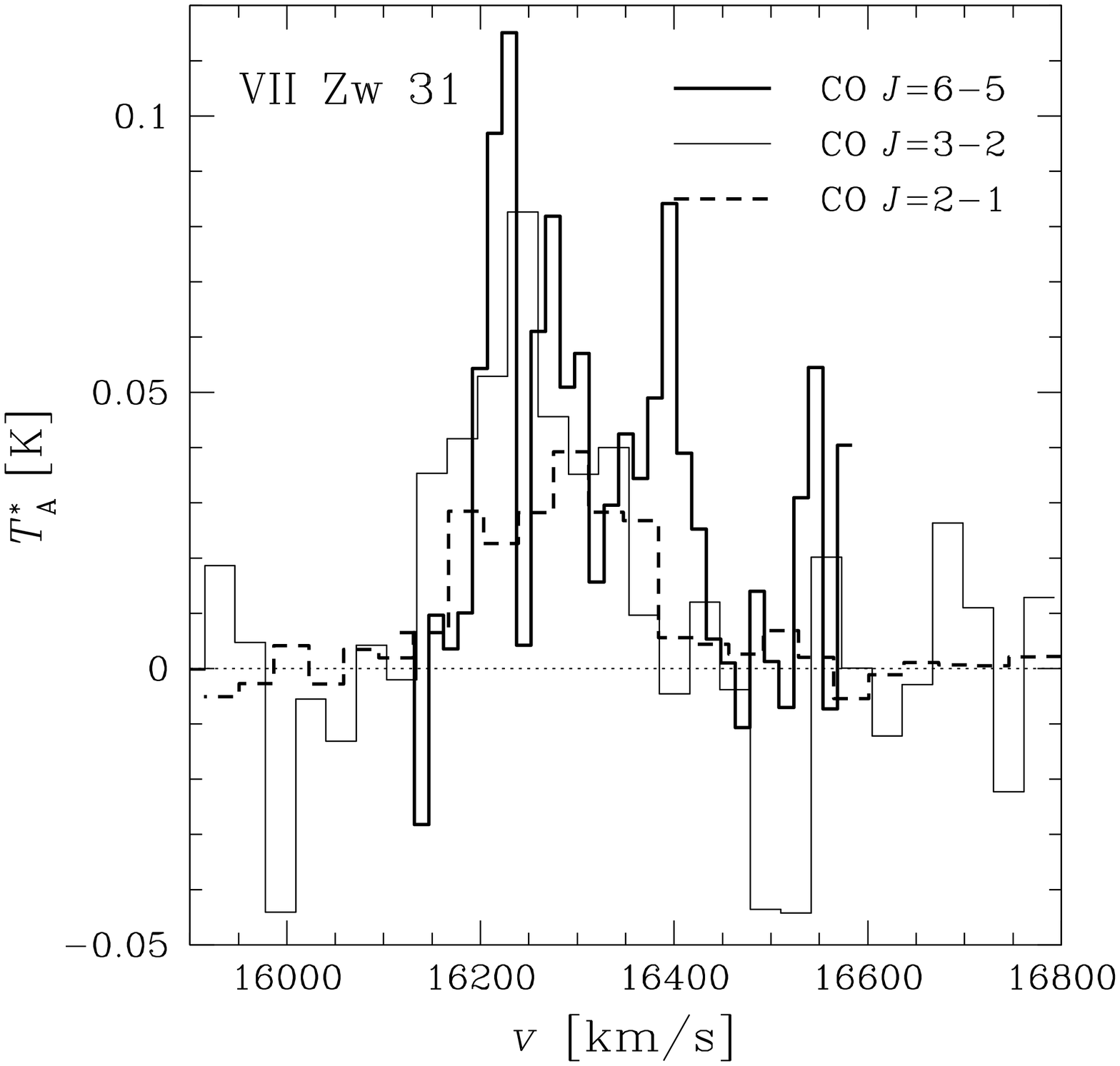}      % CO21=36km/s, CO32=31km/s, CO65=15km/s
\plotone{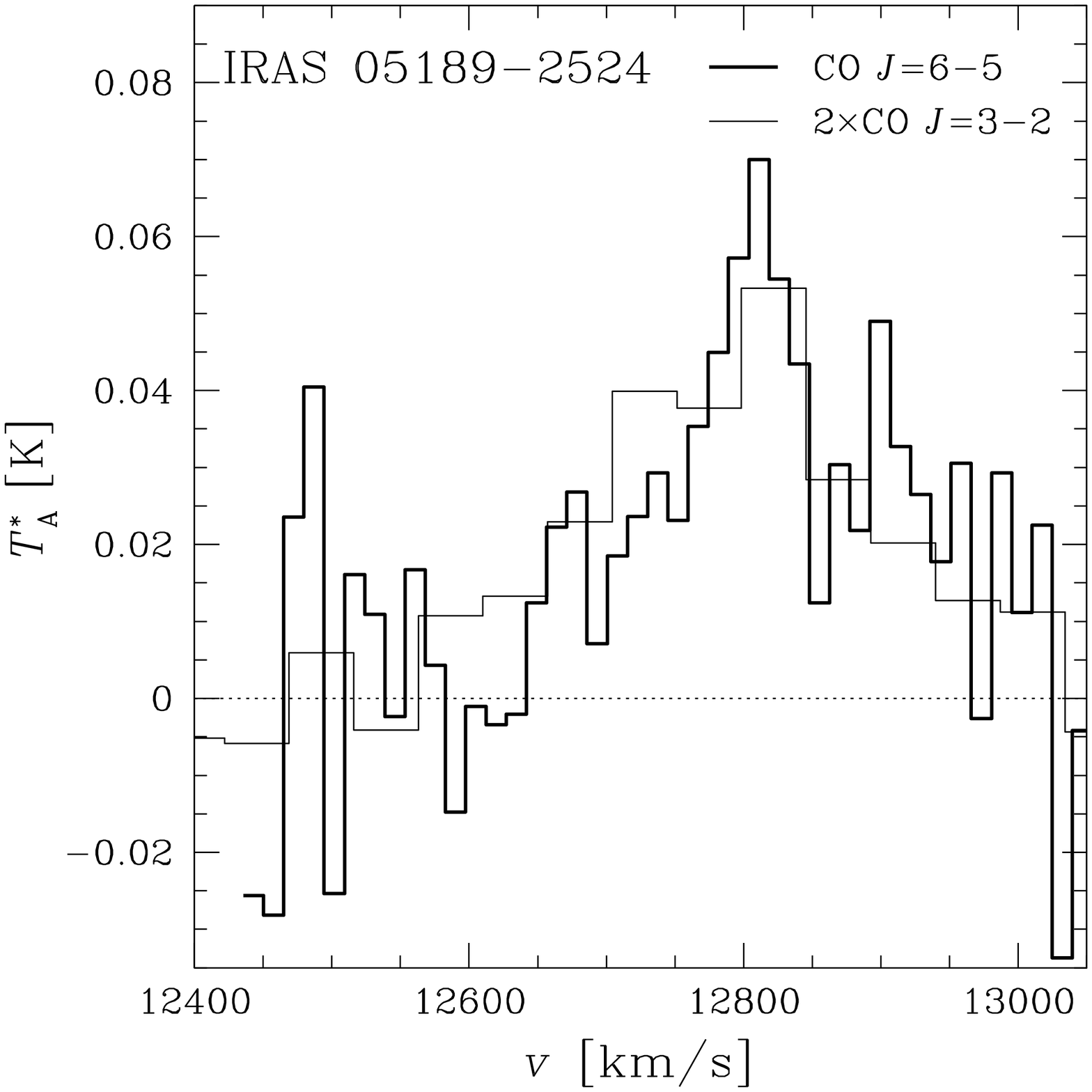} % CO32=47km/s, CO65=15km/s
\plotone{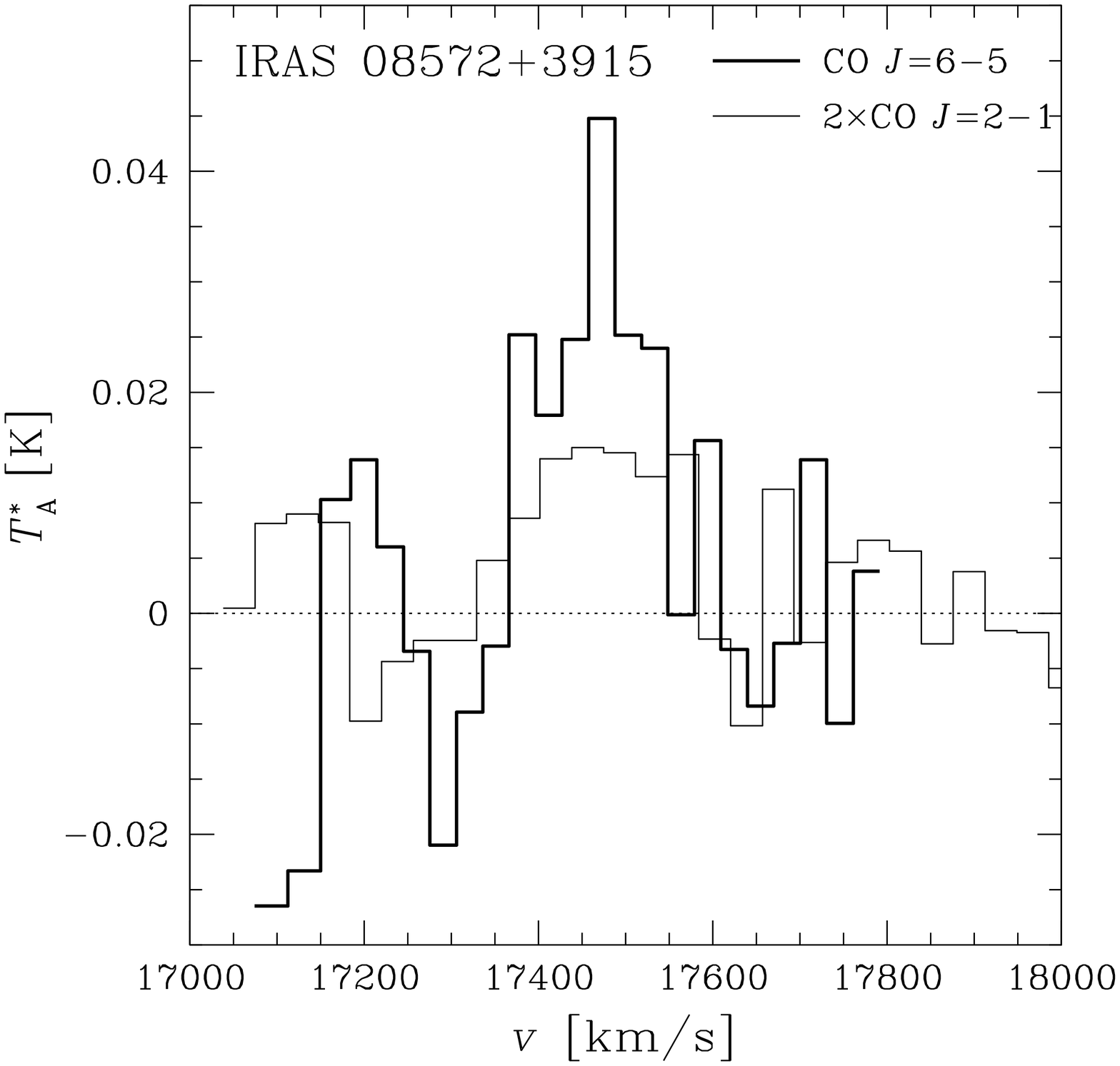} % CO21=36km/s, CO65=30km/s

\plotone{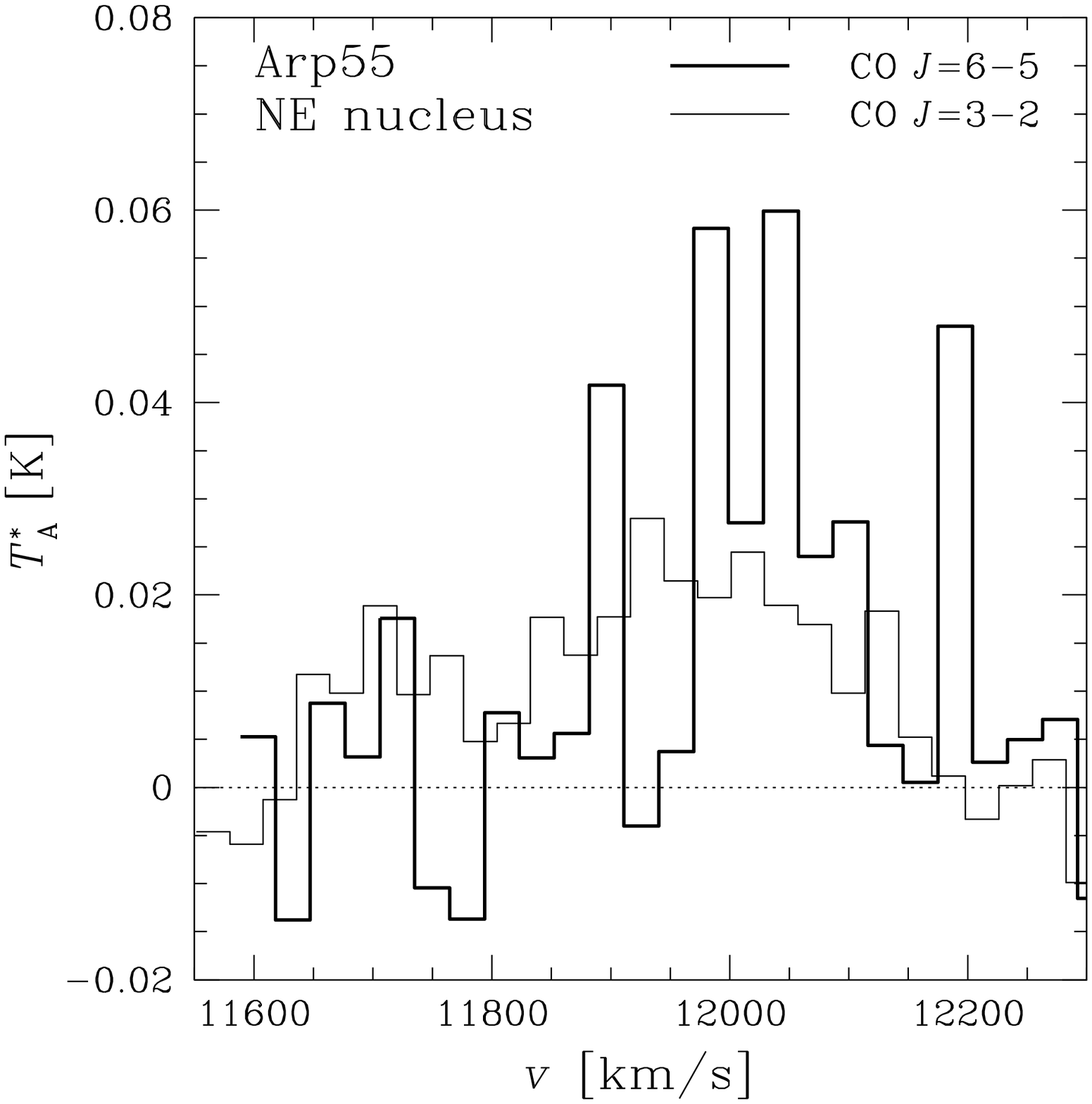}   % CO32=28km/s, CO65=29km/s
\plotone{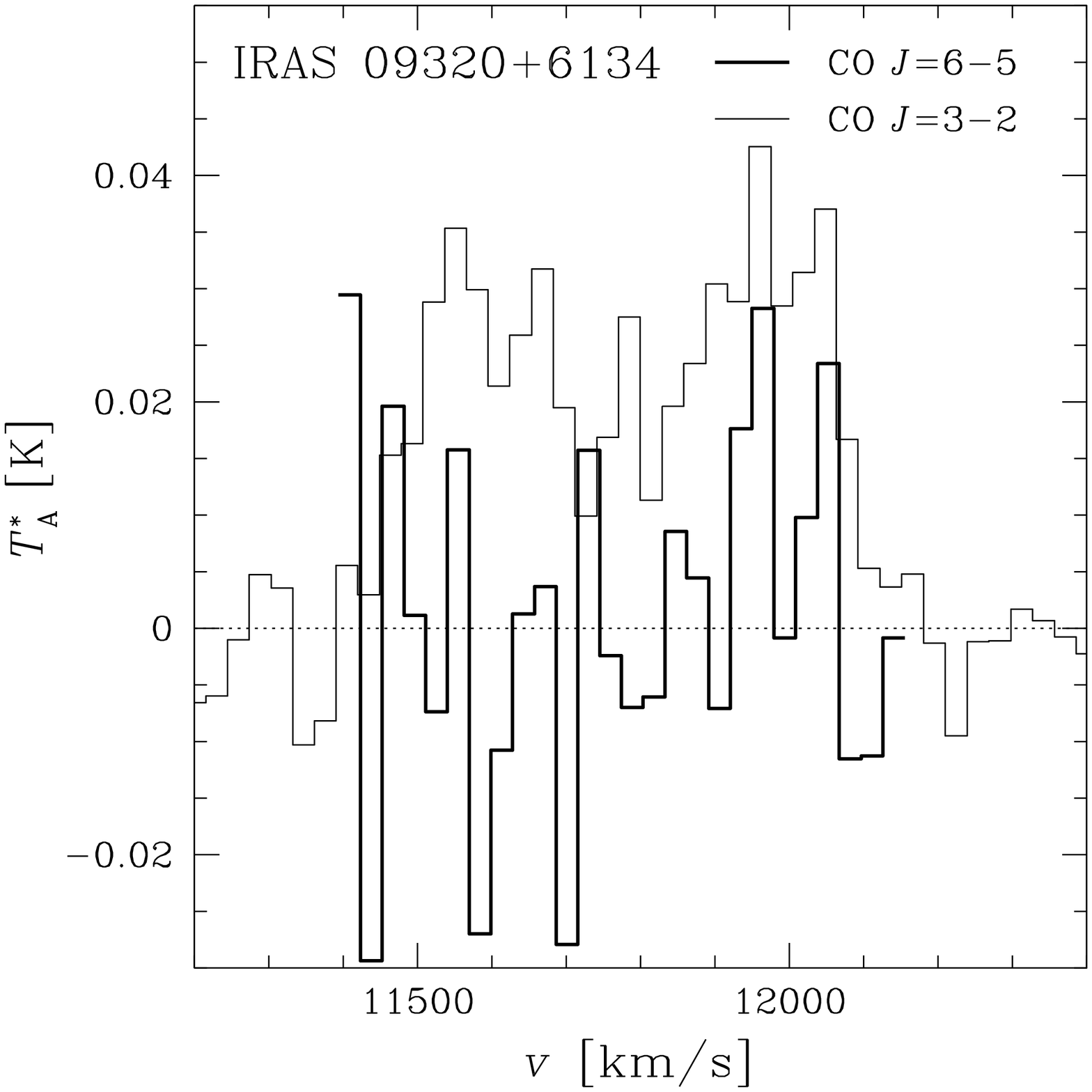} % CO32=CO65=29km/s  
\plotone{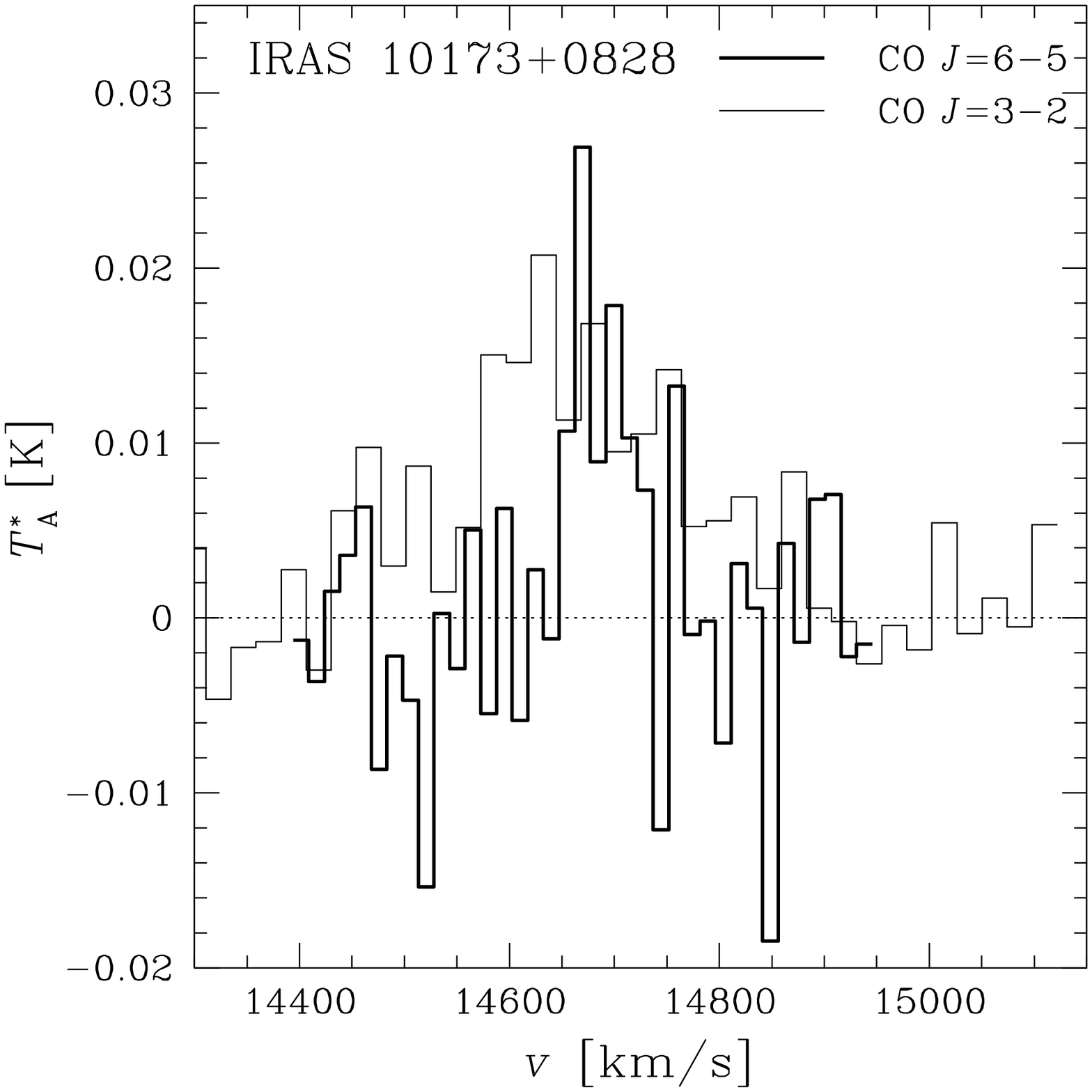} % CO32=24km/s, CO65=15km/s

\plotone{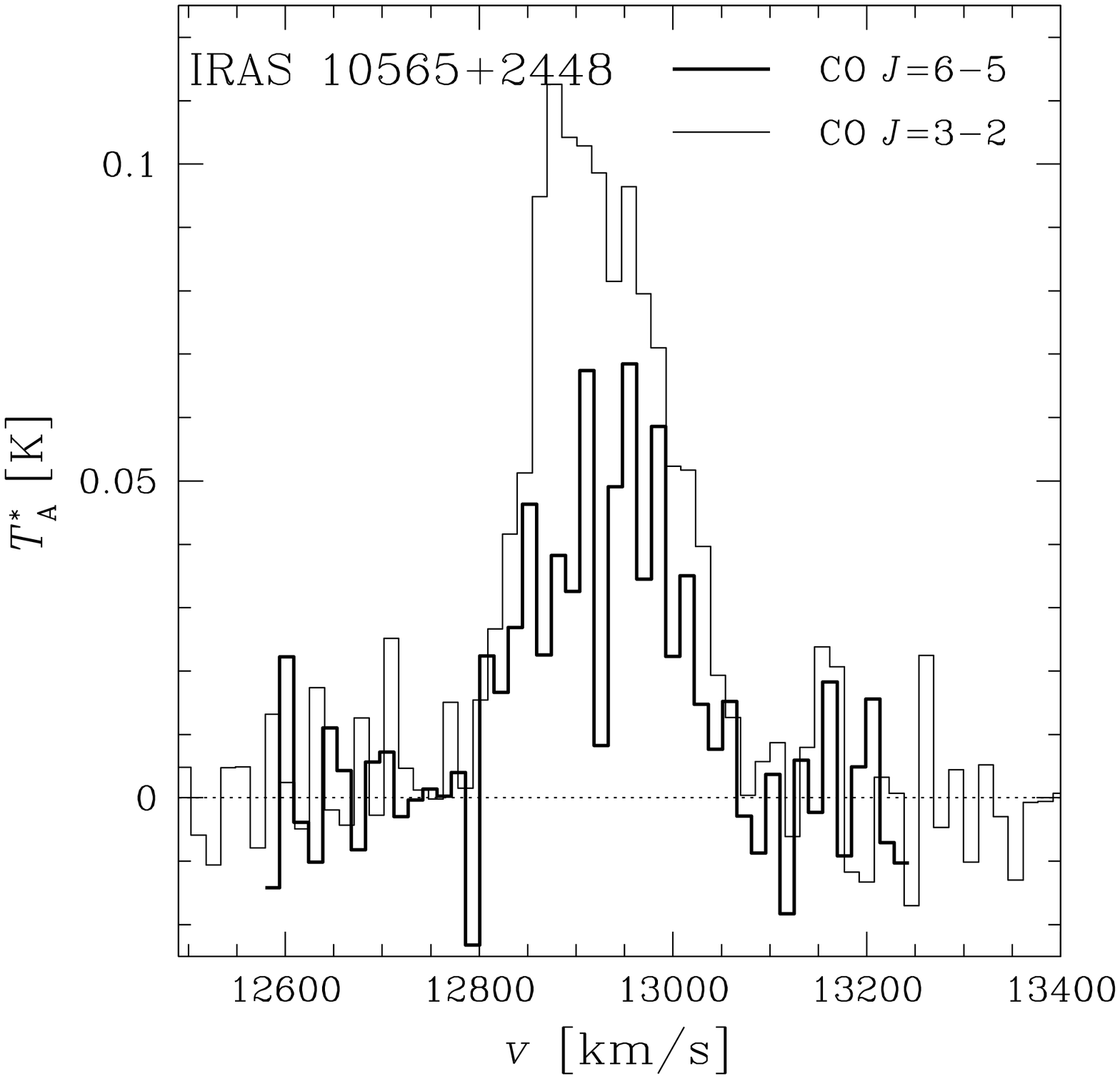} % CO32=CO65=15km/s
\plotone{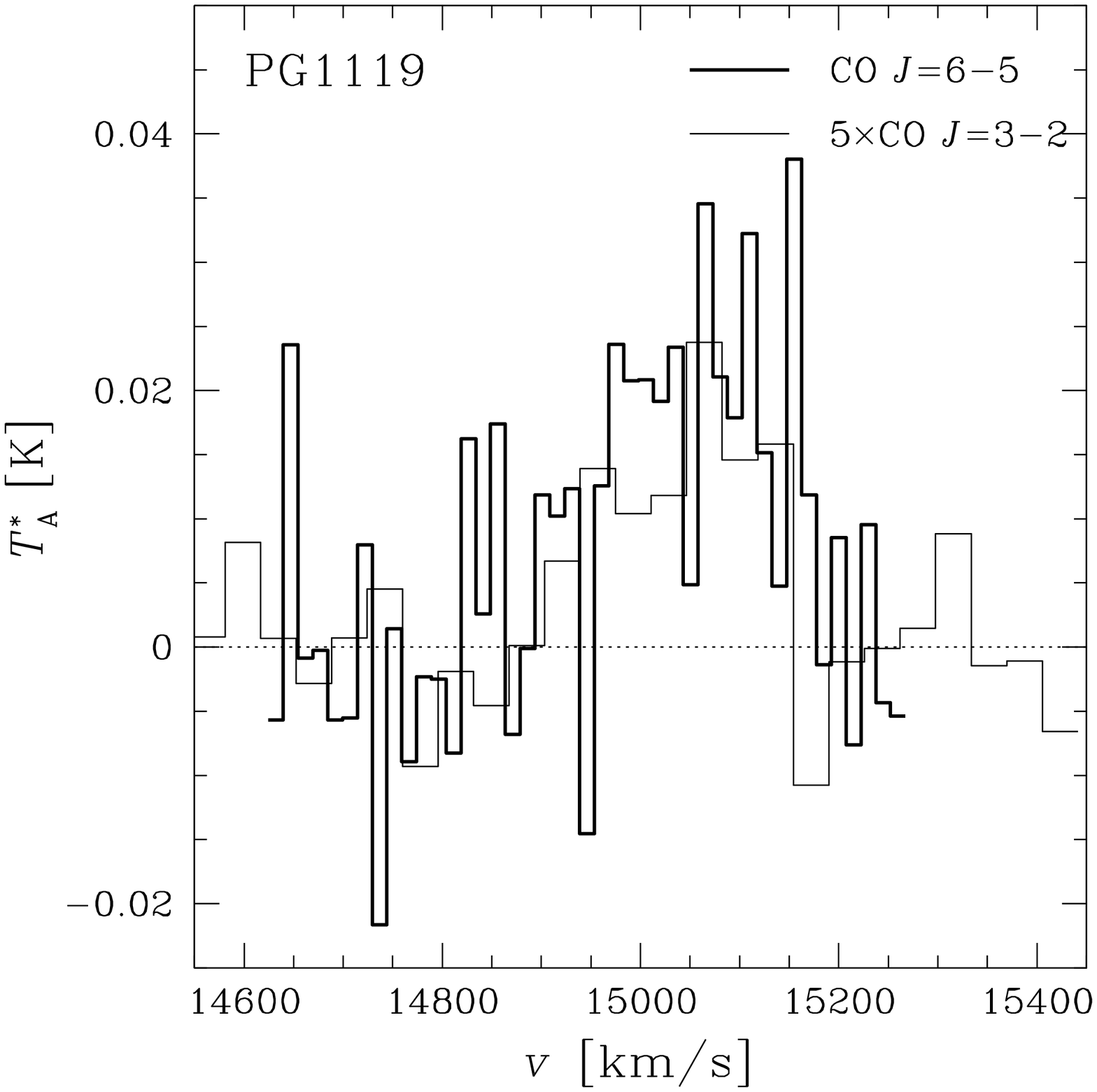}       % CO32=36km/s, CO65=15km/s
\plotone{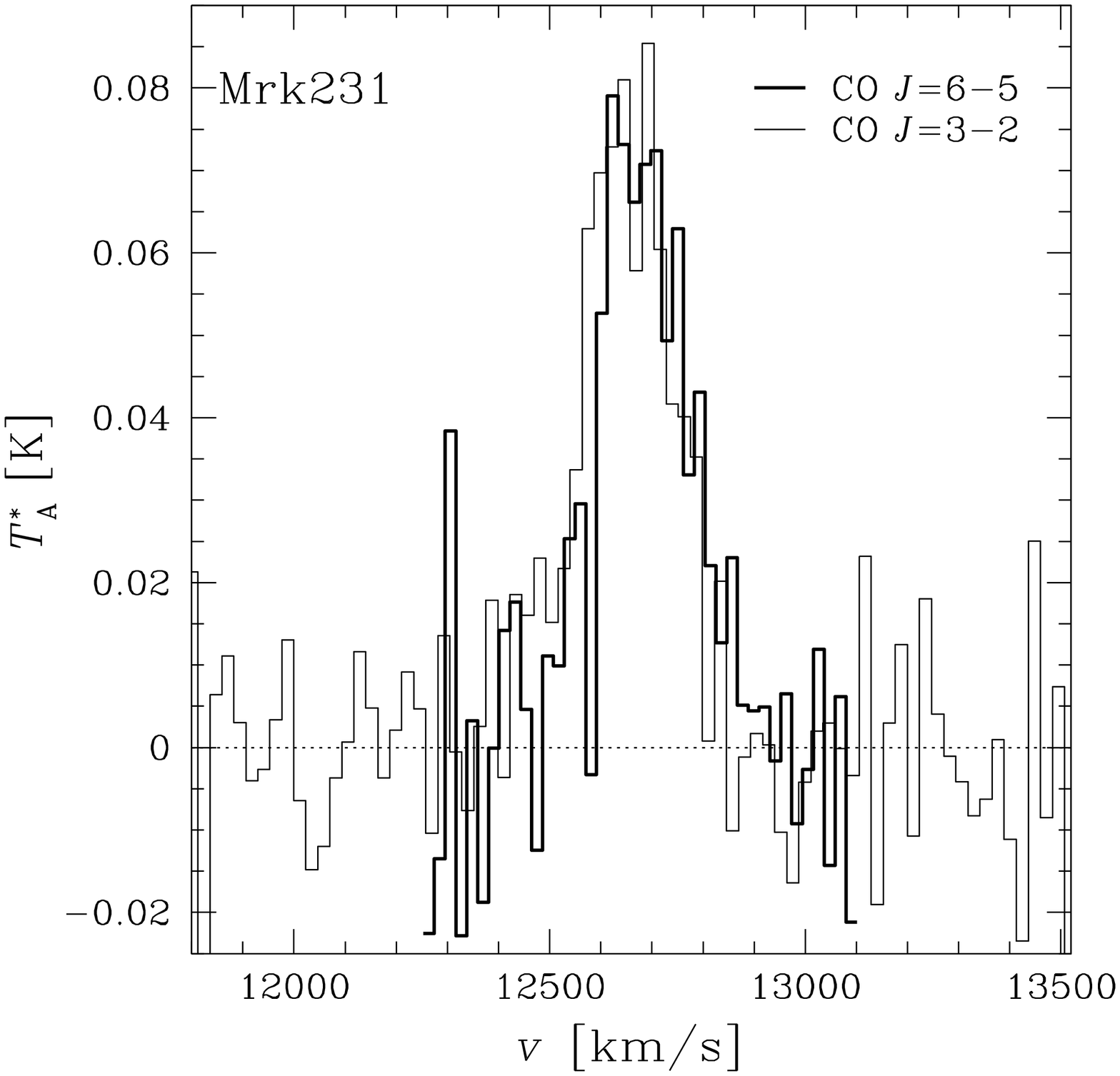}       % CO32=23km/s, CO65=22km/s

\caption{CO J=3--2,  6--5 spectra  for the galaxies  in Table  1.  The
velocity scale is LSR(cz) and typical velocity resolutions $\rm \Delta
V_{ch}$$\sim   $20--35\,km\,s$^{-1}$.  For   VII\,Zw\,31   the  J=2--1
transition   is   also   included   for  more   clarity,   while   for
IRAS\,08572+3915 the latter transition is the only one available.}
\end{figure}
\addtocounter{figure}{-1}
\begin{figure}

\plotone{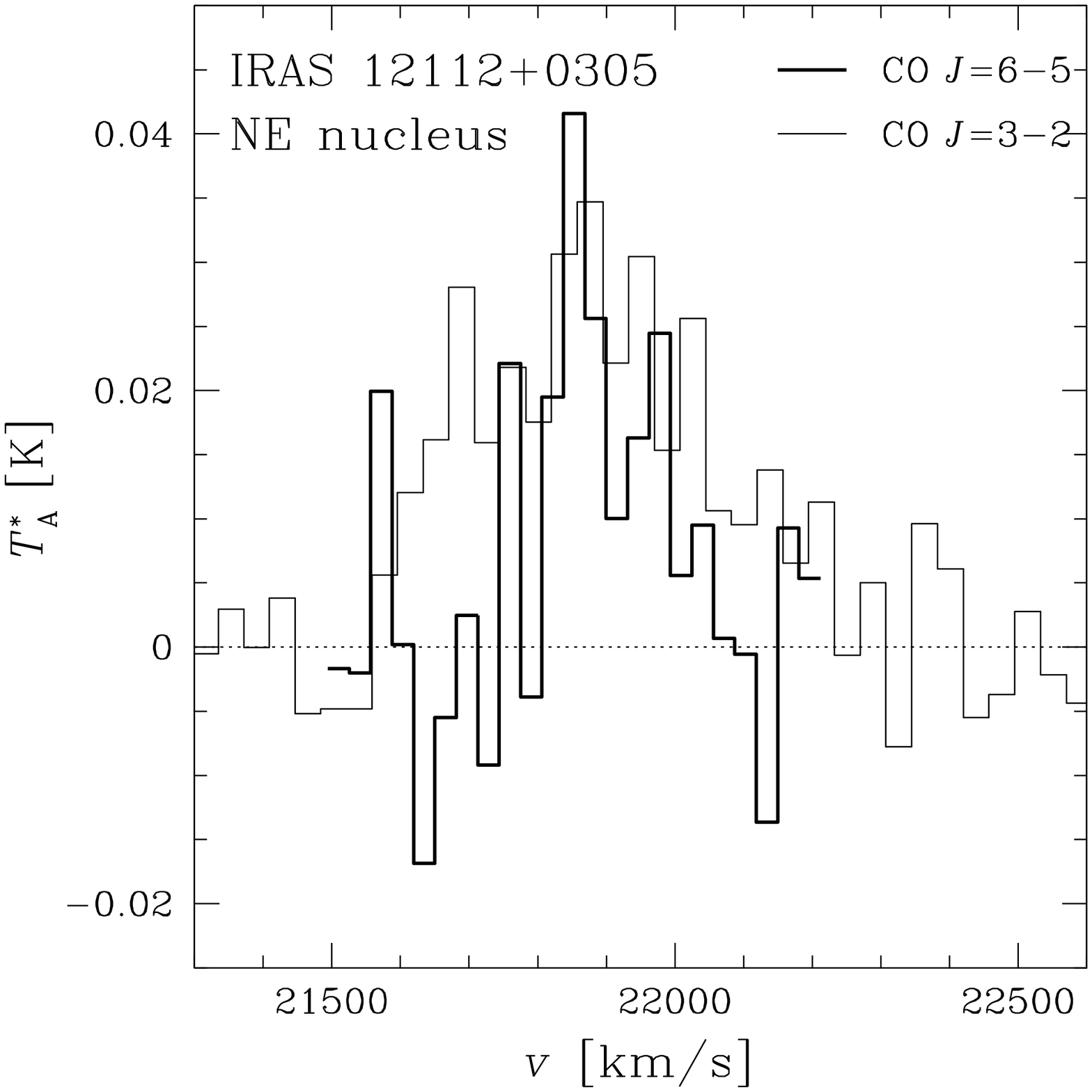} % CO32=37km/s, CO65=31km/s
\plotone{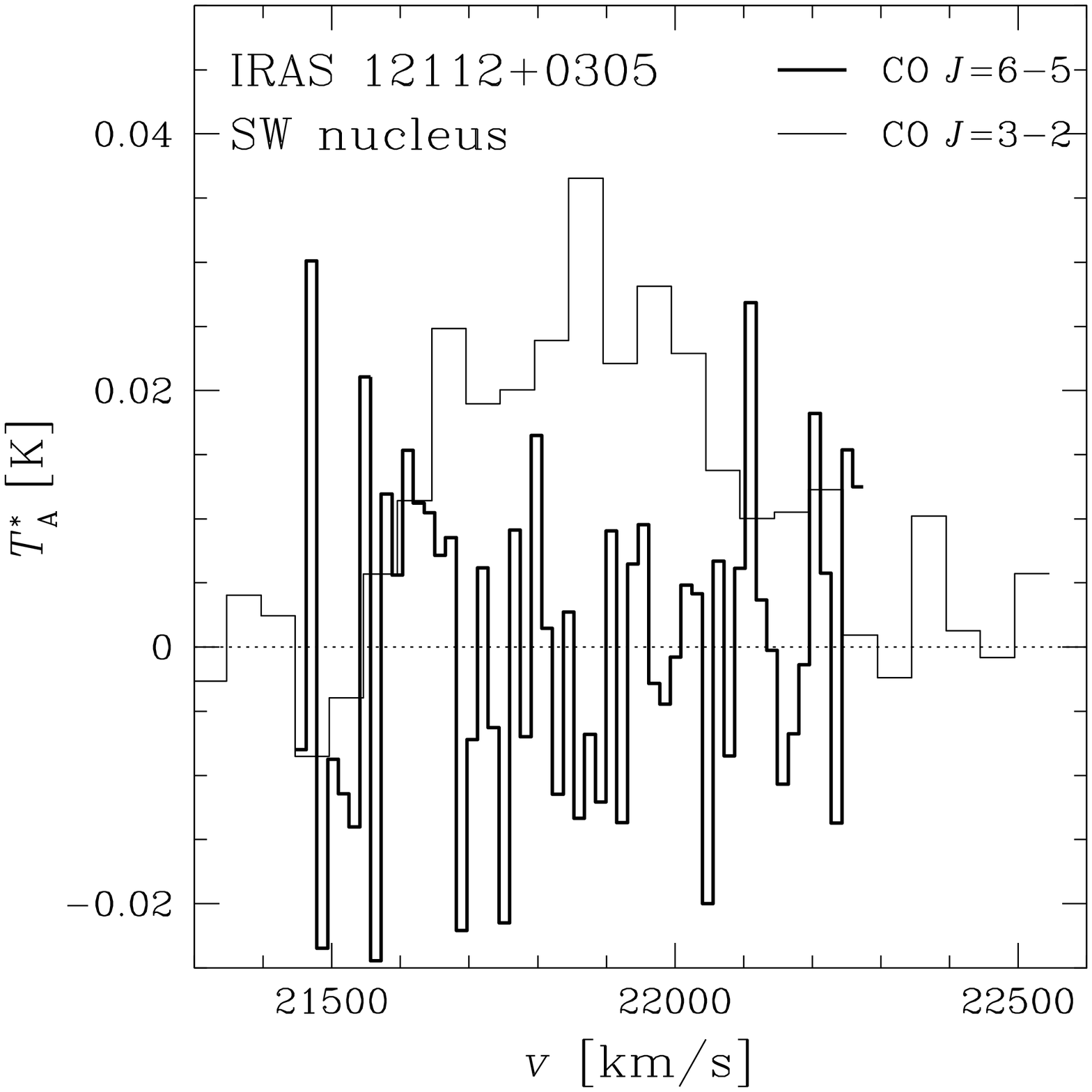} % CO32=50km/s, CO65=16km/s
\plotone{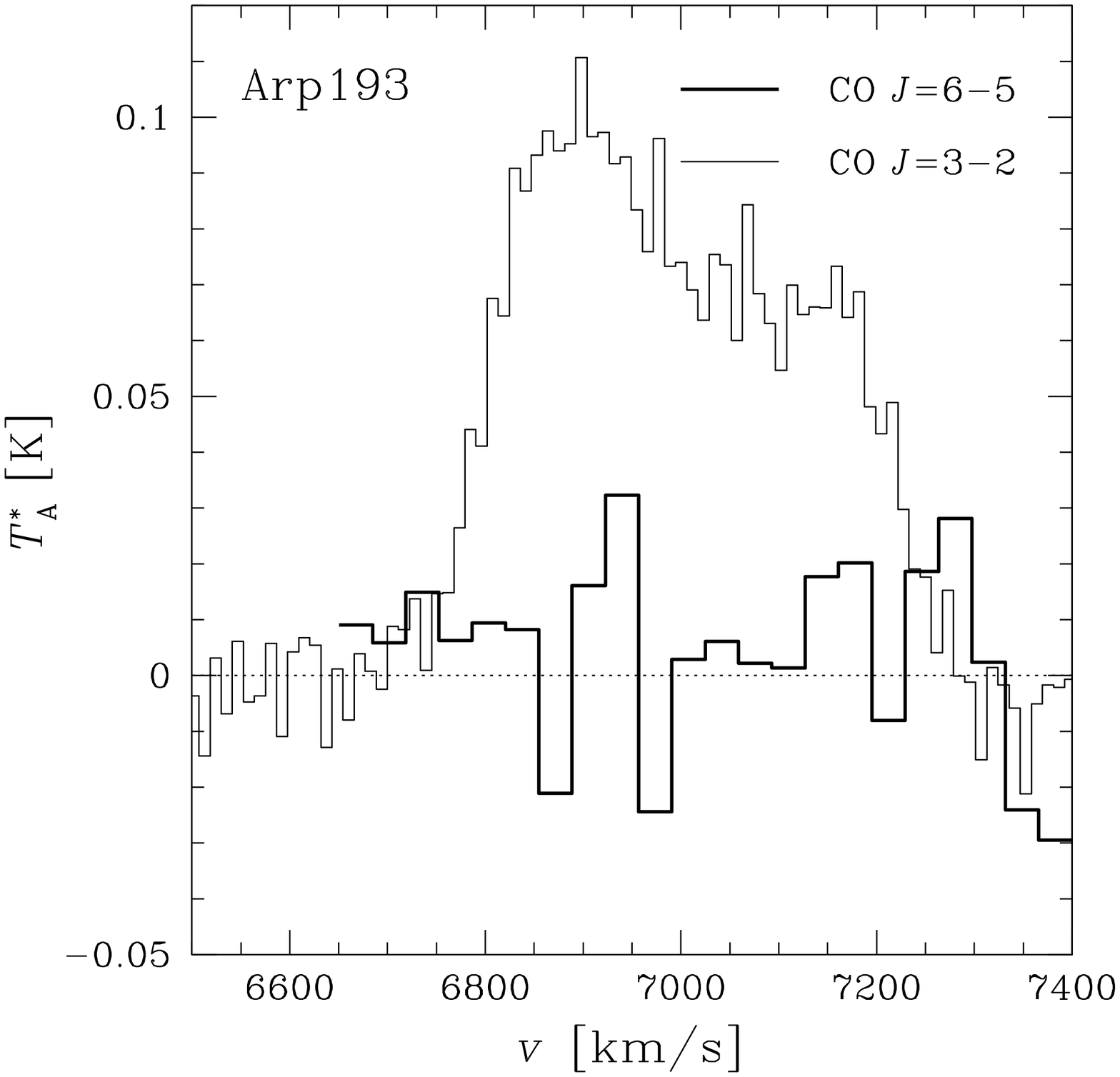}         % CO32=11km/s, CO65=34km/s

\plotone{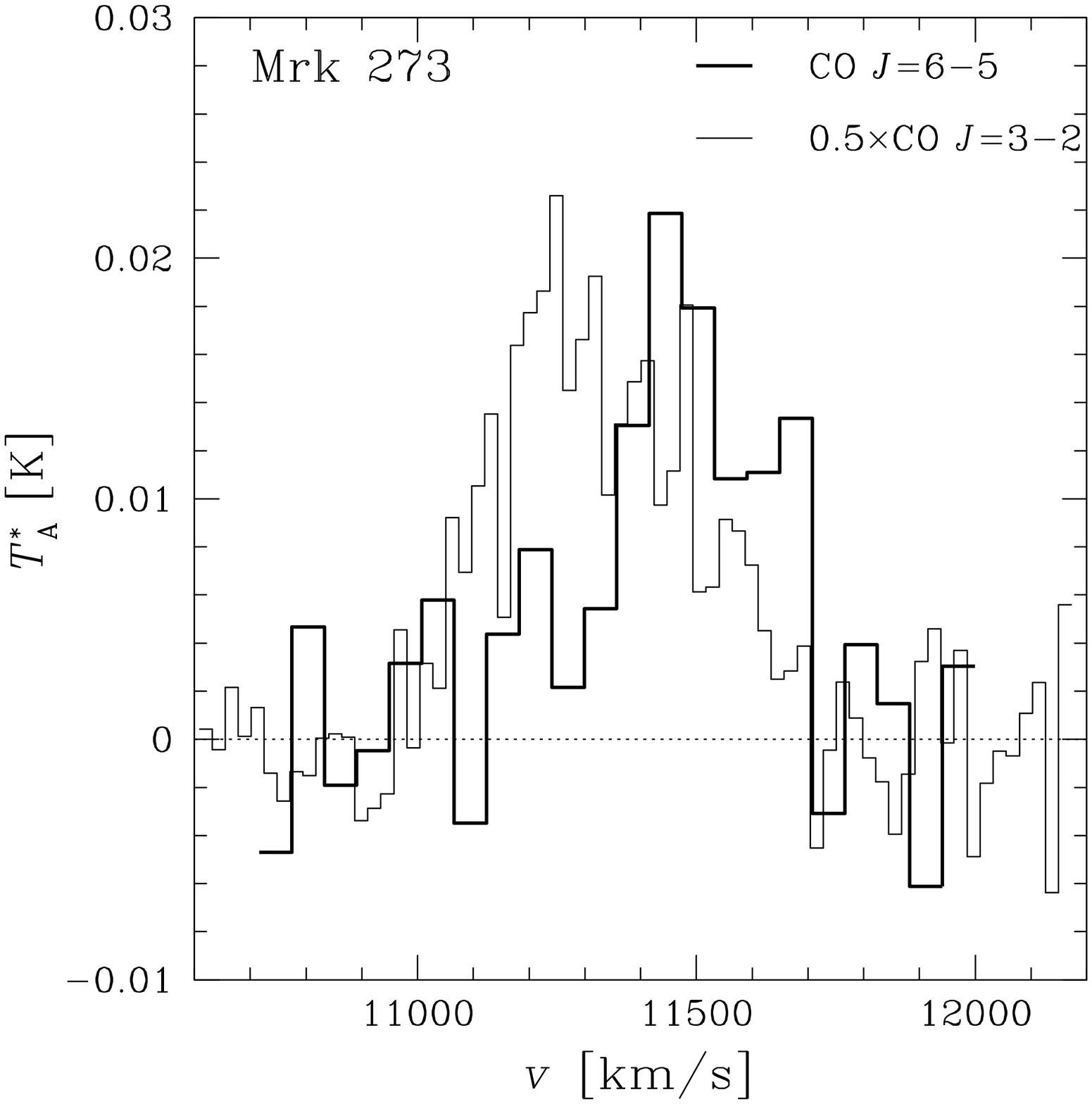}    % CO32=22km/s, CO65=58km/s
\plotone{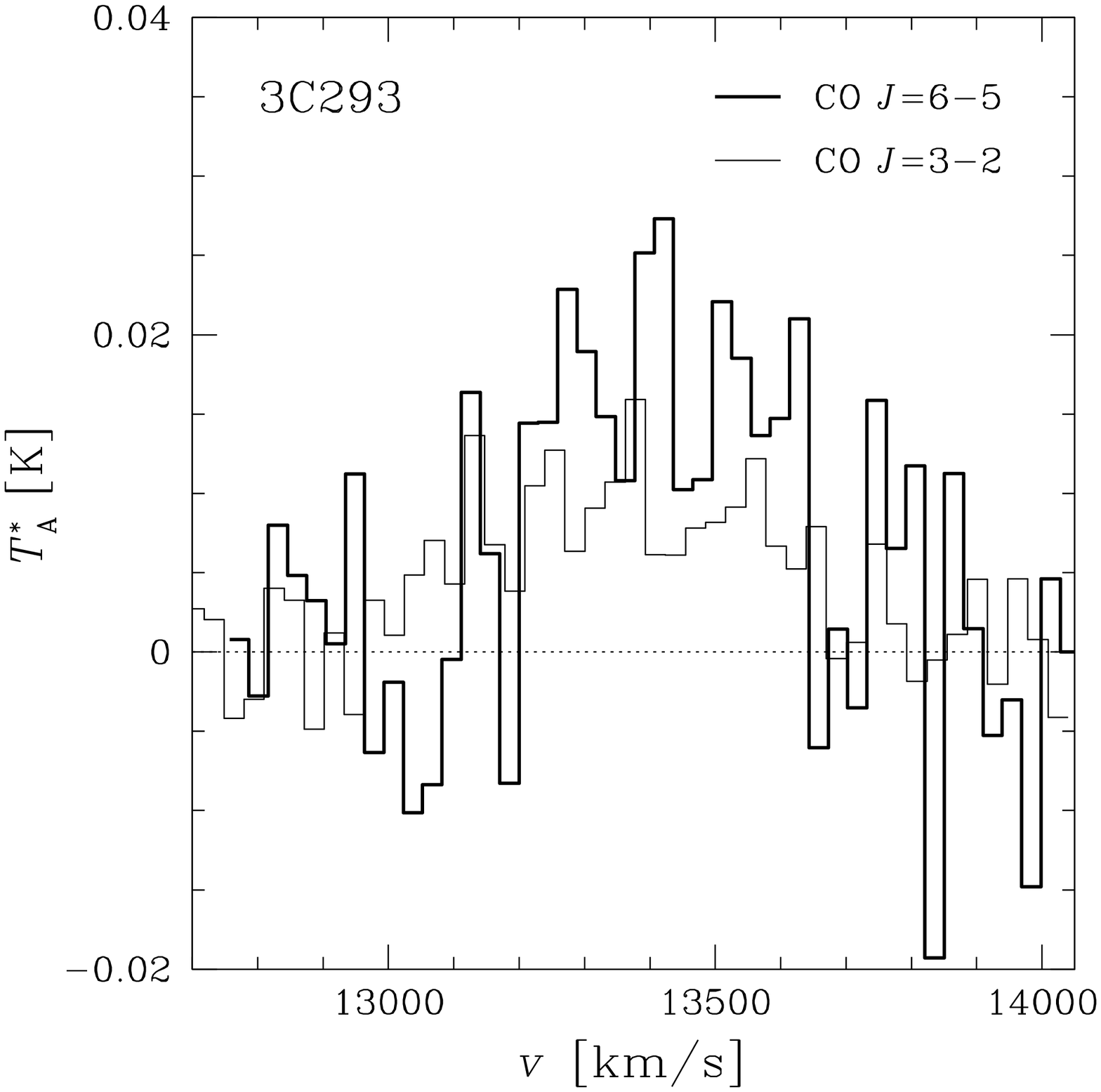}          % CO32=CO65=30km/s
\plotone{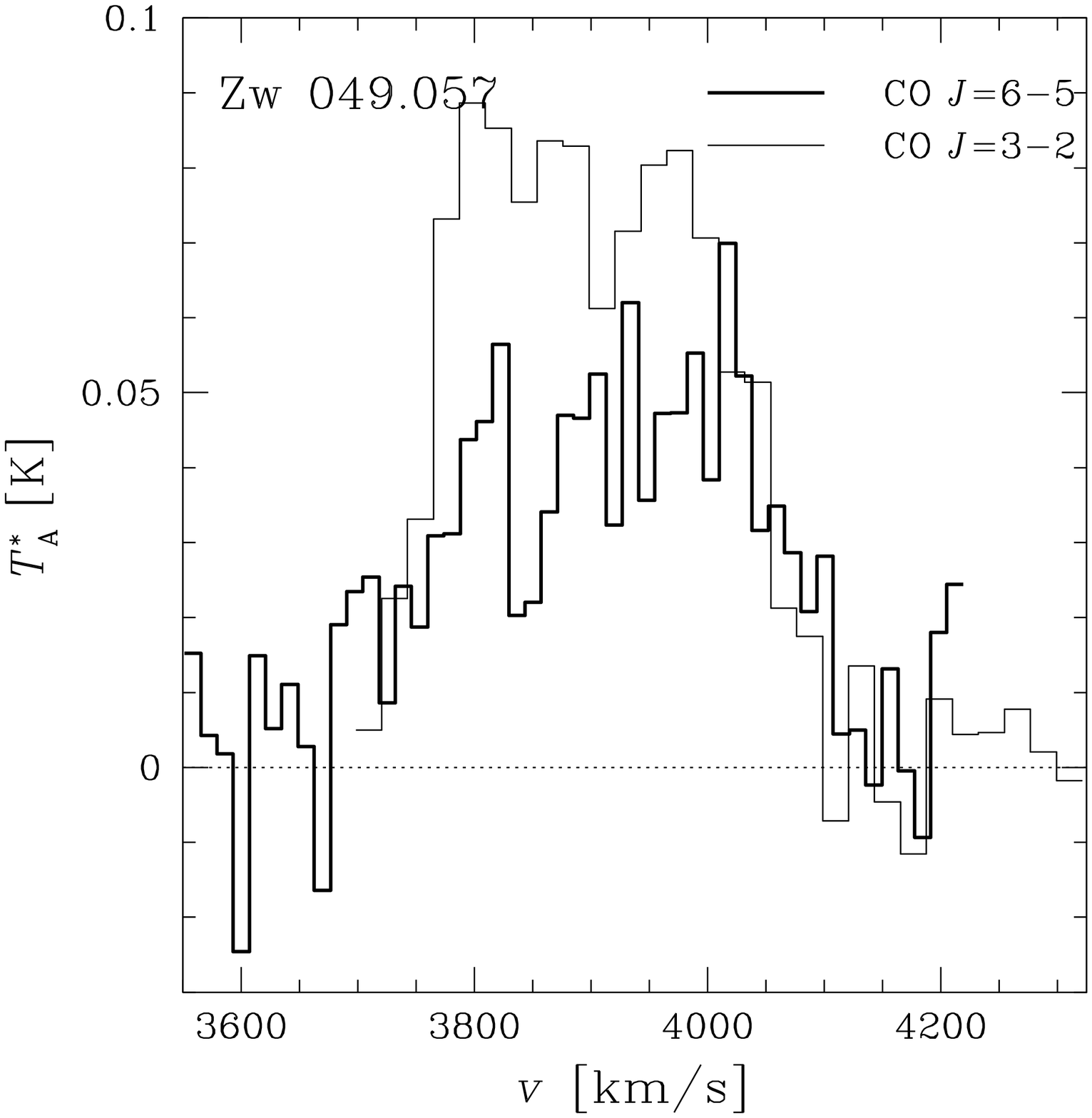}          % CO32=22km/s, CO65=14km/s

\plotone{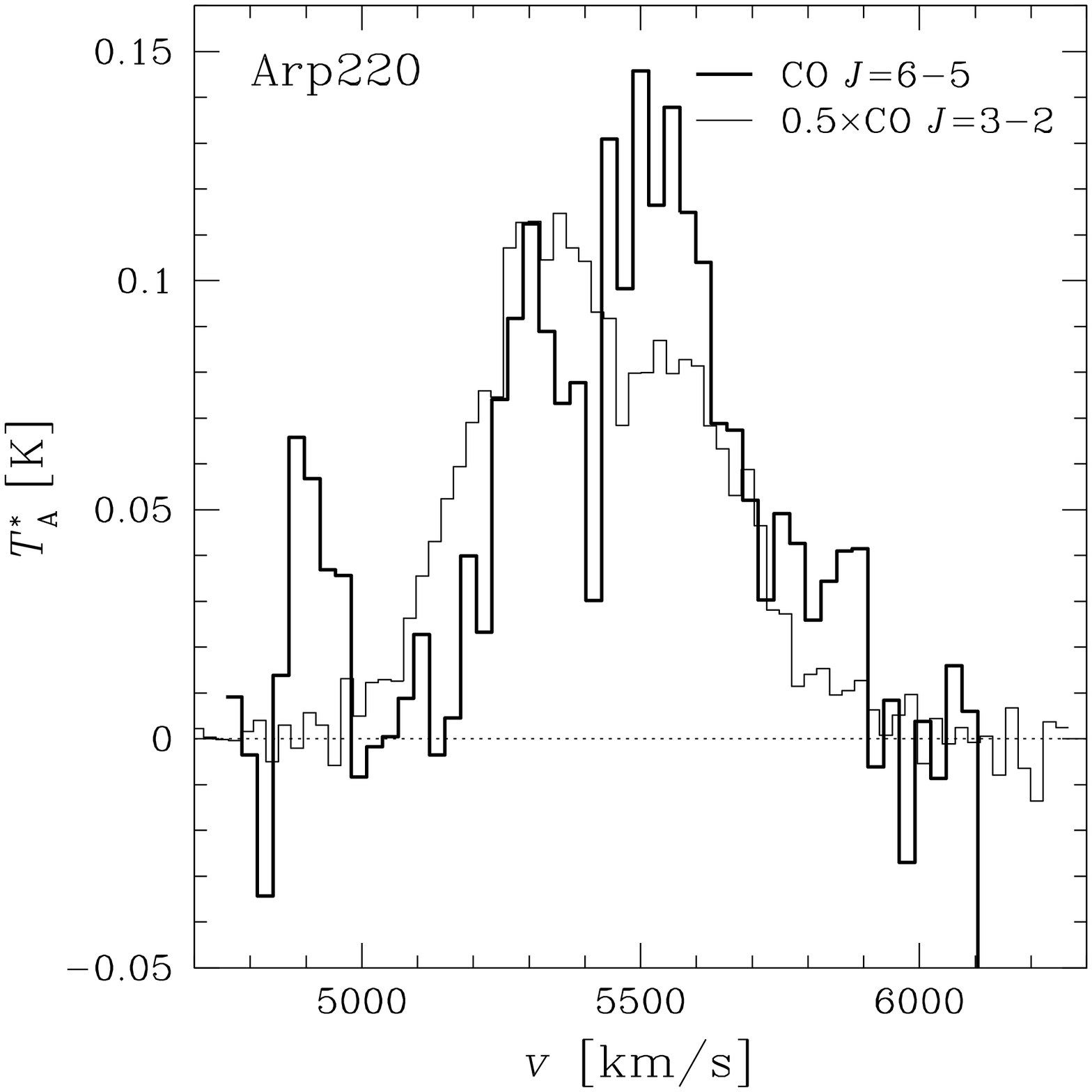} % CO32=22km/s, CO65=28km/s
\plotone{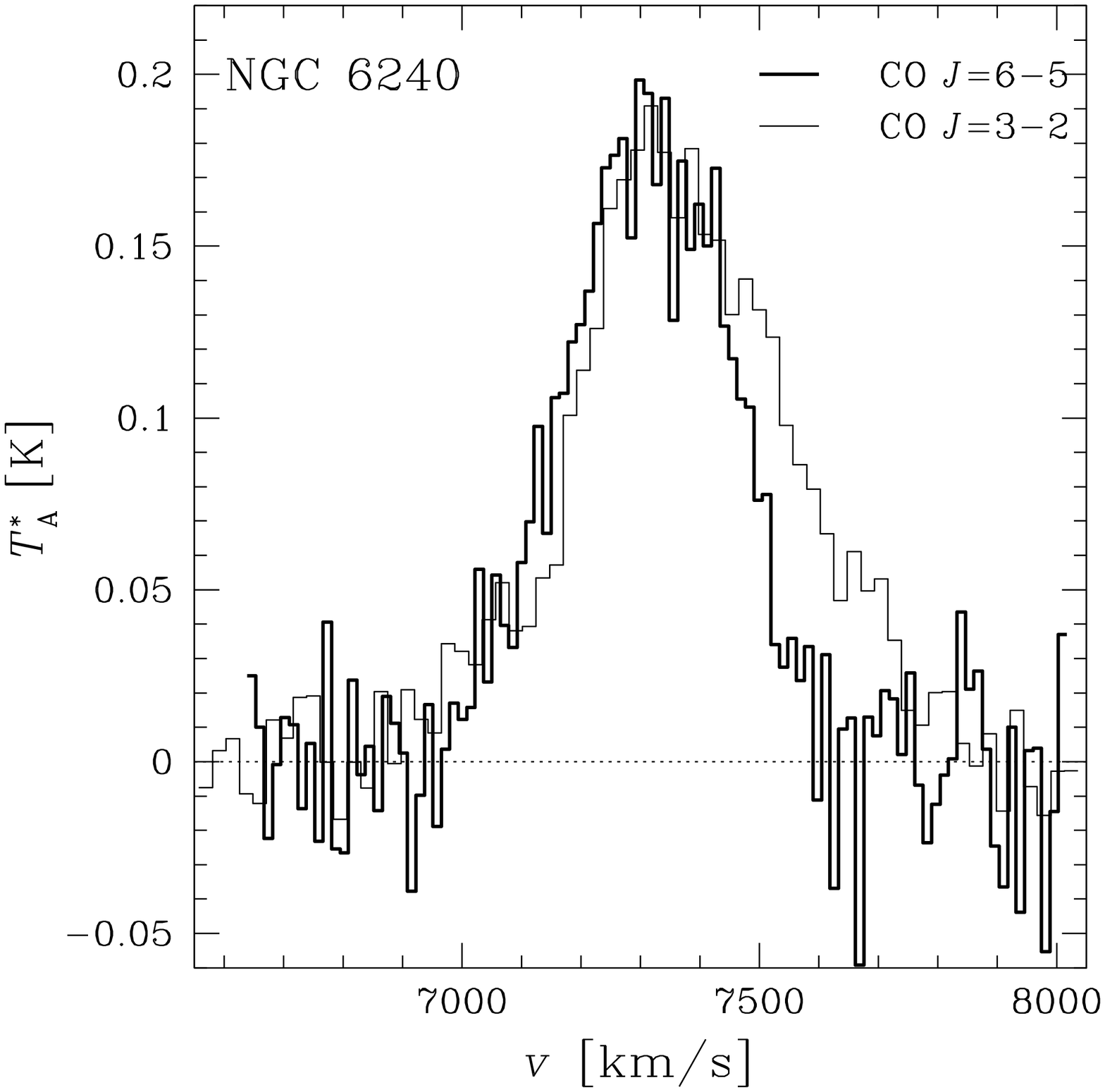}      % CO32=23km/s, CO65=14km/s
\plotone{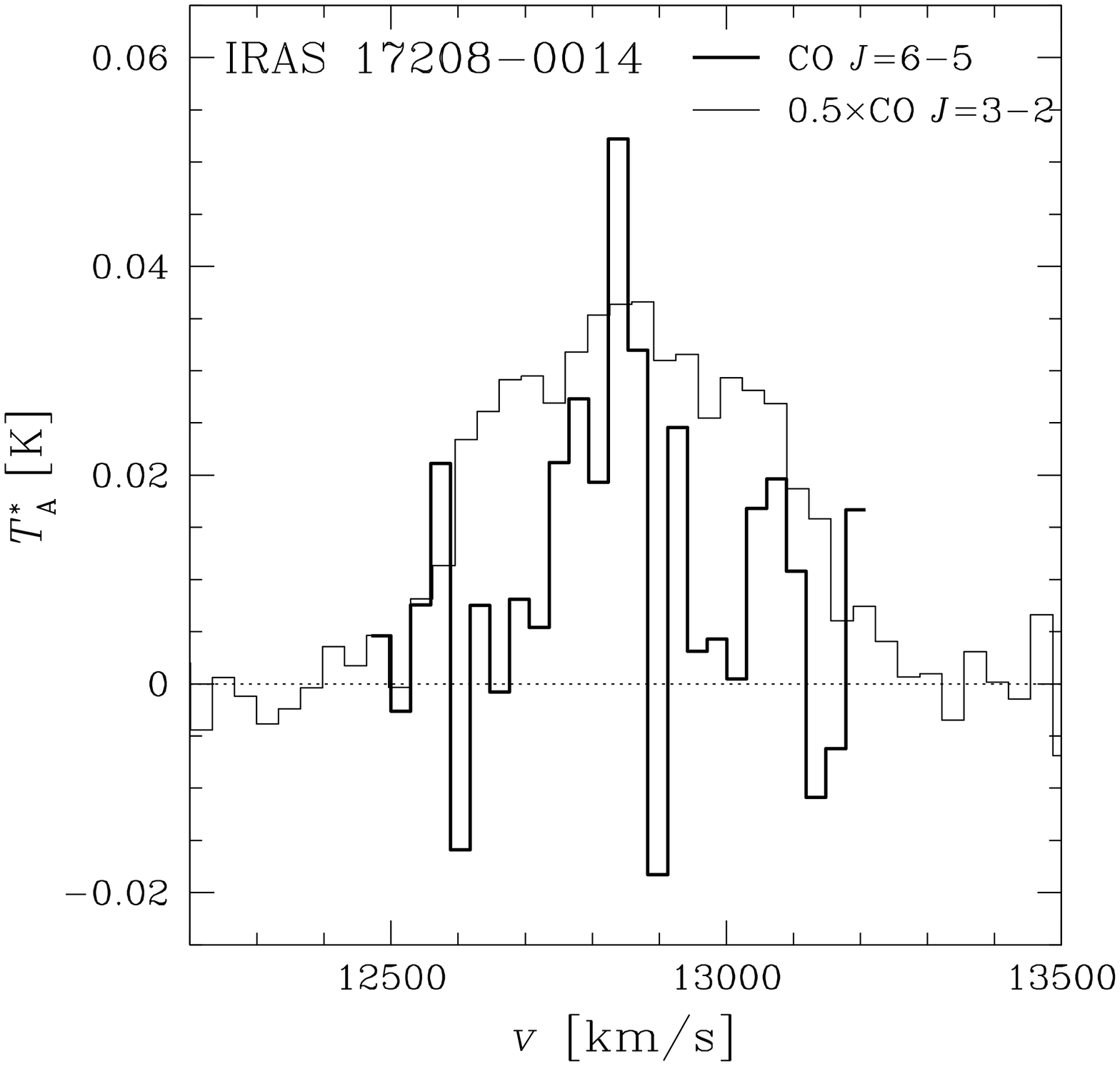}      % CO32=30km/s, CO65=29km/s
%\plotone{IR171208-0014.ps}

\plotone{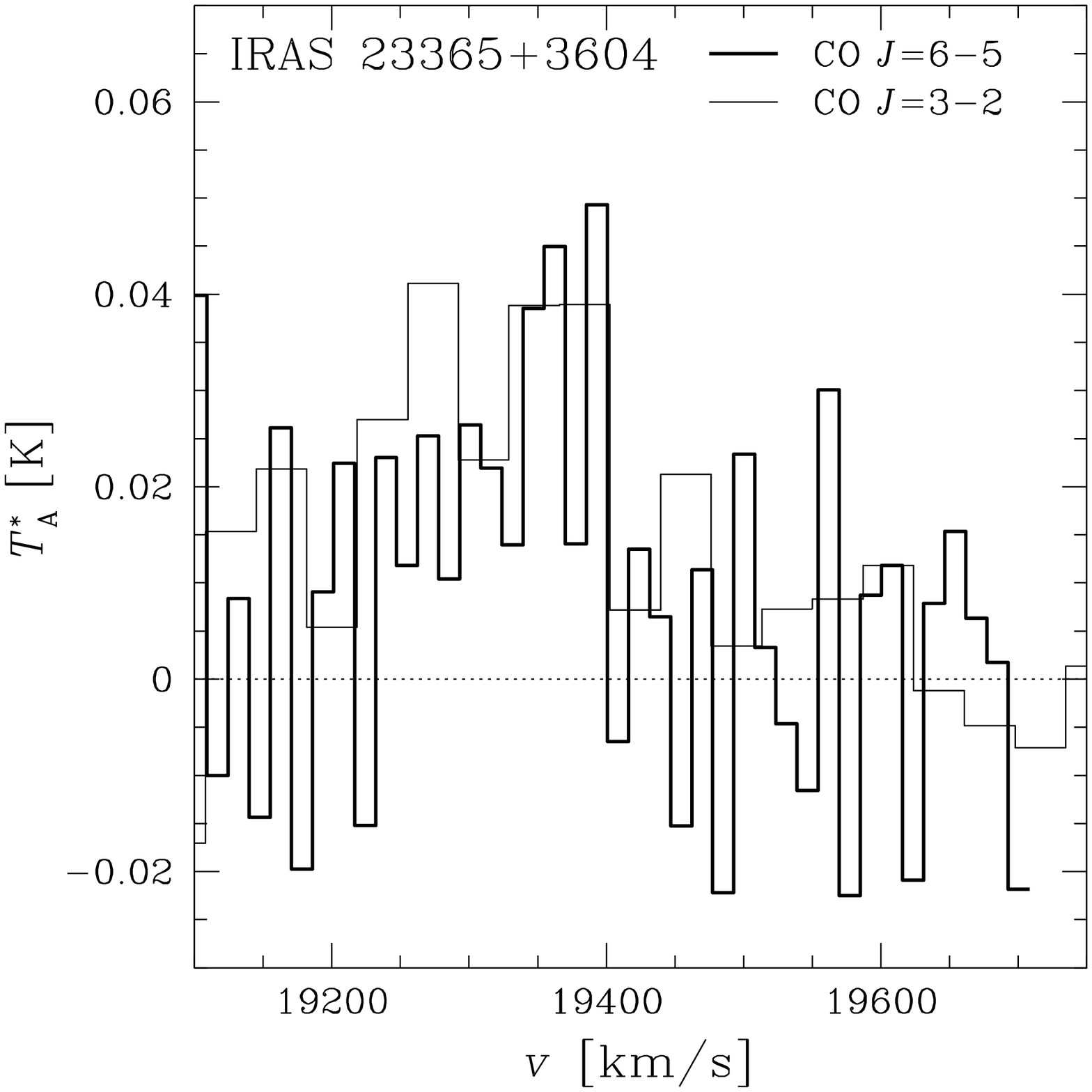}       % CO32=37km/s, CO65=15km/s
\plotone{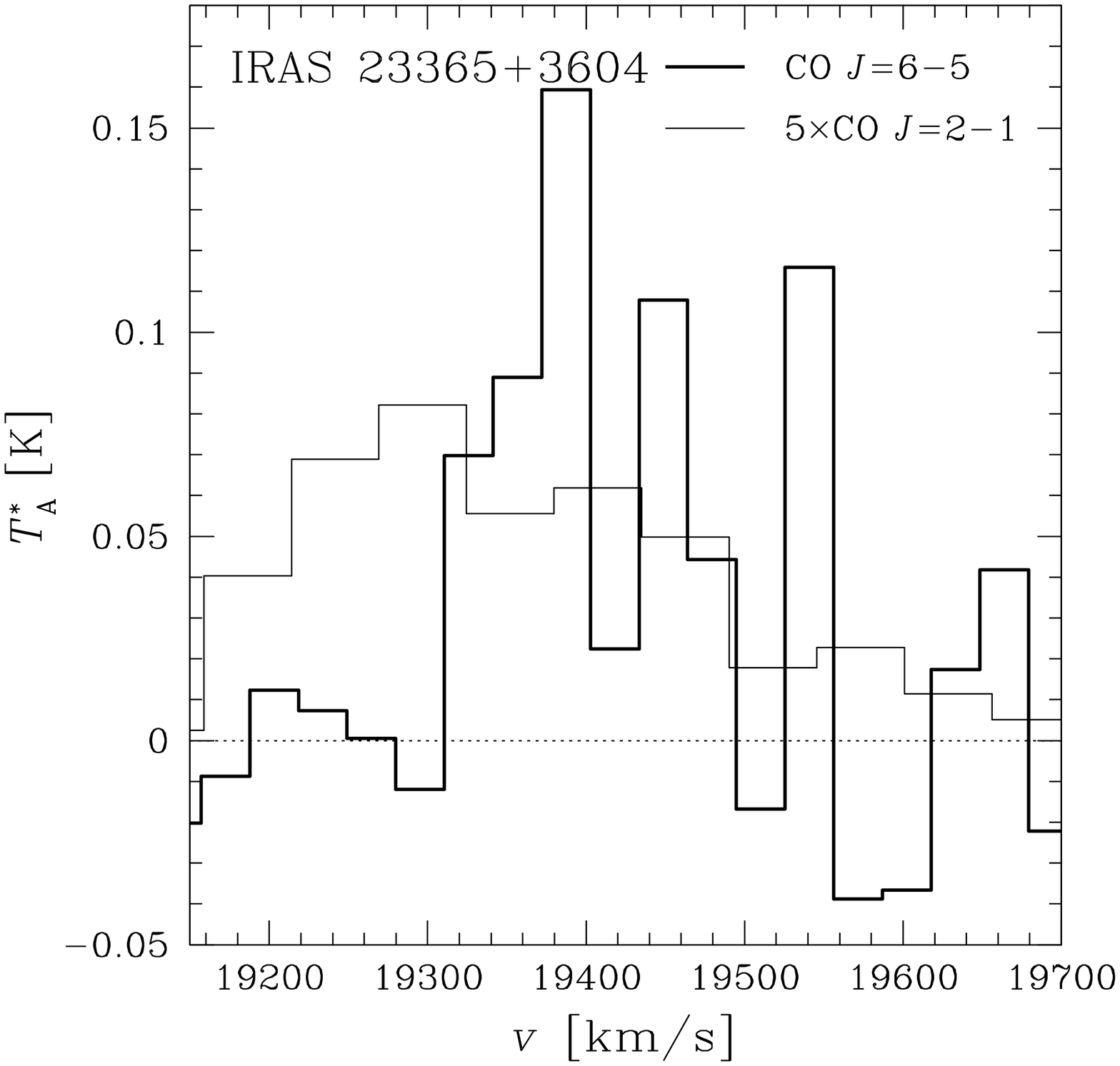} % CO21=55km/s, CO65=30km/s

\vspace*{-0.4cm}
\caption{(cont.) CO J=3--2, 6--5 spectra  for the galaxies in Table 1,
with   typical   velocity   resolutions   $\rm   \Delta   V_{ch}$$\sim
$20--30\,km\,s$^{-1}$, and in special cases up to 50--60\,km\,s$^{-1}$
(CO 6--5 in Mrk\,273, CO 2--1 in IRAS\,23365+3604). The CO J=6--5 line
in IRAS\,23365+3604  is shown overlaid separately with  the J=2--1 and
J=3--2  transitions for  clarity, if  detected this  line seems  to be
present over a narrower FWZI than in the two lower-J transitions. }
\end{figure}

\newpage

\begin{figure}
% Figure 3
\epsscale{1.0}
\plotone{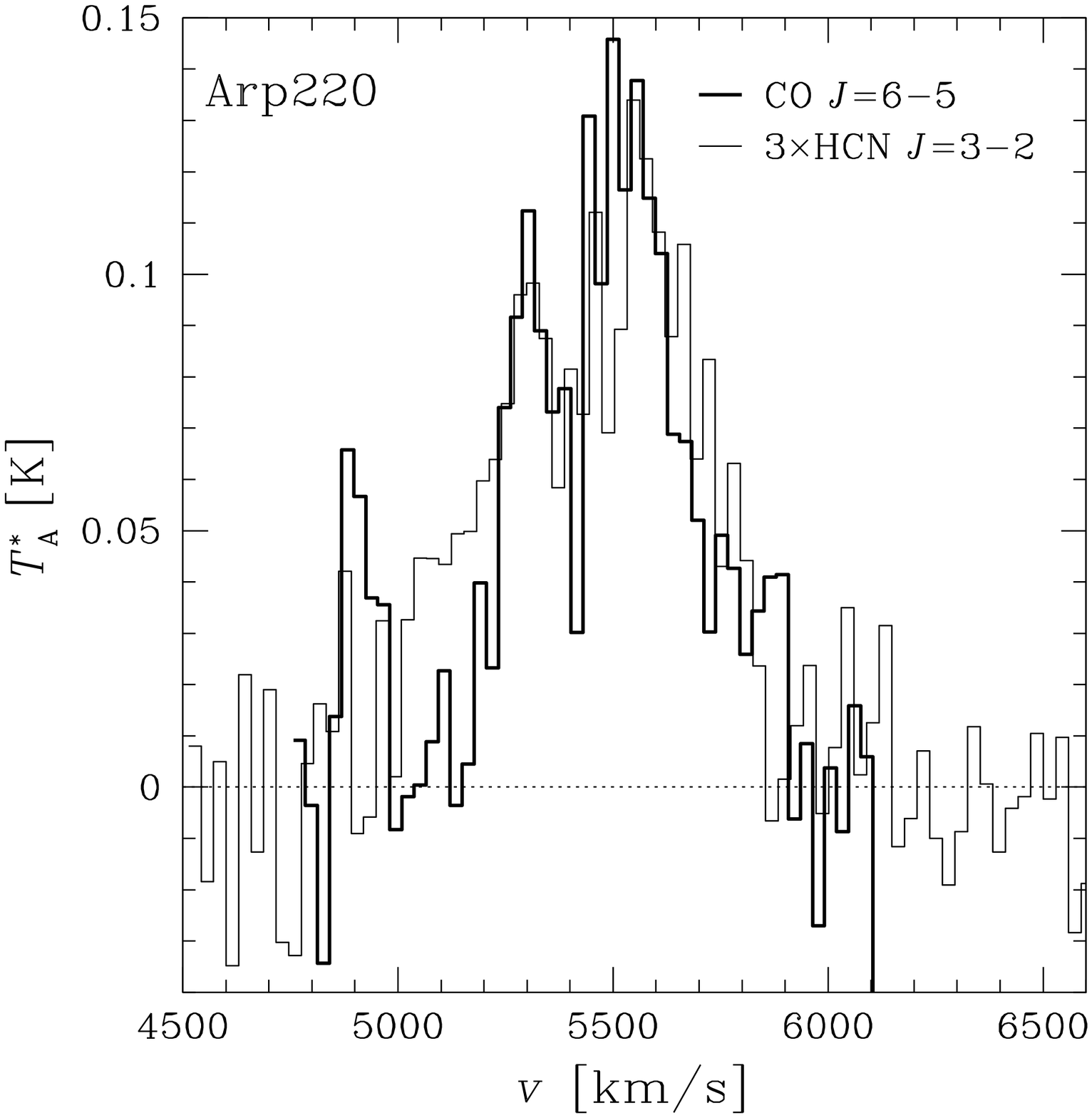}
\caption{The new CO J=6--5  spectrum of Arp\,220 overlayed on 3$\times
$HCN(3--2) from the  JCMT.  Both spectra are clearly  rising at higher
velocities associated with the  eastern nucleus that contains the bulk
of the  high density gas as indicated  also by Greve et  al. 2009. The
velocity  frame  is LSR(cz),  and  the  velocity  resolution for  both
spectra is 28\,km\,s$^{-1}$.}
\end{figure}

\newpage

\begin{figure}
%Figure 4
\epsscale{1.1}
\plotone{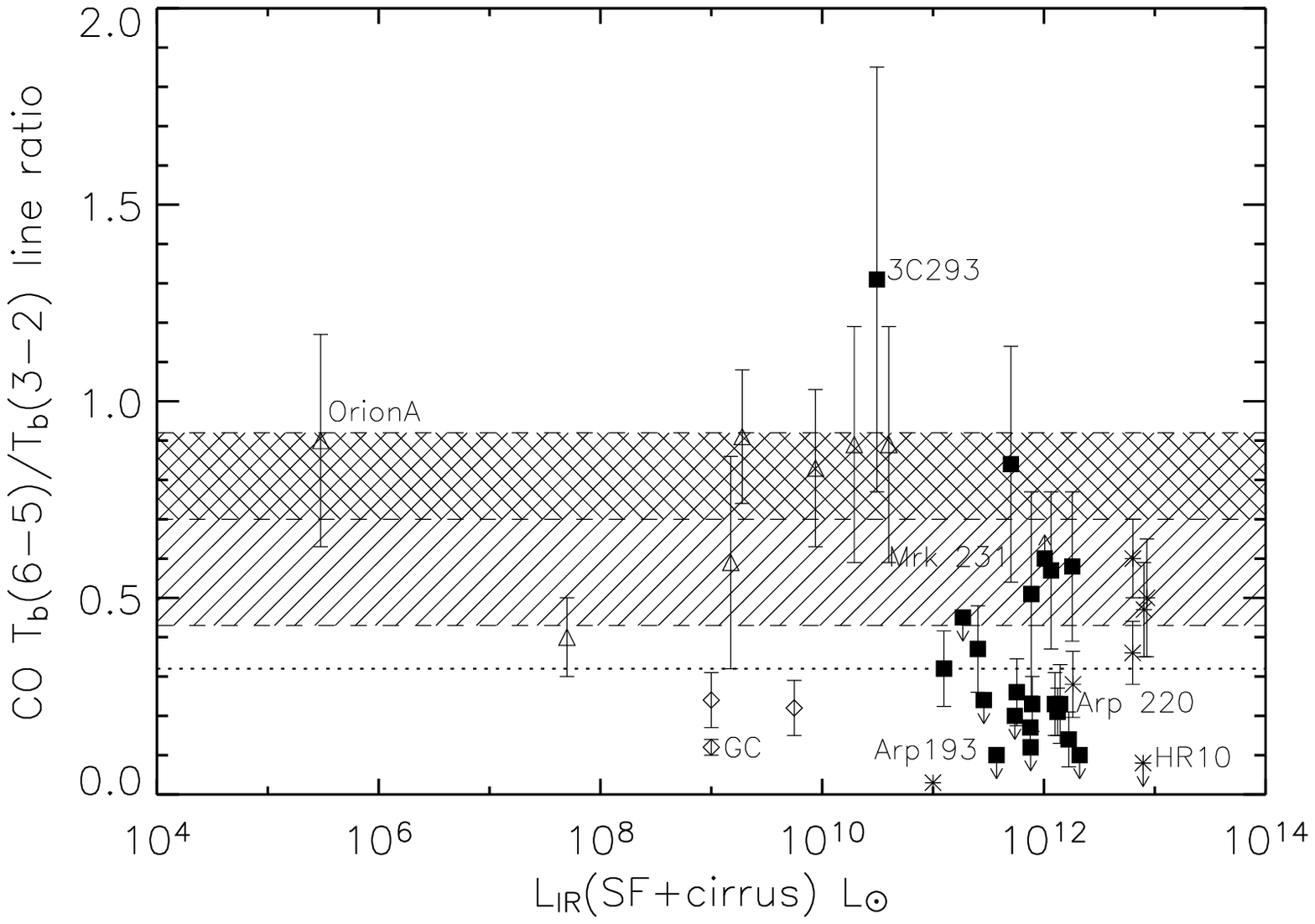}
\caption{The  CO  (6--5)/(3--2)  ratio  versus  non-AGN  IR-luminosity
(=SF+cold  cirrus contributions, listed  as $\rm  L_{IR}$ in  Table 2)
for:  a)   quiescent  (diamonds)  and   starburst  (triangles)  nearby
galaxies, b) our  sample of LIRGs (solid boxes),  and c) SMGs (stars).
We have also included a few nearby low luminosity systems (Bradford et
al.   2003; Bayet  et  al.  2006;  Mao  et al.   2000),  SMGs at  high
redshifts (Tacconi et  al.  2006; Papadopoulos \& Ivison  2002), and a
Ly-break galaxy (Baker et al.  2004).  The star-forming cloud Orion\,A
and  the Galactic  Center  (GC, SF-quiescent)  are  also marked.   The
shaded areas  mark the expected  range of $\rm R_{65/32}$  for typical
conditions of  the star-forming gas in ULIRGs  (cross-hatched) and the
less extreme star-forming environments of LIRGs (hatched).  The dotted
line marks the lowest possible $\rm R_{65/32}$ value for the dense and
cold GMC cores ($\rm n(H_2)$$>$10$^4$\,cm$^{-3}$, and T$\rm _k$=10\,K)
(see  5.2),  lower  values  are   possible  only  for  gas  with  $\rm
n(H_2)$$<$10$^4$\,cm$^{-3}$.}
\end{figure}

\clearpage

\newpage

\begin{figure}
% Figure 5
\epsscale{1.0}
\plotone{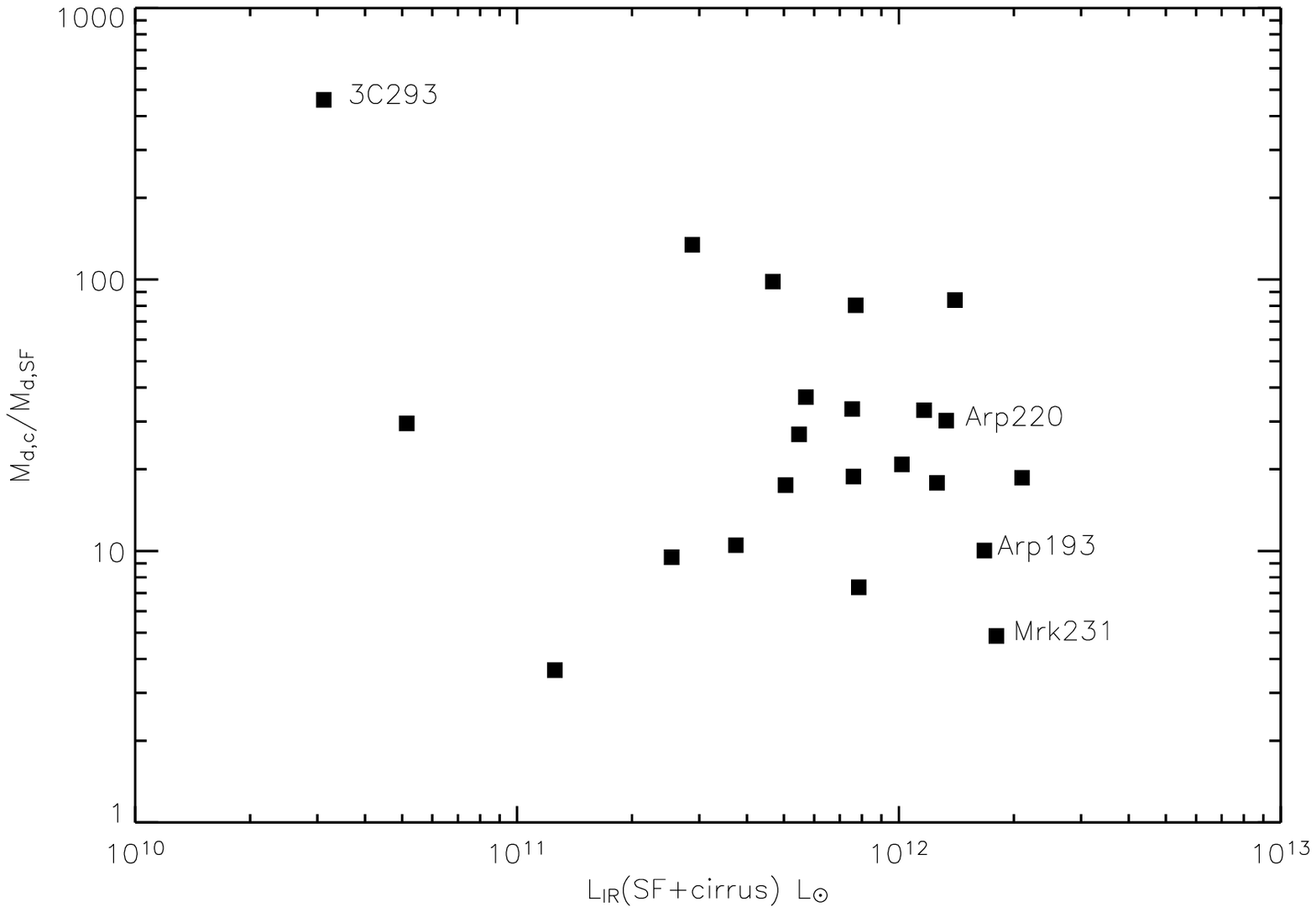}
\caption{The cold versus warm dust  mass for our sample estimated from
typical multi-component dust  SED fits (Eq.  7). In  all LIRGs most of
the  dust mass is  in the  cold phase  $\rm T_{d,c}$$\sim$(20--25)\,K,
irrespective  of IR  luminosity, even  in extreme  starbursts  such as
Arp\,220 and the starburst/QSO system Mrk\,231.}
\end{figure}

\newpage

\begin{figure}
%Figure 6
\epsscale{1.0}
\plotone{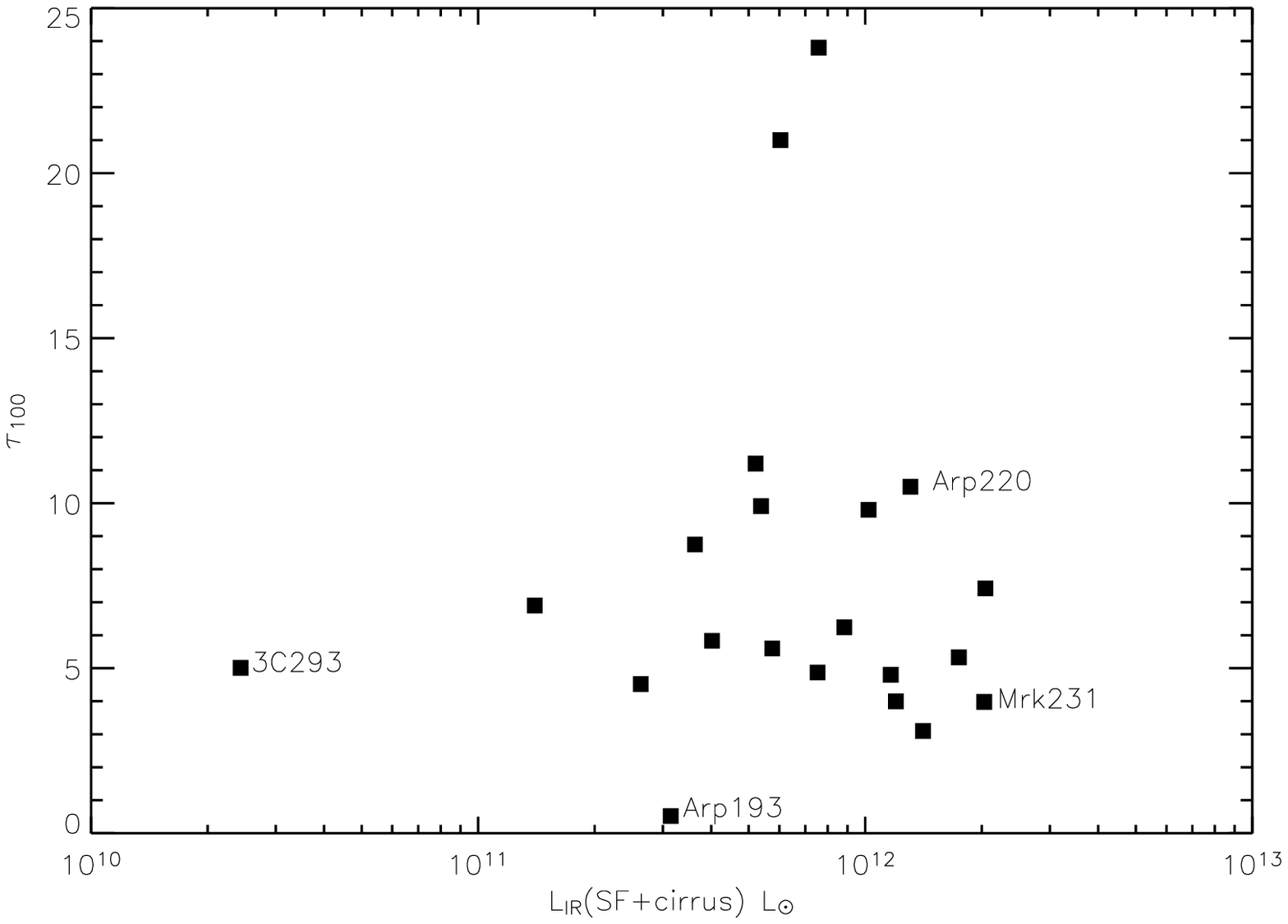}
\caption{The distribution  of dust  optical depths $\tau  _{100\mu m}$
deduced by assuming SEDs with non-negligible far-IR/submm dust optical
depths  (Equation 9) versus  IR luminosities  (with the  AGN component
subtracted  when possible:  Table  2 second  column).   The cold  dust
emission and  its asscociated masses  shown in Fig.~5 are  now equally
well reproduced by  high far-IR dust optical depths.}
\end{figure}

\newpage

\begin{figure}
%Figure 7
\epsscale{1.0}
\plotone{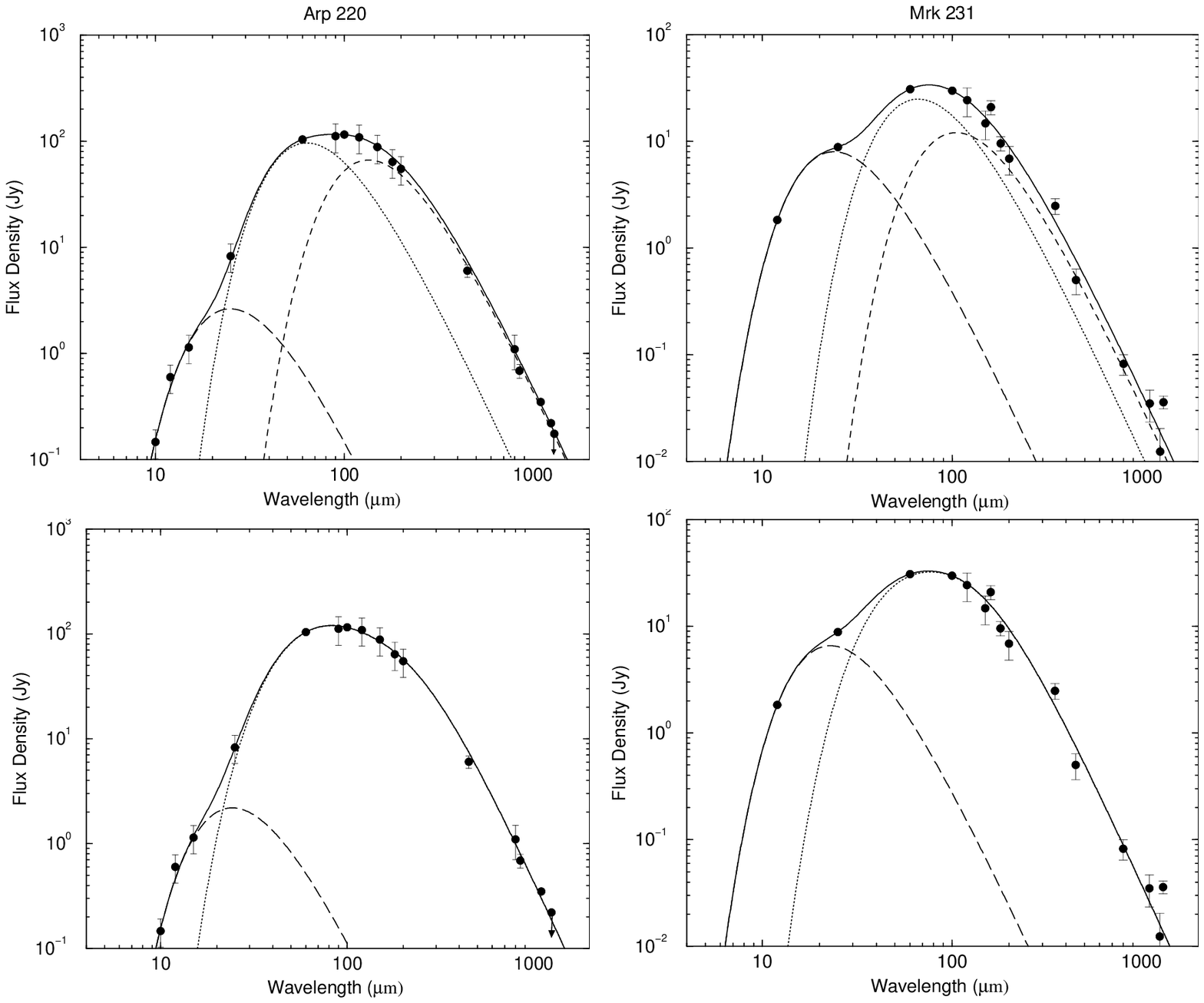}
\caption{Top panels:  Dust emission of the Arp\,220  and Mrk\,231 SEDs
 using a  classic 2-component fit  (warm dust from SF-sites,  and cold
 cirrus)  of its  non-AGN emission  (section 4,  Equation  7).  Bottom
 panels: Fits using the  assumption of significant dust optical depths
 (Equation   9).    In  the   first   case   large   ratios  of   $\rm
 M_{d,c}/M_{d,SF}$=30(Arp\,220),5(Mrk\,231) are deduced, while for the
 optically       thick       fits:       $\tau      _{\rm       100\mu
 m}$=10(Arp\,220),4(Mrk\,231).  An   AGN  hot dust  emission   fit  is
 included in all cases (small curve peaking at short wavelengths).}
\end{figure}

\newpage

\begin{figure}
% Figure 8
\epsscale{1.0}
\plotone{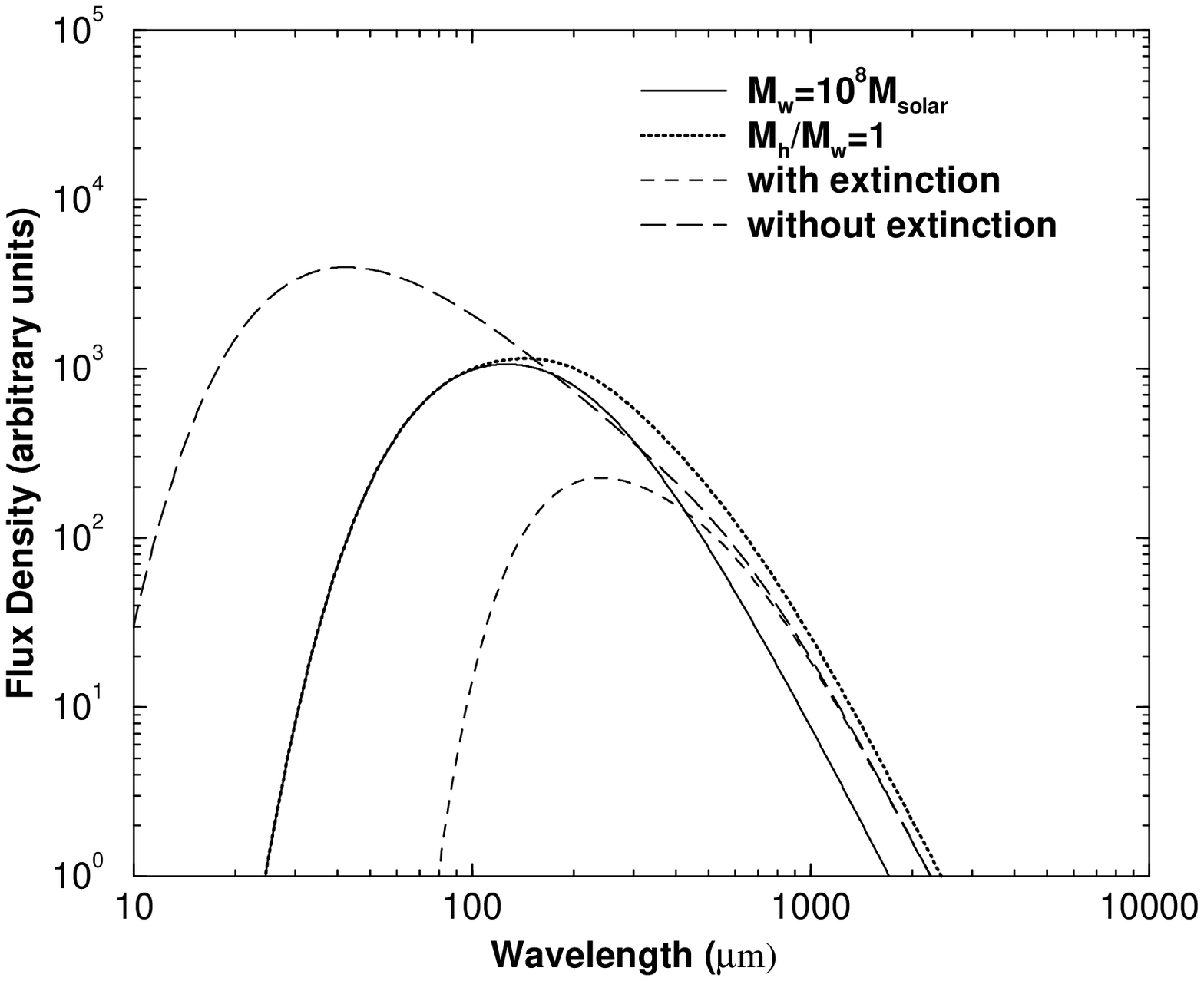}
\caption{The  dust SED  for  a  configuration akin  to  that found  in
 Arp\,220  (section  4.2,  Equation  10). The  unattanuated  hot  dust
 emission from compact starburst regions with significant dust optical
 depths at  submm wavelengths (long dashed line),  when surrounded and
 absorbed  by  a  more  extended  and cooler  dust  distribution  with
 significant optical depths at  IR wavelengths (its emission indicated
 by the  solid line), remains  virtually incospicuous, while  its peak
 shifts  at  longer wavelengths  (short-dashed  line).  The  remaining
 small contribution of the now attenuated hot component then occurs at
 long wavelengths where it can easily mask as cold dust emission.  The
 emergent {\it total}  dust SED remains dominated by  the outer colder
 dust component  (solid line)  up to the  wavelength beyond  which the
 attenuated  emission from the  hot dust  starts making  a significant
 contribution (marked by the continuation  of the total SED curve as a
 dotted line).}
\end{figure}

\newpage

\begin{figure}
%Figure 9
\centering
\epsscale{1.0}
\plotone{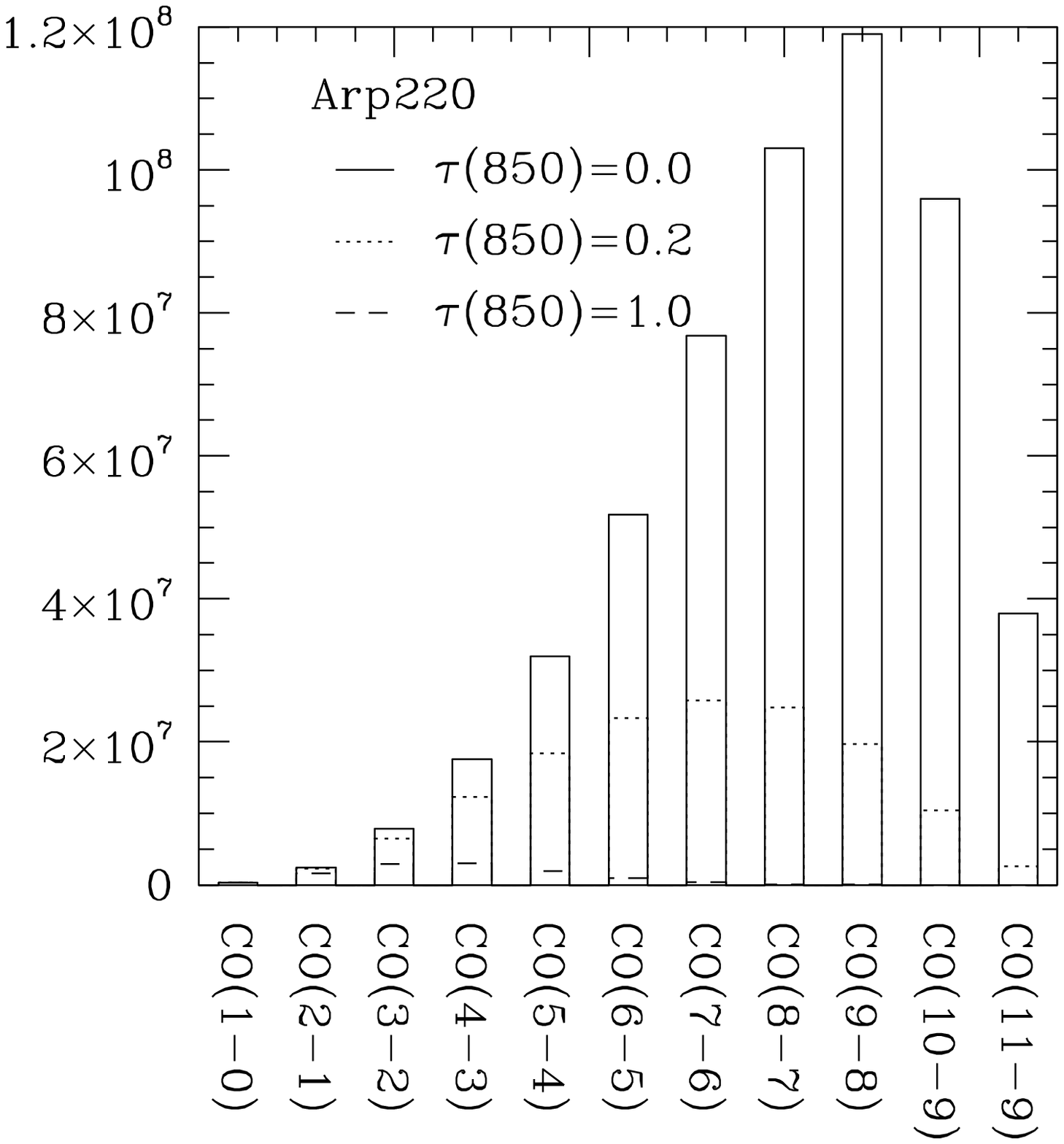}
\caption{The CO  SLED of the ULIRG  Arp\,220, and the  effects of dust
(see section 5). The  ``intrinsic'' ($\tau _{850}$=0) SLED is produced
from the LVG model of the HCN, HCO$^+$ and CS ratios reported by Greve
et al.   2009. It is obvious that  even for the small  values of $\tau
_{850}$ deduced for this system  (Papadopoulos et al. 2010, this work)
the effect on the CO SLED  beyond J=5--4 can be substantial, making it
appear much ``cooler'' than the intinsic one.}
\end{figure}

\newpage

\begin{figure}
%Figure 10
\centering
\epsscale{1.3}
\plottwo{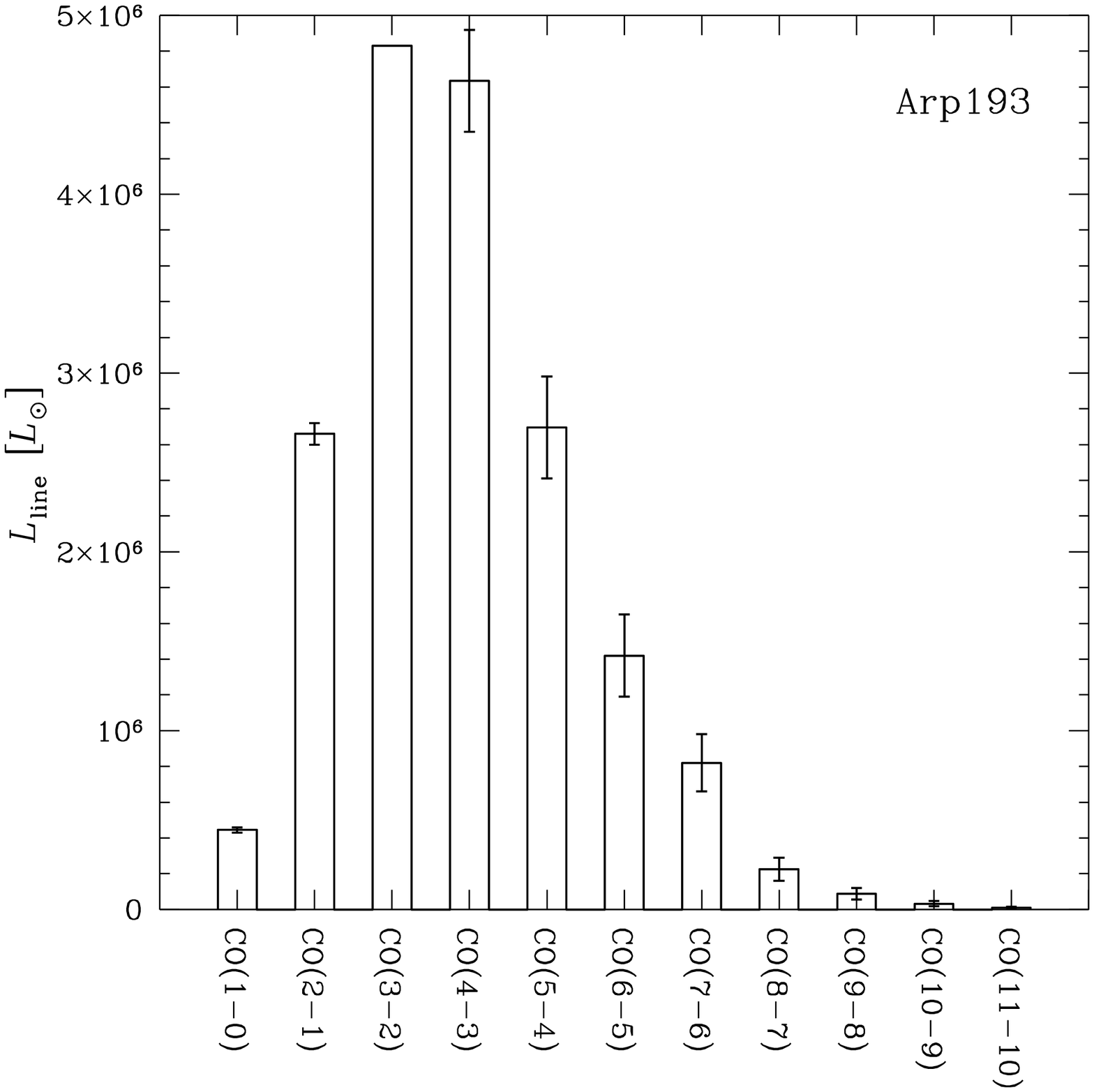}{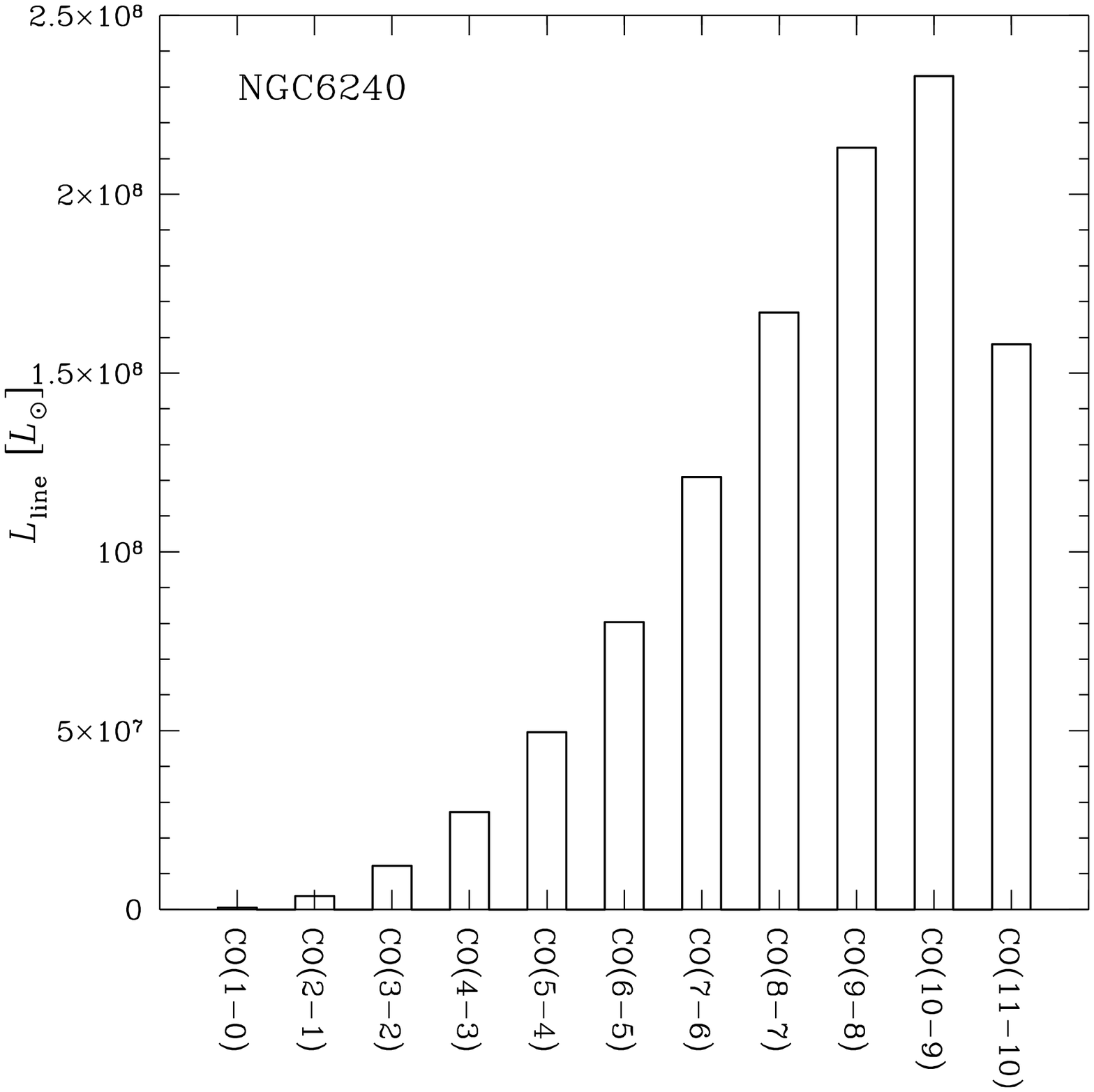}
\caption{The  CO SLEDs deduced  for the  HCN-bright (and  in principle
dense and star-forming) molecular gas phase in Arp\,193 and NGC\,6240,
two  LIRGs with  similar  total $\rm  M(H_2)$,  $\rm L^{(SF)}  _{IR}$,
star-formation efficiency  SFR/M(H$_2$), and HCN/CO  J=1--0 luminosity
ratios  (section 5.1).   Surprisingly the  physical conditions  of the
molecular  gas in  Arp\,193  are compatible  with  a total  lack of  a
massive dense (i.e. $\rm n(H_2)$$>$10$^4$\,cm$^{-3}$) gas phase, quite
unlike NGC\,6240 where  it dominates its total molecular  gas mass and
presumably  fuels its vigorous  star-formation.  A  wider range  of CO
luminosities  for  the  HCN-bright  phase  is  expected  for  Arp\,193
(denoted with  the bars,  except in the  observed CO  transitions) than
NGC\,6240.}
\end{figure}

\newpage

\begin{figure}
%Figure 11
\centering
\epsscale{1.0}
\plotone{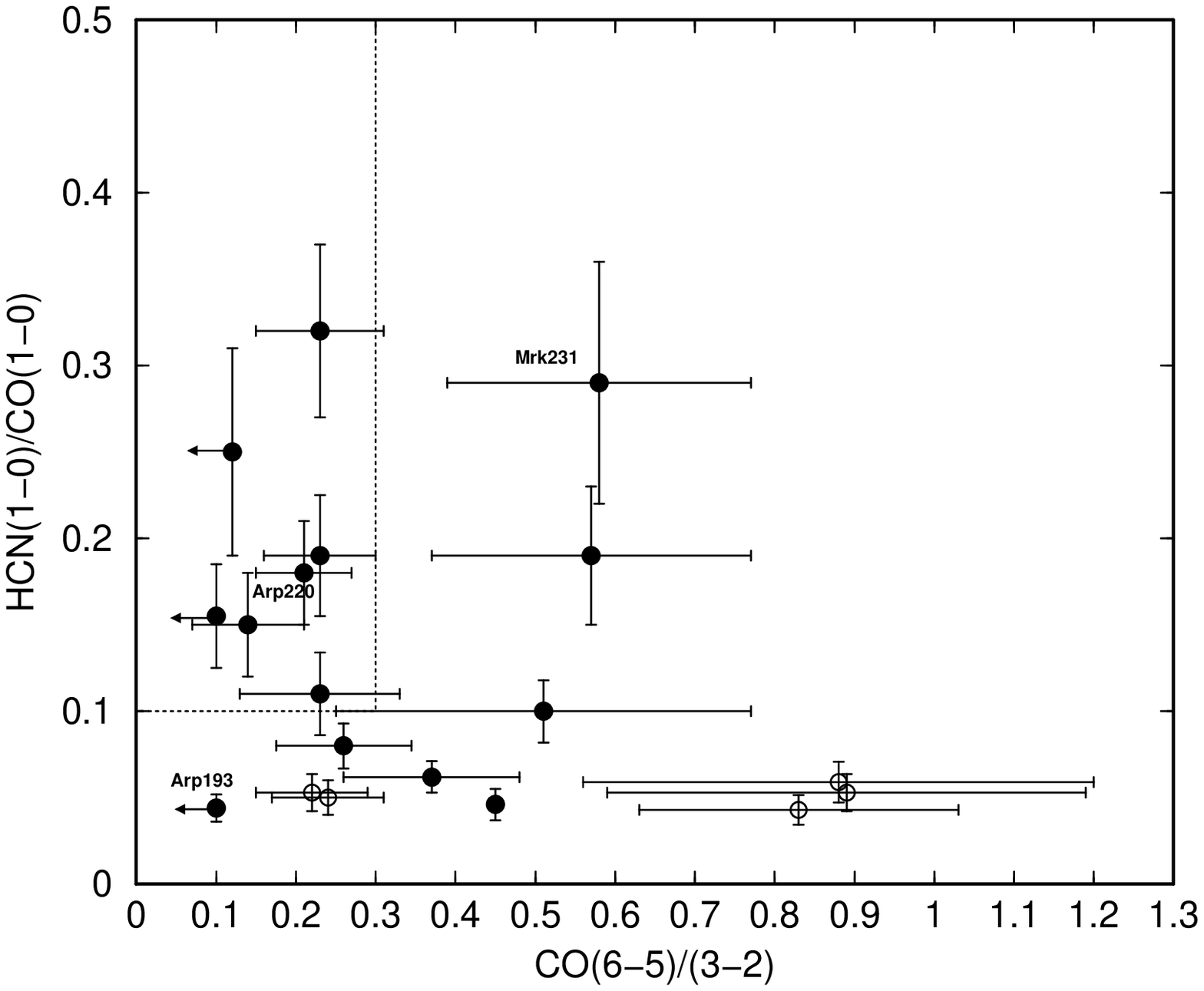}
\caption{The     CO(6--5)/(3--2)      ($\rm     R_{65/32}$)     versus
HCN(1--0)/CO(1--0)  ($\rm R_{HCN/CO}$)  brightness  temperature ratios
for our  sample (filled circles)  and nearby LIRGs (open  circles) for
which   these  ratios   are   available.   The   galaxies  with   $\rm
R_{HCN/CO}\ga 0.1$  but $\rm R_{65/32}\la 0.30$ (marked  by the dotted
window) are candidates for having  the high frequency part of their CO
SLEDs suppressed by dust (see section 5.2).}
\end{figure}

\newpage

\begin{figure}
\centering
\epsscale{1.0}
\plotone{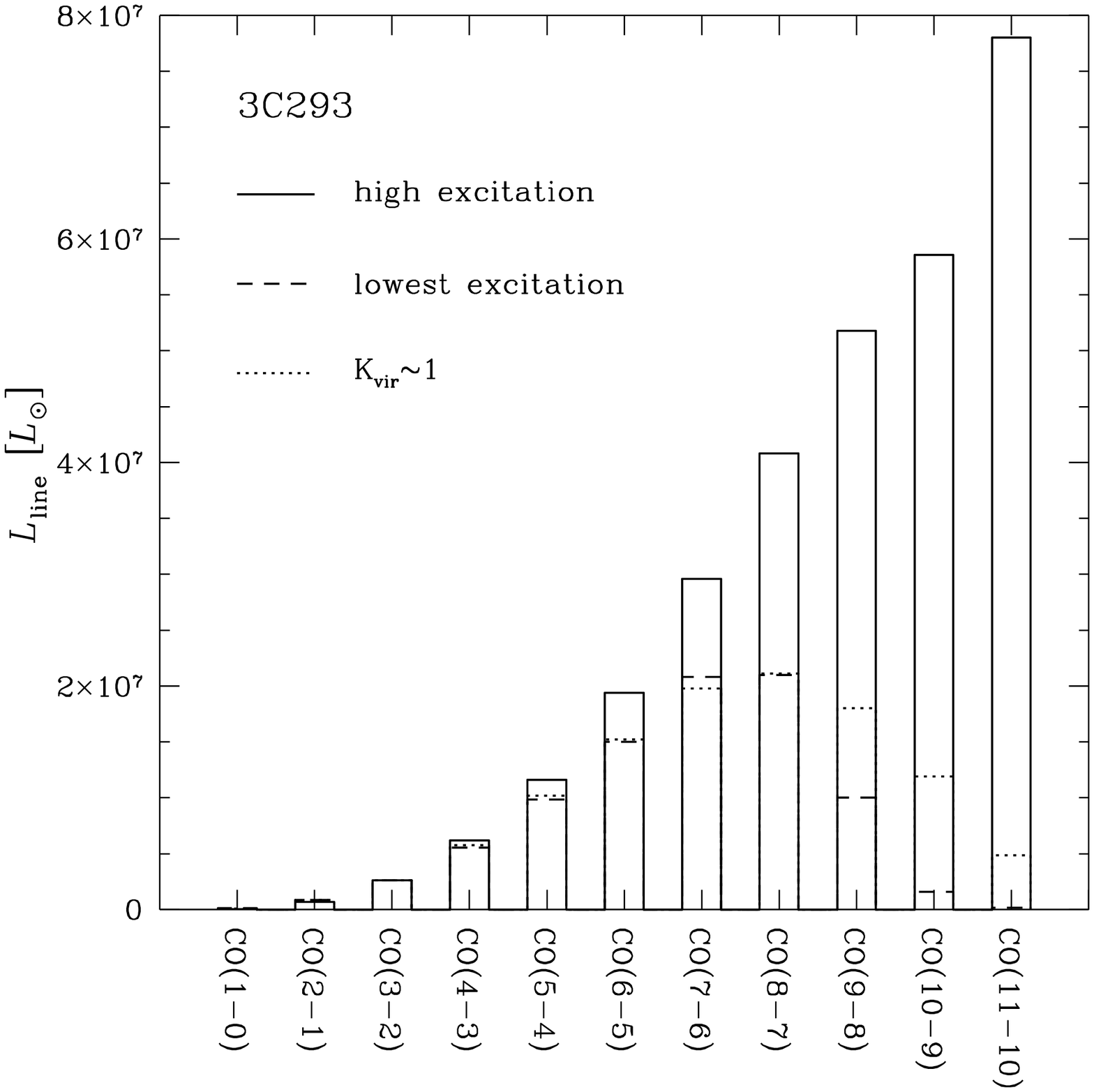}
\caption{The  CO  SLEDs possible  for  the  massive and  shock-excited
molecular gas  reservoir in the  FR\,II powerful radio  galaxy 3C\,293
where  a jet-ISM  interaction rather  than far-UV  photons  drives the
gaseous ISM energetics (see section 5.4).}
\end{figure}

\newpage

\begin{deluxetable}{lllll}
\tablecolumns{5}
\tablewidth{0pt}
\tablecaption{LIRGs: Observational parameters and CO J=6--5 line data}
\tablehead{\colhead{Galaxy\tablenotemark{a}} & \colhead{RA (J2000)\tablenotemark{b}} &
\colhead{Dec (J2000)\tablenotemark{b}} & \colhead{$\rm D_L(z)$\tablenotemark{c}} &
\colhead{ $\rm \int S_{CO(6-5)}$\,dV\tablenotemark{d}}}
\startdata
IRAS\,00057+4021 &  00 08 20.58 & $+$40 37 55.5  & 194.5(0.0445) & $\la 253$(1.29,G)\\
IRAS\,02483+4302 &  02 51 36.01 & $+$43 15 10.8  & 225.8(0.0514) & $401\pm 131$(1.27,G)\\
IRAS\,04232+1436 &  04 26 04.94 & $+$14 43 37.9  & 356.4(0.0795)& $\la 230 $(1.21,G)\\ 
IRAS\,05083+7936 (VII\,Zw\,031) & 05 16 46.51 & $+$79 40 12.5  & 239.0(0.0543) & $688\pm 207$(1.22,G)\\
IRAS\,05189--2524&  05 21 01.41 & $-$25 21 44.9  & 186.4(0.0427) & $585\pm 200$(1.20,G)\\
IRAS\,08572+3915(NW nucleus) &  09 00 25.4 & $+$39 03 54.0 & 256.9(0.0582) & $370\pm 145$(1.21,G)\\
IRAS\,09126+4432 (Arp\,55\,NE) &09 15 55.63&$+$44 19 58.0 &173.3(0.0398)&$\la 500$(1.9)\tablenotemark{e}\\
IRAS\,09320+6134 (UGC 05101) & 09 35 51.53 &$+$61 21 11.6 & 171.1(0.0393) & $\la 281$(1.35,G)\\
IRAS\,10173+0828 & 10 20 00.19 & $+$08 13 34.5 & 214.4(0.0489) & $\la 104$(1.27,G)\\
IRAS\,10565+2448 & 10 59 18.15 & $+$24 32 34.4 & 188.2 (0.0428) & $506\pm 148$(1.35,G)\\
IRAS\,11191+1200 (PG\,1119+120) & 11 21 47.12 & $+$11 44 18.3 & 219.4(0.0500) & $261\pm 82$(1.14,K$_c$)\\
IRAS\,12112+0305 (NE)\tablenotemark{f} & 12 13 46.00 & $+$02 48 41.0 & 324.4(0.0727) & $327\pm 143$(1.22,G)\\
 ``\ \ \ \ \  ``  \hspace*{2.2cm}(SW)\tablenotemark{f} & 12 13 45.90 & $+$02 48 39.0 & 324.4(0.0727)& $\la 82$(1.21,G)\\
IRAS\,12540+5708 (Mrk\,231)\tablenotemark{g}& 12 56 14.21 & $+$56 52 25.1&184.1(0.0422)& $1320\pm 400$(1.29,G)\\
IRAS\,13183+3423 (Arp\,193)     &  13 20 35.32 & $+$34 08 22.2 & 100.2(0.0233) & $\leq 420$ (1.28,G)\\
IRAS\,13428+5608 (Mrk\,273)     &  13 44 42.12 & $+$55 53 13.5 & 163.9(0.0378) & $447\pm 143$(1.45,G)\\
IRAS\,F13500+3141 (3C\,293)     &  13 52 17.77 & $+$31 26 46.1 & 194.9(0.0446) & $1086\pm 347$(1.29,K$_c$)\\
IRAS\,15107+0724 (Zw\,049.057)  &  15 13 13.07 & $+$07 13 32.0 & 55.0(0.0129)  & $904\pm 241$(1.14,K$_c$) \\
IRAS\,15237+2340 (Arp\,220)     &  15 34 57.24 & $+$23 30 11.2 & 78.0(0.0182)  & $3130\pm 810$(1.21,G)\\
IRAS\,16504+0228 (NGC\,6240)    &  16 52 59.05 & $+$02 24 05.8 & 104.6(0.0243) & $3321\pm 860$(1.21,G)\\ 
IRAS\,17208--0014               &  17 23 21.92 & $-$00 17 00.7 & 186.8(0.0428) & $340\pm 140$(1.17,G)\tablenotemark{h}\\
IRAS\,23007+0836 (NGC\,7469)    &  23 03 15.60 & $+$08 52 26.3 & 69.7(0.0163)  & $2355\pm 590$\tablenotemark{i} \\
IRAS\,23365+3604                &  23 39 01.25 & $+$36 21 08.4 & 285.6(0.0644) & $270\pm 95$(1.37,G)\tablenotemark{h}\\
\enddata
\tablenotetext{a}{IRAS source name and the most common alternative}
\tablenotetext{b}{The source coordinates used for the observations.}
\tablenotetext{c}{The luminosity distance in Mpc, and the corresponding redshift  in the parenthesis.}
\tablenotetext{d}{Velocity-integrated line flux density (in Jy\,km\,s$^{-1}$) with
                 the (value,type) of any applied corrections\\
\hspace*{0.5cm} reported in the parenthesis (see 2.2). All upper limits are at the 2$\sigma$ level.}
\tablenotetext{e}{Possible detection for the NE nucleus only,  highly uncertain because of a pointing
                 offset of \\ 
\hspace*{0.5cm}  $\sim $3.85$''$ (thus a beam-shift point-source correction factor of 1.90 has been applied).}
\tablenotetext{f}{Tentative detection of the bright NE nucleus ($\sim $75\% of the CO emission
                 of this double nuclei\\
              \hspace*{0.5cm} system (Evans et al. 2002). The SW nucleus was not detected, but this particular
                  observation\\
\hspace*{0.5cm}   may have suffered from a telescope pointing offset.}
\tablenotetext{g}{From Papadopoulos et al. 2007 but for $\rm n^*_a=0.25$, and a G correction applied (see 2.2.1)}
\tablenotetext{h}{Very tentative detections}
\tablenotetext{i}{Obtained with ZEUS at the CSO, van der Werf et al. 2010.}
\end{deluxetable}

\newpage

\begin{deluxetable}{lcccc}
\tablecolumns{5}
\tablewidth{0pc}
\tablecaption{LIRGs: IR luminosities, CO (6--5)/(3--2), HCN/CO J=1--0 line ratios}
\tablehead{\colhead{Galaxy} & \colhead{$\rm log\left(L_{IR}/L_{\odot }\right)$\tablenotemark{a}} &
\colhead{$\rm log(L^{(SF)} _{IR}/L_{\odot })$\tablenotemark{a}} &
\colhead{$\rm R_{65/32}$\tablenotemark{b}}  &
\colhead{$\rm R_{HCN/CO}$\tablenotemark{c}}}
\startdata
IRAS\,00057+4021 &  11.46 & 11.46   & $\leq 0.24$                     & \nodata  \\ 
IRAS\,02483+4302 &  11.70 & 11.54   & $0.84\pm 0.30$                  & \nodata  \\ 
IRAS\,04232+1436 &  11.88 & 11.87   & $\leq 0.17$                     & \nodata \\
VII\,Zw\,031     &  11.89 & 11.69   & $0.51\pm 0.26$                  & $0.10\pm 0.018$\\
IRAS\,05189--2524&  12.06 & 11.95   &$0.57\pm 0.20$                   &  $0.19\pm 0.040$\\
IRAS\,08572+3915 &  12.01 & 11.99   & $\geq 0.6$                      & \nodata \\
%Arp\,55 (SW+NE) & 11.67 & 11.35 & $\leq 0.17$                       & $0.046\pm 0.009$\\
%Arp\,55 (SW)    & \nodata& \nodata& $\la 0.088$                     & \nodata \\
Arp\,55 (NE nucleus)\tablenotemark{d}& 11.27 & 10.95  & $\leq 0.45$  & $0.046\pm 0.009$\\
IRAS\,09320+6134 & 11.88  & 11.59 & $\leq 0.12$                       & $ 0.25\pm 0.06$\\
IRAS\,10173+0828 & 11.74  & 11.62 & $\leq 0.20$                       & \nodata \\
IRAS\,10565+2448 & 11.89  & 11.81 & $0.23\pm 0.07$                    & $0.19\pm 0.035$\\
PG\,1119+120    & 10.71  & 10.46 & $3.1\pm 1.2$                      & \nodata \\
IRAS\,12112+0305(NE nucleus)& 12.22 & 12.08 & $0.14\pm 0.07$         & $0.15\pm 0.03$\\ 
Mrk\,231        & 12.25  & 12.14 & $0.58\pm 0.19$                    & $0.29\pm 0.07$\\
Arp\,193        & 11.57  & 11.34 & $\leq 0.10$                       & $0.044\pm 0.008$\\
Mrk\,273        & 12.08  & 11.92 & $0.23\pm 0.08$                    & $0.32\pm 0.05$\\
3C\,293         & 10.49  & 10.06 & $1.31\pm0.54 $                    & \nodata \\
Zw\,049.057     & 11.10  & 11.07 & $0.32\pm 0.096$                   & \nodata \\
Arp\,220        & 12.12  & 12.00 & $0.21\pm  0.06$                   & $0.18\pm 0.03$\\
NGC\,6240       & 11.76  & 11.58 & $0.26\pm 0.085$                   & $0.08\pm 0.013$\\
IRAS\,17208--0014&12.32  & 12.19 & $\leq 0.10$                       & $0.15\pm 0.03$\\
NGC\,7469       & 11.40  & 11.21 & $0.37\pm 0.11$                    & $0.062\pm 0.009$\\
IRAS\,23365+3604& 12.15  & 11.79 & $0.23\pm 0.10$                    & $0.11\pm 0.024$\\
\enddata
\tablenotetext{a}{The SF+cold cirrus (=L$_{\rm IR}$(total)-L$_{\rm IR}$(AGN))  and SF-related
 IR luminosities, obtained from their respective SED components (see section 4) integrated over
 $\lambda $=(8--1000)\,$\mu $m.}
\tablenotetext{b}{The $\rm R_{65/32}$=$\rm L^{'} _{CO(6-5)}/L^{'} _{CO(3-2)}$ 
ratio ($\rm L^{'}$ estimated from Equation 3), with CO J=3--2 data obtained from Papadopoulos
et al. 2010b (all upper limits are at the 2$\sigma $ level).}
\tablenotetext{c}{The $\rm R_{HCN/CO}$=$\rm L^{'} _{HCN(1-0)}/L^{'} _{CO(1-0)}$ 
ratio (cf. Equation 3). The HCN J=1--0 line fluxes are from Gao \& Solomon 2004 and
Gracia-Carpio et al. 2006. In cases of disagreement (Arp\,193,  IRAS\,23365+3604, 
and IRAS\,17208--0014) we adopted the values from the latter as more reliable.  The 
 CO J=1--0 fluxes used are  averages from literature data compiled in Papadopoulos
 et al. 2010b. For NGC\,6240 and Arp\,220 we used the values reported by
 Greve et al.~2009.}
\tablenotetext{d}{The $\rm R_{65/32}$ ratio is highly uncertain. We assumed that
$\sim 2/5$ of the total IR luminosity of this double-source system emanates from
the NE nucleus (same fraction as its CO J=3--2 luminosity (Leech et al. 2010).}
\end{deluxetable}

\newpage

\begin{deluxetable}{lccccc|ccc}
\tablecolumns{9}
\rotate
\tablewidth{0pc}
\tablecaption{Dust emission SED parameters, gas/dust ratios\tablenotemark{a}}
\tablehead{
\colhead{Galaxy\hspace{2.5cm} }& \colhead{$\rm T_{d,SF}, T_{d,c}$} & \colhead{$\rm R_{c/w}$\tablenotemark{b}} 
  &  \colhead{$\rm M_{d}$} & \colhead{$\rm M(H_2)$} & \colhead{$\rm M(H_2)/M_{d}$} 
  & \colhead{$\rm \tau _{\circ}, T_{d,\tau}(K)$} & \colhead{$\rm M_{d,\tau}$} 
  & \colhead{$\rm M(H_2)/M_{d,\tau}$} \\
  &    (K), (K)    &   & $\rm (10^8\,M_{\odot})$ & $\rm (10^9\,M_{\odot})$ &  &  & $\rm (10^7\,M_{\odot})$ &  }
\startdata
IRAS 00057+4021\tablenotemark{*} &  37, 6 &  134&  25.7&  4.1  & 1.5 &  8.7, 65 &  0.4& 1057\\
IRAS 02483+4302                  &  36, 20&  17 &  4.6 &  3.6  & 8   &  11,  49 &  2.0& 176\\
IRAS 04232+1436                  &  35, 9 &  33 & 23.7 &  10   & 4   &  24,  69 & 0.80& 1220 \\
VII\,Zw\,031                     &  51, 23&  80 &  3.6 &  11.5 & 32  &  21, 55  &  11 & 107\\
IR\,05189--2524                  &  60, 28&  33 &  1.1 &  3.9  & 35  & 4.8, 72  &  0.7& 557\\
%Arp\,55 (NE nucleus)&  53,  25&  98&  1.6&  6.8&  41&  5.8,  47.3&  1.84& 370\\
IRAS 08572+3915                  & 52, 19 &  21 &  8.2 & 1.6   & 1.9 & 10, 85   & 6.5 & 24\\
IRAS 09320+6134                  & 40, 25 &  19 &  2.9 & 4.8   &  16 &  5,  47  & 3.7 & 131\\
IRAS 10173+0828                  & 48, 23 &  27 &  1.5 & 6.0   &  39 &  10,  64 & 0.60& 993\\
IRAS 10565+2448                  & 41, 23 &  7  &  1.9 & 6.4   &  32 &  6.2, 61 & 1.3 & 516\\
PG\,1119+120                     &  41, 22&  29&  0.3& 0.5     &  16 &\nodata,\nodata & \nodata&\nodata \\
IRAS 12112+0305 (NE nucleus)     &  43, 25&  10&  3.3& 7.5     &  22 &  5.3, 60&  2.8& 272\\
Mrk\,231                         &  44, 28&  5 &  1.9& 7.0     &  36 &  4,  66 &  2.1& 336\\    
Arp\,193                         &  42, 27&  10&  0.7& 4.7     &  59 &  0.5, 38&  3.9& 119\\  
Mrk\,273                         &  52, 29&  18&  1.3& 5       &  38 &  4, 65  &  1.3& 385\\
3C\,293                          &  57, 22&  459& 0.2& 4.6     &  177&  5, 38&  0.3& 1398\\     
Zw\,049.057                      &  33, 17&  4 &  0.6& 0.9     &  13 &  7, 52&  0.4& 214\\    
Arp\,220                         &  46, 22&  30&  5.7& 6.0     &  11 &  10.5, 62& 1.8& 339\\      
NGC\,6240                        &  52, 26&  37&  8.0& 8       &  10 &  5.6,  59& 0.8& 976\\
IRAS 17208--0014                 &  46, 24&  19&  5.3& 13      &  24 &  7.4,  60& 3.2 & 416\\
NGC\,7469                        &  42, 27&  9 &  0.4& 3.5     &  72 &  4.5,  55&  0.6& 602\\
IRAS\,23365+3604                 &  60, 30&  84&  8.2& 7       &  8  &  3, 54   &  3.0 &223\\  
\enddata
\tablenotetext{a}{The full  dataset used in these SED fits will be presented
                  in an upcoming paper (Papadopoulos et al. 2010, in preperation).}
\tablenotetext{b}{The $\rm M_{d,c}/M_{d,SF}$ cold/warm dust mass ratio.}
\tablenotetext{*}{Very sparsely sampled dust SED, highly uncertain fit.}
\end{deluxetable}

%NOTE TO MANOLIS: For Arp\,55 assume that 60% of L, M's you find are associated with
% the M(H2), CO ratios etc, and for IRAS 12112+0305 that should be 75%

\newpage 

\begin{deluxetable}{lllll}
\tablecolumns{5}
\tablewidth{0pc}
\tablecaption{Physical conditions of the molecular gas in 3C\,293}
\tablehead{
\colhead{(4--3)/(3--2), (6--5)/(3--2)\tablenotemark{a}} & \colhead{\rm $\rm T_k$} & 
\colhead{$\rm n(H_2)$}& \colhead{$\rm K_{vir}$\tablenotemark{b}}& \colhead{$\chi ^2$ \tablenotemark{c}}\\
                               & \ \ \ (K)       & \ \ \ ($\rm cm^{-3}$)      &              &           }
\startdata
2.3, 1.3 (measured)            & 15--125    & (1--3)$\times 10^5$  & 0.027--0.154$^*$ & 2.21--1.6 \\
                               & 130--310   & $10^5$               & 15.4         & 1.6--1.23\\
1.27, 0.7 (minimum)            & 15--30&(0.3--3)$\times 10^5$ &0.027--0.154$^*$ & 1.16--0.74\\
                               & 40--55     & $3\times 10^3$       & 0.27$^*$        & 0.7\\
                               & 75--115    & $3\times 10^4$       & 27.4            & 0.63\\
                               & 135--255   & $10^4$               & 4.9--15.4       & 0.68--0.59\\
                               & 260--310   & $10^4$               & 49              & 0.58-0.51\\
2.3\,\tablenotemark{d}, 2.34 (maximum)&15--125 &(1--3)$\times 10^5$ &0.027--0.154$^*$ & 2.42-1.94 \\
                               & 130--310   & $10^5$               & 15.4            & 1.93--1.54 \\
1.27, 1.3 (medium)             & 15--125    & (1--3)$\times 10^5$     & 0.027--0.154$^*$& 1.75--0.89\\
                               & 130--310   & $10^5$               & 15.4            & 0.88-0.21\\
1.27, 2.3 (medium)             & 15--125    & (1--3)$\times 10^5$     & 0.027--0.154$^*$& 2.11-1.56\\
                               & 130-310    & $10^5$               & 15.4            & 1.56-1.15\\
2.3\,\tablenotemark{d}, 0.7 (medium)&15--30 &(0.3--3)$\times 10^5$ &0.027--0.154$^*$ &1.66--1.4\\
                               & 40--55     & $3\times 10^3$       & 0.27$^*$        & 1.36\\
                               & 80--130    & $3\times 10^4$       & 27.4            & 1.31--1.33\\
                               & 145--245   & $10^4$               & 4.9--15.4       & 1.35--1.30\\
                               & 250--310   & $10^4$               & 49              & 1.30--1.25\\
\enddata
\tablenotetext{a}{CO line luminosity ratios $\rm (L^{'} _x\pm \sigma_x)/(L^{'} _y\pm \sigma_y)$ (Eq. 3), 
with (type) denoting the $\rm L\pm \sigma$ values used, for example:
(measured)=($\rm L^{'} _{43}/L^{'} _{32}$, $\rm L^{'} _{65}/L^{'} _{32}$),\\
 (minimum)= [$\rm (L^{'} _{43}-\sigma _{43})/(L^{'} _{32}+\sigma_{32})$,
 $\rm (L^{'} _{65}-\sigma_{65})/(L^{'} _{32}+\sigma _{32})$] etc}
\tablenotetext{b}{The minimum virial parameter of the LVG solutions (Eq. 6 for $\alpha =1$).
 Values $\rm K_{vir}<1$ are forbidden (see 3.1) and are marked with an asterisk.}
\tablenotetext{c}{The $\rm \chi ^2 = \sum _{k}1/(\sigma_k)^2\left[R_{k}(LVG)-R_{k}(obs)\right]^2$ 
of the LVG fit of the observed line ratios $\rm R_{k}(obs)$, when a range of $\chi ^2$ values is
reported, the low values always correspond to the highest temperatures.}
\tablenotetext{d}{For $\rm R_{43/32}$, the observed value is so
               high that is effectively identical to the LTE, optically thin limit: $(4/3)^2=1.77$,
               thus we do not consider the $+\sigma$ case.} 
\tablenotetext{*}{Kinematically unphysical solutions (gas motions slower than virial).}
\end{deluxetable}


\newpage



\begin{thebibliography}{}
\bibitem[]{} Aalto S., Booth R. S., Black J. M.,\& Johansson L. E. B. 1995 A\&A, 300, 369{}{}
\bibitem[]{} Baker A. J., Tacconi L. J., Genzel R., Lehnert M. D., \& Lutz D. 2004, ApJ, 604, 125{}{}
\bibitem[]{} Bayet, E., Gerin, M., Phillips, T.G., \& Contursi, A. 2006, A\&A, 460, 467{}{}
\bibitem[]{} Bryant P. M., \& Scoville N. Z. 1999, AJ, 117, 2632{}{}
\bibitem[]{} Condon J. J., Huang Z.-P., Yin Q. F., \& Huan T. X. 1991, ApJ, 378, 65{}{}
\bibitem[]{} Condon J. J. 2001, in {\it Single-dish Radio Astronomy: Techniques and Applications}
             ASP Conf. Series, Vol. 278, pg. 160{}{}
\bibitem[]{} Cox P., \& Laureijs R. 1989, in {\it The Center of the Galaxy}: IAU  Symposium No 136
             Kluwer Academic Publishers, Dordrecht, p.121{}{}
\bibitem[]{} Daddi E., Dannerbauer H., Elbaz D.,  et al. 2008, ApJ, 673, L21{}{}
\bibitem[]{} Dannerbauer H., Daddi E., Riechers D. A. et al. 2009, ApJ,  698, L178{}{}
\bibitem[]{} De Breuck C., Downes D., Neri R., van Breugel W., Reuland M., Omont A., \&
             Ivison R. 2005, A\&A, 430, L1{}{} 
\bibitem[]{} Devereux N. A., \& Young J. S. 1990, ApJ, 359, 42{}{}
\bibitem[]{} Downes D., \& Solomon, P. M. 1998, ApJ, 507, 615{}{}
\bibitem[]{} Downes D., \& Eckart A. 2007, A\&A, 468, L57{}{}
\bibitem[]{} Elmegreen B. G., Klessen R. S., \&  Wilson C. D. 2008, ApJ, 681, 365{}{}
\bibitem[]{} Evans A. A., Sanders D. B., Surace J. A., \& Mazzarella J. M. 1999, ApJ, 511, 730{}{}
\bibitem[]{} Evans A. S. Surace J. A., \& Mazzarella J. M. 2000, ApJ, 529, L85{}{}
\bibitem[]{} Evans A. S., Frayer D. T., Surace J. A., \& Sanders D. B. 2001, AJ, 121, 3286{}{}
\bibitem[]{} Evans A. A., Mazzarella J. M., Surace J. A., \& Sanders D. B. 2002, ApJ, 580, 749{}{}
\bibitem[]{} Evans A. A., Mazzarella J. M., Surace J. A., Frayer D. T. Iwasawa K., \& Sanders D. B.
             2005, ApJS, 159, 197{}{}
\bibitem[]{} Emonts B. H. C., Morganti R., Tadhunter C. N., Oosterloo T. A., Holt J., 
             \& van der Hulst J. M. 2005, MNRAS, 362, 931{}{}
%\bibitem[]{} Eales S. A., Wynn-Williams C. G., \& Duncan W. D. 1989, ApJ, 339, 859{}{}
%\bibitem[]{} Emerson J. P., Clegg P. E., Gee G., et al. 1984, Nature, vol. 311, pg. 237{}{}
\bibitem[]{} Floyd D. J. E., Perlman E., Leahy J. P. et al. 2006, ApJ, 639, 23{}{}
\bibitem[]{} Gao Y. \& Solomon P. M. 2004, ApJ, 606, 271{}{}
\bibitem[]{} Goldsmith P. F. 2001, ApJ, 557, 736{}{}
\bibitem[]{} Graci\'a-Carpio J., Garc\'ia-Burillo S., Planesas P., \& Coline L. 2006, ApJ, 640, L135{}{}
\bibitem[]{} Graci\'a-Carpio J., Garc\'ia-Burillo S., Planesas P., Fuente A., \& Usero A.
             2008, A\&A, 479, 703{}{}
\bibitem[]{} Greve T. R.,  Bertoldi F., Smail Ian, et al. 2005, MNRAS, 359, 1165{}{} 
\bibitem[]{} Greve T. R., Papadopoulos, P. P., Gao Y., \& Radford S. J. E. 2009, ApJ, 692, 1432{}{}
\bibitem[]{} G\"usten R., Serabyn E., Kaseman C., et al. 1993, ApJ, 402, 537{}{}
\bibitem[]{} Hughes D. H., Serjeant S., Dunlop J., et al. 1998, Nature, 394, 241{}{}
\bibitem[]{} Iono D., Ho P. T., Yun M. S., Matsushita S., Peck A. B., Sakamoto K. 2004, ApJ, 616, L63{}{}
\bibitem[]{} Iono D., Wilson C. D., Yun M. S., et al. 2009, ApJ (in press, arXiv:0902.0121){}{}
\bibitem[]{} Jansen D. J.,  1995, PhD Thesis, Leiden Observatory{}{}
\bibitem[]{} Klamer I. J., Ekers R. D., Sadler E. M., \& Hunstead R. W. 2004, ApJ, 612, L97{}{}
\bibitem[]{} LeFloc'h E., Aussel H.,  Ilbert O. et al. 2009, ApJ, 703, 222{}{}
\bibitem[]{} Lisenfeld U., Isaak K. G., \& Hills R. 2000, MNRAS, 312, 433{}{}
\bibitem[]{} Loenen E. 2009, PhD Thesis, Univ. of Groeningen, pg. 113{}{}
\bibitem[]{} Luhman M. L., Satyapal S., Fischer J. et al. 1998, ApJ, 504, L11{}{}
\bibitem[]{} Luhman M. L., Satyapal S.,  Fischer J., et al. 2003, ApJ, 594, 758{}{}
\bibitem[]{} Mao R. Q., Henkel C., Schulz A., et al. 2000, A\&A, 358, 433{}{}
\bibitem[]{} Marrone D. P., Battat J., Bensch F. et al., 2004, ApJ, 612, 940{}{}
\bibitem[]{} Matsushita S., Iono D., Petitpas G., et al. 2009, ApJ, 693, 56{}{}
\bibitem[]{} Meijerink R., \&  Spaans M. 2005, A\&A, 436, 397{}{}
%\bibitem[]{} Meijerink R., \&  Spaans M. \& Israel F. P. 2006, ApJ, 650, L103{}{}
\bibitem[]{} Morganti R., Oosterloo T. A., Emonts B. H. C., van der Hulst J. M., 
             \& Tadhunter C. N. 2003, ApJ, 593, L69{}{}
\bibitem[]{} Omont A. 2007, Rep. Prog. Phys. 70, 1099{}{} 
\bibitem[]{} Ogle P.,  Antonucci R., Appleton P. N., \&  Whysong D. 2007, ApJ,  668, 699{}{}
\bibitem[]{} Papadopoulos P. P., \& Seaquist E. R. 1999, ApJ, 516, 114{}{}
\bibitem[]{} Papadopoulos P. P., \& Ivison, R. J. 2002, ApJ, 564, L9
\bibitem[]{} Papadopoulos P. P., Kovacs A., Evans A. S. \& Barthel P. 2008, A\&A, 483, 487{}{}
\bibitem[]{} Papadopoulos P. P., Isaak K. G., \& van der Werf P. P. 2007, ApJ, 668 815{}{}
\bibitem[]{} Papadopoulos P. P., Isaak K. G., \& van der Werf P. P. 2010, ApJ, 771, 757{}{}
\bibitem[]{} Papadopoulos P. P., Isaak K. G., van der Werf P. P., \& Xilouris E. M. 2010, (in preparation){}{}
\bibitem[]{} Pierce-Price D., Richer J. S., Greaves J. S. 2000, ApJ, 545, L121{}{}
\bibitem[]{} Planesas P., Mirabel I. F., \& Sanders D. B. 1991, ApJ, 370, 172{}{}
\bibitem[]{} Rodr\'igez-Fern\'andez N. J. Mart\'in-Pintado J., Fuente A., et al. 2001, A\&A, 365, 174{}{}
\bibitem[]{} Sakamoto K., Scoville N. Z., Yun M. S., Crosas M., Genzel R., 
             \& Tacconi L. J. 1999, ApJ, 514, 68{}{}
\bibitem[]{} Sakamoto K., Wang J., Wiedner M. C., et al. 2008, ApJ, 684, 957{}{}
\bibitem[]{} Sanders D. B., Scoville N. Z., Sargent A. I., \& Soifer B. T. 1988, ApJ, 324, L55{}{}
\bibitem[]{} Sanders D. B., \& Mirabel I. F. 1989, ApJ, 1989, 340, L53{}{}
\bibitem[]{} Sanders D. B., Scoville N. Z., \& Soifer B. T. 1991, ApJ, 370, 15 8{}{}
\bibitem[]{} Sanders D. B., \& Mirabel I. F. 1996, ARA\&A, 34, 749{}{}
\bibitem[]{} Sanders D. B., Scoville N. Z., \& Soifer B. T. 1991, ApJ, 370, 158{}{}
\bibitem[]{} Sanders D. B., Mazzarella J. M., Kim D.-C., Surace J. A., \& Soifer B. T. 2003, 
             AJ, 126, 1607{}{}
\bibitem[]{} Sanders D. B., \& Ishida C. M., 2004, in {\it The Neutral ISM in Starburst Galaxies}
             ASP Conference Series, Vol. 320, pg. 230{}{}
\bibitem[]{} Scoville N. Z., Yun M., Bryant P. M. 1997, ApJ, 484, 702{}{}
\bibitem[]{} Scoville N. Z., 2004,  in {\it The Neutral ISM in Starburst Galaxies}
             ASP Conference Series, Vol. 320, pg. 253{}{}
\bibitem[]{} Schleicher D. R. G., Spaans M.,\& Klessen R. S. 2010, A\&A, (in press, arXiv:1001.2118){}{}
\bibitem[]{} Sodroski T. J., Bennett C., Boggess N. et al. 1994, ApJ, 428, 638{}{}
\bibitem[]{} Solomon P. M., Radford S. J. E. \& Downes D. 1990, ApJ, 348, L53{}{}
\bibitem[]{} Solomon P. M., Downes D., \& Radford S. J. E. 1992, ApJ, 387, L55{}{}
\bibitem[]{} Solomon P. M., Downes D., Radford S. J. E., \& Barrett J. W. 1997, ApJ, 478, 144{}{}
\bibitem[]{} Solomon P. M., \& Vanden Bout P. A., 2005, ARA\&A, 43,  677{}{}
\bibitem[]{} Soifer B. T., Neugebauer G., Helou G. et al. 1984, ApJ, 283, L1{}
\bibitem[]{} Soifer B. T., Sanders D. B., Madore B. F., et al. 1987, ApJ, 320, 238{}{}
\bibitem[]{} Soifer B. T., Boehmer L., Neugebauer G., \& Sanders D. B. 1989, AJ, 98, 766{}{}
\bibitem[]{} Smail  I., Ivison R. J., \& Blain A. 1997, ApJ, 490, L5{}{}
\bibitem[]{} Tacconi L. J., Neri R., Chapman S. C., Genzel R., Smail I., Ivison R. J., 
             Bertoldi F., Blain A., Cox P., Greve T. R., \& Omont A. 2006, ApJ, 640, 228{}{}
\bibitem[]{} Thomas H. C., Clemens M. S., Alexander P., Green D. A., Eales S., \& Dunne L. 2001, 
             in {\it Gas and Galaxy Evolution}, ASP Conference Proceedings, Vol. 240. p. 224
\bibitem[]{} Tinney C. G., Scoville N. Z., Sanders D. B., \& Soifer B. T. 1990, ApJ. 362, 473{}{}
\bibitem[]{} Walter F., Carilli C., Bertoldi F. et al. 2004, ApJ, 615, L17{}{} 
\bibitem[]{} Walter F., \& Carilli C. 2008, in {\it Science with the Atacama Large Millimeter Array}
             Ap\&SS, 313, 313{}{}
\bibitem[]{} Wang Z., Scoville N. Z., \& Sanders D. B. 1991, ApJ, 368, 112{}{}
%\bibitem[]{} Weiss A., Walter F., \& Scoville N. Z. 2004, in 
%            \it The Neutral ISM in Starburst Galaxies} ASP Conference Series, Vol. 320, pg. 142{}{}
\bibitem[]{} Weiss A., Downes D., Walter F., \& Henkel C. 2007a, in {\it From Z-machines to ALMA:
             (Sub)Millimeter Spectroscopy of Galaxies}, ASP Conference Series, Vol 375, pg. 25{}{} 
\bibitem[]{} Weiss A., Downes D., Neri R., Walter F., Henkel C., Wilner D. J., Wagg J., Wiklind T.
             2007b, A\&A, 467, 955{}{}
%\bibitem[]{} Wild W., Harris A. I., \& Eckart A. et al. 1992, A\&A, 265, 447{}{}
\bibitem[]{} Wilson C. D., Petitpas G. R., Iono D., et al. 2008, ApJS, 178, 189{}{}
\bibitem[]{} Wright E. L. 2006, PASP, 118, 1711{}{}
\bibitem[]{} Wu J., Evans N. J. II.,  Gao Yu, et al. 2005, ApJ, 635, L173{}{}
\bibitem[]{} Yao L., Seaquist E. R., Kuno N., \& Dunne L. 2003, ApJ, 588, 771{}{}
%\bibitem[]{} Young J. S.,  Kenney J., Lord S. D., Schloerb F. P. 1984, ApJ, 287, L65{}{}
\end{thebibliography}
\end{document}